\documentclass[twocolumn,prl,nofootinbib,superscriptaddress,amssymb,amsmath,amsmath,floatfix]{revtex4-1}
\usepackage{graphicx}
\usepackage{epstopdf}
\usepackage{comment}
\usepackage{bm}
\usepackage{wrapfig}
\usepackage{times}
\usepackage[colorlinks=true,linkcolor=blue,urlcolor=blue,citecolor=blue]{hyperref}
\usepackage{float}
\usepackage{dsfont}
\usepackage{amssymb}
\usepackage{braket}
\usepackage[utf8]{inputenc}

\begin{document}
\title{Ultracold high-spin $\Sigma$-state polar molecules for new physics searches}

\author{Alessio Ciamei $^\ddagger$}
\email{ciamei@lens.unifi.it}
\affiliation{Istituto Nazionale di Ottica del Consiglio Nazionale delle Ricerche (CNR-INO), 50019 Sesto Fiorentino, Italy}
\affiliation{\mbox{European Laboratory for Non-Linear Spectroscopy (LENS), Universit\`{a} di Firenze, 50019 Sesto Fiorentino, Italy}}

\author{Adam Koza $^\ddagger$}
\author{Marcin Gronowski}
\affiliation{Faculty of Physics, University of Warsaw, Pasteura 5, 02-093 Warsaw, Poland}
\author{Micha{\l} Tomza}
\email{michal.tomza@fuw.edu.pl}
\affiliation{Faculty of Physics, University of Warsaw, Pasteura 5, 02-093 Warsaw, Poland}
\def\thefootnote{$^\ddagger$}\footnotetext{These authors contributed equally to this work}\def\thefootnote{\arabic{footnote}}

\date{\today}

\begin{abstract}
We propose high-spin $\Sigma$-state polar molecules assembled from ultracold atoms to probe charge-parity violating physics beyond the Standard Model.
We identify YbCr as a prime candidate to search for the electric dipole moment of the electron. We show that the combination of relativistic ytterbium and high-spin chromium, amenable to magneto-association, leads to molecules with easy-to-polarize parity doublets and large intramolecular electric fields. Based on \textit{ab initio} results for molecular constants, we predict a sensitivity of $\delta d_{\textrm{e}}= ( 6 \times 10^{-31} / \sqrt{n_{\mathrm{day}}})\,e\,\mathrm{cm}$ via standard spin-precession measurements, we assess the experimental feasibility, and discuss potential extensions to more advanced quantum control as well as searches of the nuclear magnetic quadrupole moment. This work paves the way to next-generation searches for new physics with ultracold molecules in both the leptonic and hadronic sectors.
\end{abstract}

\maketitle

%==============================================================================
\paragraph*{Introduction} --- Detection of a non-zero electric dipole moment (EDM) of a fundamental particle with today's experimental sensitivities would unequivocally signal charge-parity violation (CPV) beyond the Standard Model (SM)
%originating from beyond Standard Model (SM) physics 
\cite{RevModPhys.90.025008,RevModPhys.91.015001}, and possibly explain matter-antimatter asymmetry in Nature \cite{RevModPhys.76.1,Canetti_2012,Sakharov1967,Patrignani_2016}. As for the electron EDM (eEDM), atomic-molecular-optics experiments exploit paramagnetic polar molecules, which, thanks to their large internal electric field, offer extreme sensitivity. The most stringent limit $|d_{\textrm{e}}|<4.1\times10^{-30}~e$\,cm was obtained on trapped HfF$^+$ ions \cite{doi:10.1126/science.adg4084}, improving by a factor of 2.5 on the previous bound set by a cryogenic molecular beam of metastable ThO \cite{Andreev2018}. On the one hand, assuming maximal symmetry violation, these experiments already restrict the mass of conjectured beyond-SM particles leading to CPV above the $4-40$~TeV range \cite{Andreev2018,doi:10.1126/science.adg4084}, which is well beyond the direct reach of current particle colliders. On the other hand, releasing this assumption but focusing on a certain mass range, they might hint, in combination with particle colliders, to very small CPV phases and the possibility of spontaneous symmetry breaking \cite{eEDMWhitePaper}.

While experimental searches have so far used relatively hot molecules, orders of magnitude of improvement in sensitivity is expected by more advanced quantum control at the ultracold temperatures achievable in atomic, molecular and optical physics experiments \cite{eEDMWhitePaper,DeMille2024,PhysRevLett.132.190001}. Over the last two decades, these platforms have demonstrated exquisite control over neutral atoms, down to the single quantum state and even single particle control, leading on the one hand to quantum simulation and computing, and on the other to high-precision measurements. More recently, use of optical lattices and tweezers has opened the way to entanglement resources for quantum computing \cite{Saffman_2016,Henriet2020quantumcomputing,Browaeys2020,Kaufman2021,Daley2022,doi:10.1073/pnas.2304294120,TakahashiArxiv} and metrology \cite{Ludlow2019,RevModPhys.90.035005,Malia2022,Bornet2023,Eckner2023,Robinson2024}. Extension of this experimental toolbox to eEDM-sensitive molecular species would allow longer coherence times for improved sensitivity in standard-quantum limit (SQL) spin-precession measurements, as well as application of recently proposed entanglement-enhanced metrology protocols \cite{PhysRevLett.131.193602}. However, the strategy to get there is unclear \cite{DeMille2024}. 

Despite tremendous progress to directly laser cool and trap heavy, eEDM-sensitive species, including diatomic radicals \cite{PhysRevLett.120.123201,Fitch_2021,Kogel_2021} and even polyatomic symmetric top molecules \cite{doi:10.1126/science.adg8155,PhysRevA.106.052811}, the only strategy so far able to deliver molecular gases at high phase-space density relies on the assembly from pre-cooled atoms via magnetic Feshbach resonances (FRs). In this case, colliding atom pairs are associated with a magnetic-field sweep across a FR \cite{RevModPhys.78.1311,RevModPhys.82.1225} and later transferred to the ro-vibrational ground state via stimulated Raman adiabatic passage (STIRAP) \cite{RevModPhys.89.015006}. This has been experimentally demonstrated on a number of bialkalis (AA) and has led to the creation of molecular quantum gases \cite{Valtolina2020,Duda2023,Bigagli2024}, and their application to many-body physics, ultracold chemistry and quantum computation \cite{Kaufman2021,Karman2024,Cornish2024,holland}. Nonetheless, these species feature a singlet electronic ground state without unpaired electron spins, and are thus eEDM insensitive. A potential improvement beyond standard AAs, albeit with relatively poor eEDM sensitivities, was identified in alkali-alkaline-earth (AAE) compounds \cite{PhysRevA.80.042508}. Despite the experimental discovery of magnetic FRs in these systems \cite{Barbé2018}, their ultranarrow character has so far hindered magnetoassociation. More recently, diatomics isovalent to AAE have been proposed, notably YbAg \cite{PhysRevLett.125.153201,PoletJCP24} and RaAg \cite{Fleig_2021}, which thanks to the high electron affinity of Ag feature a high sensitivity. However, this strategy presents several challenges both for the preparation of the ultracold atomic mixture and for molecule formation due to FRs of similar character as those in AAEs.

In this Letter, we propose a comprehensive strategy to realize and probe ultracold molecular gases of eEDM sensitive molecules assembled from orbitally-isotropic transition-metal chromium ($^7S_3$) and closed-shell ytterbium ($^1S_0$). Thanks to the high electron spin of Cr and the heavy, highly-relativistic Yb core, this species meets all experimental  requirements, regarding both molecule production and internal structure. 
While we provide the projected sensitivity for spin-precession experiments within the SQL, extensions to more advanced protocols as well as searches for the nuclear quadrupole magnetic moment are discussed.
Finally, this strategy can be applied to other isovalent species, which will further allow probing of hadronic CPV effects.

\paragraph*{Experimental requirements} --- How the choice of the molecular species is crucial for eEDM searches is illustrated by considering the quantum-projection noise limited sensitivity valid for SQL spin-precession measurements \cite{Khriplovich1997,PhDVutha,Baron_2017}:
\begin{equation}
\label{eq:SQLsensitivity}
\delta d_{\textrm{e}} = \frac{\hslash}{2 E_{\mathrm{int}} |\langle\hat{\Sigma} \rangle| \tau \sqrt{N}}.
\end{equation}
Here $E_{\mathrm{int}}$ is the (state-dependent) internal electric field of the molecule aligned along its internuclear axis $\bold{n}$, $|\langle\hat{ \Sigma} \rangle|=|\langle \bold{S}\cdot\bold{n} \rangle|$ is the (state-dependent) projection of the total electron spin $\bold{S}$ onto $\bold{n}$, $\tau$ is the coherent interrogation time, and $N$ is the total number of detected particles. The effective internal electric field is conveniently defined as $E_{\mathrm{eff}}=E_{\mathrm{int}} |\langle\hat{ \Sigma} \rangle|$ \cite{Baron_2017}. Assuming a long $\tau$ by virtue of the attainable ultracold temperatures, three main experimental requirements need to be considered: (i) large molecule number $N$, which implies efficient molecule production; (ii) strong $E_{\mathrm{int}}$, which requires at least one heavy nucleus and one unpaired electron in an $s-p$ hybridized orbital; (iii) high $|\langle\hat{ \Sigma} \rangle|$ can be easily achieved in species with high electric polarizability featuring parity-doubled states. As we will describe later, the existence of such states also allows for robust rejection of systematic errors in the experiment and implementation of more advanced protocols.

\paragraph*{$\Omega$-doubling} --- The Yb$^{\delta+}$Cr$^{\delta-}$ bond between Yb ($^1S_0$) and Cr ($^7S_3$) results in a single ground state of the $^7\Sigma^+$ symmetry. The opportunity to obtain a Hund's case (a) molecule in a $\Sigma$ state is non-trivial, and arises from the presence of strong second-order spin-orbit (and spin-spin) interaction, rather than the typical first-order spin-orbit for $\Lambda>0$ states, with $\Lambda$ being the electronic orbital angular momentum projection along $\mathbf{n}$. Remarkably, this allows us to obtain Hund's case (a) diatomics from $S$-term ground-state atoms, and contrarily to a common misconception, endows them with parity doublets. The corresponding Coriolis-mixing-induced splitting decreases fast with electron-spin projection $|\Sigma|$ due to the higher perturbation order and makes large $|\Sigma|$ states highly polarizable. Following literature, we introduce the quantum number $\Omega=\Lambda+\Sigma$, here reducing to $\Omega=\Sigma$, and refer to such states as $\Omega$-like doublets. The combination of heavy Yb and high-spin Cr ensures the occurrence of this scenario. Introducing the total angular momentum $\mathbf{J}$ excluding nuclear spins, the effective rotational Hamiltonian without external fields can be constructed as follows
%in analogy with the one of SeO ($X^3\Sigma^-$), and CrN ($^4\Sigma^-$) 
\cite{Brown_Carrington_2003}:
\begin{multline}
\label{eq:Heff0}
\mathcal{H}_0 = B_0 T^1(\mathbf{J})\cdot T^1(\mathbf{J})+(\gamma_0-2B_0)T^1(\mathbf{J})\cdot T^1(\mathbf{S})+\\ (2\sqrt{6}/3)\lambda_0 T^2_0(\mathbf{S})+(B_0-\gamma_0)T^1(\mathbf{S})\cdot T^1(\mathbf{S}),
\end{multline}
where the rotational constant $B_0=0.041~\mathrm{cm}^{-1}$, the spin-spin coupling parameter  $\lambda_0=0.16~\mathrm{cm}^{-1}$, and the electron spin-rotation coupling constant $\gamma_0=9.7\times10^{-4}~\mathrm{cm}^{-1}$ are calculated with state-of-the-art \textit{ab initio} methods \cite{SupplInfo}. Being $\lambda_0/B_0 \gg 1$, the electron spin locks to the internuclear axis with projection $\Omega$ and makes ground-state YbCr a good Hund's case (a) molecule for high $\Omega$ and low $J$. Hence, the zero-field eigenstates are labelled in terms of the quantum numbers $|\Omega,J,M_J,e(f)\rangle$, where $e(f)$ denotes $\pm 1$ parity according to spectroscopic notation \cite{Brown_Carrington_2003}.

%================================================================================

\begin{figure}[t]
\centering
  \includegraphics[width=\columnwidth]{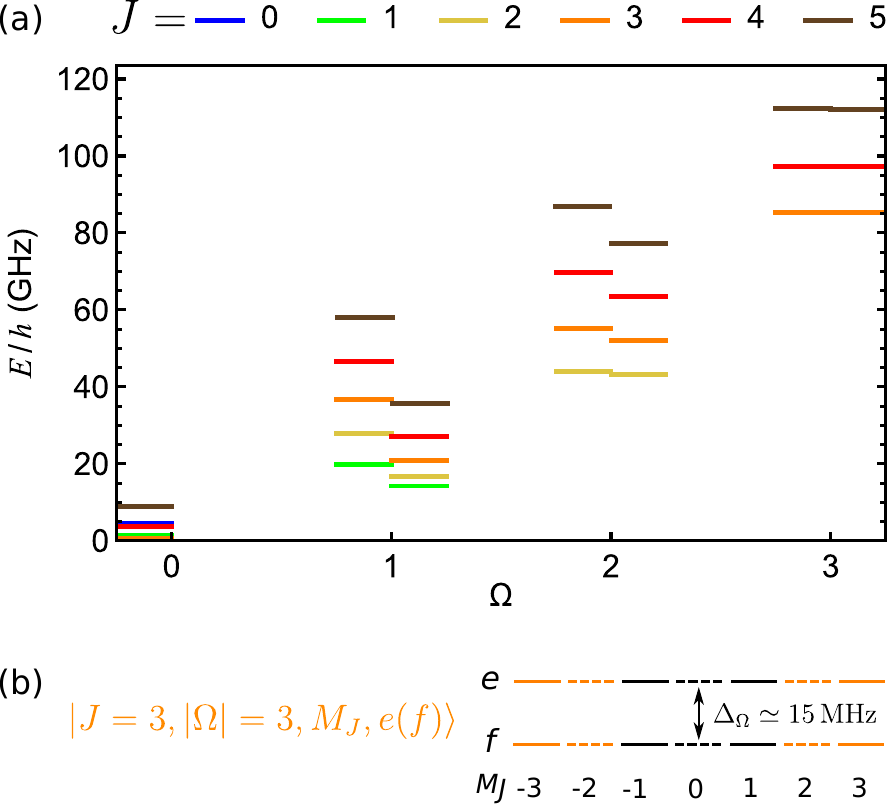}
   \vskip -5 pt
  \caption{Rotational spectrum of YbCr molecules. (a) Simulated rotational spectrum for $J\leq5$ of the ground vibrational level showing the quantization of the electronic spin onto the internuclear axis (Hund's case a). (b) Zoom onto the $J=3$, $\Omega=3$ parity doublet.}
 \vskip -5 pt
  \label{fgr:Spectrum}
\end{figure}

\begin{figure}[t]
\centering
  \includegraphics[width=\columnwidth]{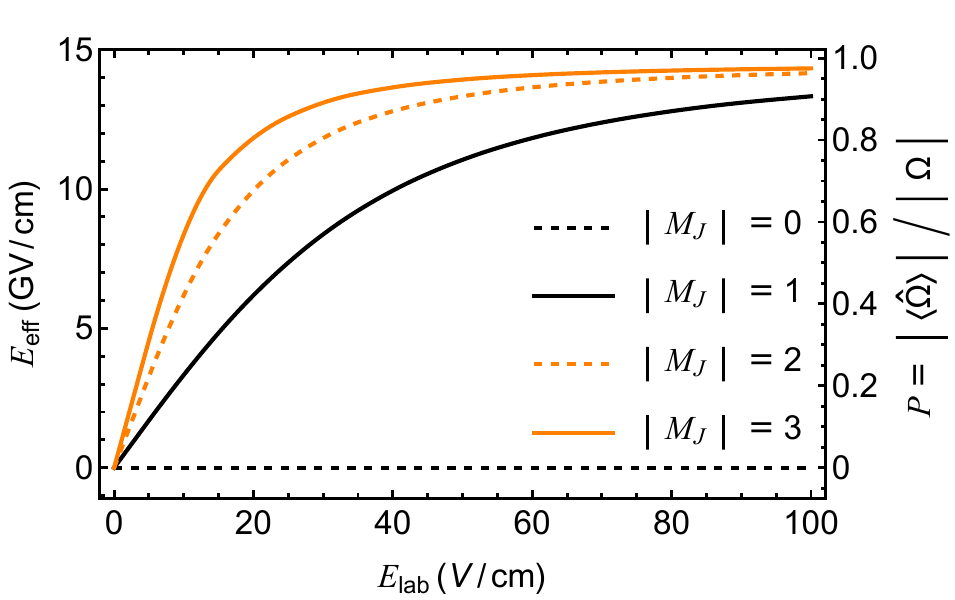}
   \vskip -5 pt
  \caption{Effective electric field $E_{\mathrm{eff}}$ and polarization $P$ induced by an external static electric field $E_{\mathrm{lab}}$ for different $|M_J|$ of $^7\Sigma^+_3\,(J=3)$.}
 \vskip -5 pt
  \label{fgr:Eeff}
\end{figure}

While appearance of Hund's case (a) rotational spectra in $\Sigma$ states was understood long ago \cite{doi:10.1139/p67-209}, and $\Omega$-doubling was more recently observed in SeO($X^3\Sigma^-$) and CrN($^4\Sigma^-$) \cite{Brown_Carrington_2003}, the possibility to seize this molecular feature has so far been overlooked in the ultracold molecule community. In passing, we note that to the best of the authors' knowledge no Hund's case (a) $^7\Sigma$ molecule has yet been recorded, as known ground-state diatomics of this symmetry, interesting to astrophysicists, fall into Hund's case (b) (MnH \cite{Halfen_2008}, MnCl \cite{10.1063/1.1824036}). The $^7\Sigma^+_3\,(J=3)$ state of YbCr features an $\Omega$-doublet with a splitting of about $15$~MHz (see Fig.~\ref{fgr:Spectrum} and Supplemental Material for more details \cite{SupplInfo}). The effects of external static electric and magnetic fields is easily computed by inclusion of the Stark and Zeeman Hamiltonians in terms of first-rank spherical tensor operators:
\begin{equation}
\label{eq:HeffFields}
\mathcal{H}_{\textrm{St}}+\mathcal{H}_{\textrm{Z}}=-T^1(\mathbf{D})\cdot T^1(\mathbf{E_{\mathrm{lab}}})-g_s \mu_{\textrm{B}} T^1(\mathbf{S})\cdot T^1(\mathbf{B}),
\end{equation}
where the first term describes the interaction of the molecule-frame electric dipole moment $\mathbf{D}$ with the electric field $\mathbf{E_{\mathrm{lab}}}$, and the second term describes the interaction of the the electron spin $\mathbf{S}$ with the magnetic field $\mathbf{B}$ through the g-factor $g_s$ and Bohr magneton $\mu_{\textrm{B}}$ \cite{SupplInfo}. Here, $\mathbf{D}$ is aligned along $\mathbf{n}$ and is 1.24~debye \cite{SupplInfo}, the rotational g-factor and the electronic magnetic anisotropy of a few $10^{-3}$ are neglected. In the case of $^7\Sigma^+_3\,(J=3)$ states, the polarization $P=|\langle \hat{\Omega} \rangle|/|\Omega|$ as a function of the applied external electric field $E_{\mathrm{lab}}\equiv|\mathbf{E_{\mathrm{lab}}}|$ is shown in Fig.~\ref{fgr:Eeff}. While $P=0$ for $E_{\mathrm{lab}}=0$ because $|\Omega,J,M_J,e(f)\rangle$ are parity eigenstates, a modest field of tens of V/cm is sufficient to strongly polarize the molecules in the laboratory frame.

\paragraph*{eEDM sensitivity} --- In order to assess the sensitivity of YbCr to the eEDM, we include the effective CPV Hamiltonian:
\begin{equation}
\label{eq:HCPV}
\mathcal{H}_{\textrm{CPV}} =\mathcal{H}_{\textrm{SP}}+\mathcal{H}_{\textrm{eEDM}}=(W_{\textrm{s}} k_{\textrm{s}}+W_{\textrm{d}} d_{\textrm{e}})\hat{\Omega},
\end{equation}
where $\mathcal{H}_{\textrm{SP}}$ describes the scalar-pseudoscalar electron-nucleon interaction via the enhancement factor $W_{\textrm{s}}$ and the dimensionless isotope-specific $k_{\textrm{s}}$ parameter, and $\mathcal{H}_{\textrm{eEDM}}$ describes the interaction of the eEDM $d_{\textrm{e}}$ with an effective internal field via the enhancement factor $W_{\textrm{s}}$. While $k_{\textrm{s}}$ and $d_{\textrm{e}}$ are the focus of the experimental search, $W_{\textrm{s}}$ and $W_{\textrm{d}}$ are computed via \textit{ab initio} methods. The case of YbCr is well described by a single-contributing electron picture, in which the non-bonding $s$ electron orbiting close to the Yb core is the main source of CPV interactions, with $W_{\textrm{s}}=6.00~h\,$kHz and $W_{\textrm{d}}=1.17\times 10^{24}~h\,\text{Hz}/({e\,\text{cm}})$ \cite{SupplInfo}. In the experimental literature, the eEDM Hamiltonian is often expressed in terms of the effective field $\mathcal{H}_{\textrm{eEDM}}=-\mathbf{d}_e \cdot \mathbf{E}_{\mathrm{eff}}$ with $\mathbf{d}_e=d_{\textrm{e}} \mathbf{S}/|\langle \hat{\Omega} \rangle|$, $\mathbf{E}_{\mathrm{eff}}=-E_\mathrm{eff} \mathbf{n}$ and $E_\mathrm{eff}=W_{\textrm{d}} |\langle \hat{\Omega} \rangle| \mathrm{sgn}(\langle \Omega \rangle)\propto P$, which clearly shows the relation with the polarizability. As shown in Fig.~\ref{fgr:Eeff}, YbCr in $^7\Sigma^+_3\,(J=3)$ states offers $E_\mathrm{eff}$ of about 15~GV/cm at laboratory fields below 100 V/cm. This exceeds by more than one order of magnitude the effective field achievable in AAE compounds \cite{PhysRevA.80.042508}, and compares well against trappable HfF$^+$ ($ \sim 23$~GV/cm), which recently set the most stringent limit on the eEDM \cite{doi:10.1126/science.adg4084}.

%==============================================================================
\paragraph*{Molecule formation} --- Bosonic Cr and Yb isotopes are primarily well-suited for the proposed application, as they can be efficiently cooled and trapped and, thanks to their spinless nuclei ($I_{\mathrm{Cr}}=I_{\mathrm{Yb}}=0$), they are free from hyperfine structure. Several abundant bosonic isotopes are available for both Yb ($^{170,172,174,176}$Yb) and Cr ($^{50,52,54}$Cr), of which $^{174}$Yb and $^{52}$Cr offer the best performance \cite{Schreck2021,He_2019,Chomaz_2023}. For Yb, the combination of laser lights addressing both the broad $^1S_0\rightarrow{}^1P_1$ and inter-combination $^1S_0\rightarrow{}^3P_1$  transitions leads to loading rates in excess of $10^9$ atoms/s and tens of $\mu$K temperatures \cite{PhysRevA.91.053405}. For Cr, the main cooling $^7S_3\rightarrow{}^7P_4$ transition and the additional $^7S_3\rightarrow{}^7P_3$ transition for gray molasses \cite{PhysRevA.99.023607}, following the all-optical strategy demonstrated in Ref.~\cite{PhysRevA.106.053318}, will yield optically trapped samples and loading times similar to the ones of Yb.

Sub-$\mu$K temperatures and even simultaneous quantum degeneracy can be readily attained by forced evaporation with no need for sympathetic cooling, between Cr optically pumped into its stretched, high-field seeking state ($m_s=-3$) and spin-less Yb. This combination is immune both to dipolar Cr losses and inter-species two-body inelastic collisions. Efficient evaporation will be guaranteed by the similar favorable dynamic polarizabilities ($\alpha_{\mathrm{Cr}}=89$ a.u. and $\alpha_{\mathrm{Yb}}=160$ a.u.) and ideal intra-species background s-wave scattering lengths of $\sim 100~a_0$ both for Yb (field-insensitive) and for Cr (narrow and sparse FRs)  \cite{PhysRevA.77.012719,PhysRevLett.94.183201,PhysRevA.81.042716}. Detrimental large inter-species scattering, if any, will be cured by zeroing-out the s-wave interactions via magnetically tunable FRs. Fast, sub-second evaporation time to quantum degeneracy with final number in excess of $10^5$ can be achieved in dynamically-shaped traps \cite{PhysRevA.93.043403}.

According to recent theoretical predictions \cite{PhysRevResearch.6.023254}, contrarily to AAE mixtures, the Yb + Cr mixture will exhibit an extremely suitable FR scenario for magnetoassociation. At least one Feshbach resonance below 250~G with magnetic field widths between 0.1 and 10~G and immune to two-body inelastic collisions. Moreover, the field-independent (quasi field-independent) intra-species scattering of Yb (Cr) in that range \cite{PhysRevLett.94.183201}, will allow to effectively tune the inter-species scattering processes without affecting the intra-species ones. This scenario is experimentally favorable from different viewpoints: (i) FR widths suitable for association \cite{RevModPhys.78.1311}, (ii) relatively low magnetic field bias compatible with with use of high-permeability magnetic shields \cite{10.1063/1.5119915,PhysRevA.97.053625,PhDFarolfi,PhDFava}, (iii) easy asignment of FRs, and (iv) ability to efficiently populate sites of 3D optical lattices with heteronuclear doublons by zeroing out inter-species scattering. Magneto-association of Cr($^7S_3 ,m_s=-3$) + Yb($^1S_0$) scattering pairs will result in YbCr molecules in the least-bound vibrational level ($v=-1$), and angular momenta $|N=2,M_N=-2,S=3,M_S=-1\rangle$ state, where $(N,M_N)$ and $(S,M_S)$ represent the rotational and spin quantum numbers, respectively \cite{PhysRevResearch.6.023254}. Such weakly-bound molecules will be subsequently transferred to the vibrational ground state via STIRAP exploiting optically excited levels supported by either $^7\Sigma^+$ or $^7\Pi$, which correlate to the Yb($^3P_1$) + Cr($^7S_3$) threshold, at convenient laser wavelengths (see Appendix and Supplemental Material \cite{SupplInfo}).

\begin{figure}[t]
\centering
  \includegraphics[width=\columnwidth]{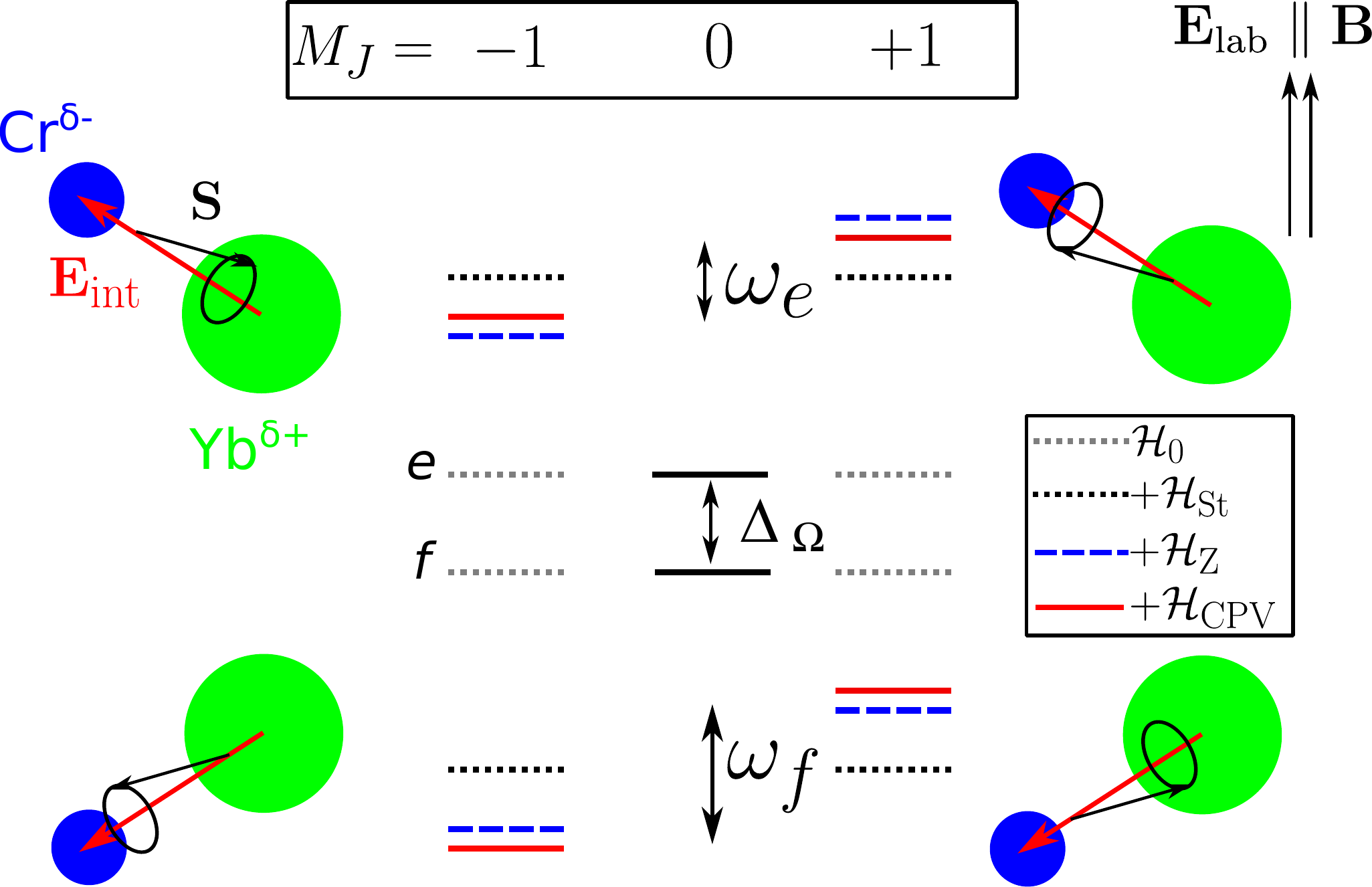}
   \vskip -5 pt
  %\caption{ Stark ($\mathcal{H}_{\mathrm{St}}$), Zeeman ($\mathcal{H}_{\mathrm{Z}}$), and eEDM ($\mathcal{H}_{\mathrm{CPV}}$) shifts from the unperturbed energies ($\mathcal{H}_{0}$) for the $M_J=0,\pm 1$ subspace of $X^7\Sigma^+_3\,(J=3)$.}
  \caption{Energy spectrum of the $M_J=0,\pm 1$ subspace of $X^7\Sigma^+_3(J=3)$ due to successive addition of the Stark ($\mathcal{H}_{\mathrm{St}}$), Zeeman ($\mathcal{H}_{\mathrm{Z}}$), and CPV ($\mathcal{H}_{\mathrm{CPV}}$) Hamiltonians to the zero-field $\mathcal{H}_{0}$.}
 \vskip -5 pt
  \label{fgr:Spectroscopy}
\end{figure}

\paragraph*{Projected SQL sensitivity}--- The expected sensitivity can be estimated in the case of standard spin-precession measurements developed by eEDM experiments on "hot" molecules featuring $\Omega$-doublets in either Hund's case (a) or (c) \cite{PhysRevA.87.052130,Andreev2018,doi:10.1126/science.adg4084}.
In the case of YbCr, we focus on the $^7\Sigma^+_3\,(J=3)$ state and consider the subspace spanned by $|3,3,M_J=\pm 1,e\rangle$ and $|3,3,M_J=\pm 1,f\rangle$, which form a top and bottom doublet, respectively, see gray dotted lines in Fig.~\ref{fgr:Spectroscopy}. Under application of a modest external electric field $\mathbf{E_{\mathrm{lab}}}$, states of same $M_J$ but opposite parity get mixed repelling each other, and are for simplicity denoted $\tilde{e} (\tilde{f})$ correlating to good parity states for vanishing field, see black dotted lines in Fig.~\ref{fgr:Spectroscopy}. As a consequence, the electron spin gets polarized along the internuclear axis and YbCr gets concurrently polarized in the laboratory frame (cf.~Fig.~\ref{fgr:Eeff}). Moreover, the application of an external magnetic field $\mathbf{B}$ induces Zeeman splittings depending only on $M_J$, see blue dashed lines in Fig.~\ref{fgr:Spectroscopy}. The Stark shift is common-mode between the components of each doublet, whereas the differential Zeeman shift is $2 g_{e,f}(E) \mu_{\textrm{B}} B$, with $g_{e,f}(0)\simeq 3/2$ with a relative difference between the doublets below $\%$-level for $E_{\mathrm{lab}}\lesssim 100~$V/cm \cite{SupplInfo}. The energy splittings $\hbar \omega_{e,f}=2 g_{e,f} \mu_{\textrm{B}} B$ can be probed by Ramsey spectroscopy and, within this approach, offer the maximum sensitivity to eEDM with minimum sensitivity to external field noise among possible $\pm M_J$ pairs in YbCr ground state. Swapping between the "top" and "bottom" doublets enables close-to-ideal \emph{spectroscopic} inversion of the external electric field, and yields a phase difference due to the sought-after P,T-odd interaction, see red solid lines in Fig.~\ref{fgr:Spectroscopy}. 

Depending on the actual implementation of the STIRAP, one can transfer the YbCr population to $|3,3,0,f\rangle$ via rapid adiabatic passage, and then drive $|3,3,0,f\rangle \rightarrow (|3,3,+1,\tilde{e}\rangle \pm |3,3,-1,\tilde{e}\rangle)/\sqrt{2}$ with a radio-frequency (RF) pulse. The molecular state subsequently evolves freely in a Ramsey scheme accumulating a phase difference $\phi$. A second RF pulse will bring a population $\cos^2(\phi)$ back to $|3,3,0,f\rangle$, after which the sensitive part of the measurement is over. Choosing the right time to sit on the maximum slope of the Ramsey fringe will yield the highest phase sensitivity. The leftover population in the doublet will be used for signal normalization. STIRAP, in combination with RF pulses if necessary, will bring YbCr populations back to the least bound vibrational level, encoding the phase information in the Cr spin degree of freedom. Finally, simultaneous dissociation across the almost degenerate FRs for different magnetic sublevels, followed by time of flight under a Stern-Gerlach force and final absorption imaging of Cr, will enable a normalized measurement of the original state populations and extraction of the phase $\phi$. In this standard approach, the coherence time will be limited by magnetic-field noise. The stability and homogeneity of the external magnetic field is crucial, as it limits the coherent interrogation time $\tau$. In the case of ultracold YbCr gases, the sample volume of about $(100\mu\mathrm{m})^3$ would be much smaller than in previous experiments on "hot" molecules, and use of magnetic-field shielding will help controlling a stable and homogeneous low bias field in the range from 1 to 10~mG which could be further monitored by co-magnetometry on unbound Cr atoms. Recent advancements in the field of spinor Bose-Einstein condensates have shown coherence times above 1~s for transitions with $\mu_{\textrm{B}}$-level sensitivity \cite{10.1063/1.5119915,PhDFarolfi}. Using the predicted $E_{\mathrm{eff}}= 14.7$~GV/cm, a coherence time of $1$~s, $10^5$ trapped molecules, 6~s cycle time, we obtain a projected statistical sensitivity of $\delta d_{\textrm{e}}= ( 6 \times 10^{-31} / \sqrt{n_{\textrm{day}}})~e\,\mathrm{cm}$.

The reported sensitivity appears robust against potential experimental limitations. Firstly, systematics that can arise from the interaction of the molecules with the trap light used to realize the optical lattices. While the wavelength can be safely chosen to be far detuned from any electric-dipole allowed transition ($\lambda>1~\mu$m), the associated vector polarizability mimics the effect of a magnetic field on the $\Omega$-doublets, which, if correlated with E-reversals, could generate false EDM signals \cite{Fitch_2021,HoekstraOL}. While this effect is ideally canceled by employing lattice beams with perfect linear polarization, it typically results in a stringent requirement on the level of degree of linear polarization. Our strategy, based on ultracold molecules, dramatically relaxes this requirement: even for sub-Hz tunneling rates, corresponding to around $100~h\,$kHz absolute well depth originating from the scalar polarizability \cite{PhysRevA.88.012519}, a suppression of the vector polarizability of $10^{-6}$ is sufficient to pull this effect below the single-molecule shot noise limit. Considering an order of magnitude suppression due to the ratio between spin-orbit splitting and frequency of the relevant $X^7\Sigma^+\rightarrow B^7\Sigma^+$ and $X^7\Sigma^+\rightarrow A^7\Pi$ transitions, this requires the low-power lattice beams to be linearly polarized to better than $10^{-5}$, safely larger than the best recorded values \cite{PhysRevLett.111.243006}. Furthermore, effects of long-range interactions should be addressed since, unlike for quantum metrology protocols, they should be avoided  \cite{Fitch_2021}. Here this is indeed the case even for nearest-neighbor molecules in optical lattice at 1064~nm wavelength \cite{SupplInfo}: the differential shift within the doublets, induced by electric dipole-dipole interaction, is zero to first order and $\lesssim 100~h\,$mHz to second order. The differential shift originating from magnetic dipole-dipole is already below this level to first order. Finally, given the small vibrational ($\omega_0\simeq 90~\mathrm{cm}^{-1}$), rotational ($B_0= 0.041~\mathrm{cm}^{-1}$) \cite{PhysRevA.88.012519,SupplInfo}, and spin-spin ($\lambda_0=0.16~\mathrm{cm}^{-1}$) constants \cite{PhysRevResearch.6.023254,SupplInfo}, together with the large energy necessary to address other electronic states via electric-dipole allowed transitions ($\sim 10^4~\mathrm{cm}^{-1}$), we expect the absorption of black-body radiation, peaking at about $600~\mathrm{cm}^{-1}$ at room temperature, to be negligible for the proposed experiments, as can be seen comparing with previous results \cite{doi:10.1080/00268970701466261}.

\paragraph*{Beyond SQL and hadronic CPV} --- While ground-state YbCr does not provide orbital-spin magnetic moment quasi-cancellation as in ThO or HfF$^+$, its internal structure will allow for the engineering of field insensitive transitions \cite{PhysRevLett.131.183003} and, thanks to the ultracold temperatures, even for the implementation of quantum metrology protocols beyond the standard quantum limit \cite{PhysRevLett.131.193602}. The latter have been recently proposed for parity-doubled molecules \emph{with} magnetic field sensitivity trapped in optical lattices or optical tweezer arrays. 

Moreover, our results can be extended to the search for CP violation in the hadronic sector. The $^{173}$Yb$^{52}$Cr isotopologue, with its high-spin Yb nucleus ($I=5/2$), appears an interesting candidate for the search of the nuclear magnetic quadrupole moment (NMQM) with $W_\mathrm{M}=9.48 \times 10^{31}{h\,\text{Hz}}/(e\,\text{cm}^2)$~\cite{SupplInfo}, while the nuclear Schiff moment (NSM), whose sensitivity is enhanced by octupole deformed nuclei, could be investigated in RaCr \cite{NickHutzlerComm,eEDMWhitePaper}. 

\paragraph*{Conclusions} --- In summary, we identify high-spin $\Sigma$-state polar molecules as ideal molecular species to realize an ultracold molecule platform for eEDM searches. In particular, YbCr satisfies all experimental requirements to potentially gain more than one order of magnitude in sensitivity with already established measurement protocols, while allowing for more advanced quantum control. More generally, these results will pave the way to a whole new class of ultracold molecules, like the isovalent Cr/Mo+Yb/Ra or even Eu/Am-containing diatomics \cite{PhysRevA.90.022514}, for next-generation CPV searches. 

\vskip 15 pt
\begin{acknowledgements}
A.C. thanks A. Vutha, N. Hutzler, G. Carugno, G. Santambrogio, and M. Zaccanti for fruitful discussions.
We gratefully acknowledge the National Science Centre Poland (grants no.~2020/38/E/ST2/00564 and no.~2024/53/N/ST2/03814) and the European Union (ERC, 101042989 -- QuantMol) for financial support and the Poland’s high-performance computing infrastructure PLGrid (HPC Center: ACK Cyfronet AGH) for providing computer facilities and support (computational grant no.~PLG/2024/017527). This work was supported by Italian Ministry of University and Research, and co-funded by the European Union — NextGenerationEU, under the Young Researcher Grants MSCA\_0000042 (PoPaMol, fellowship to A.C.).
\end{acknowledgements}

\section*{Appendix}

The presented formation and application schemes are based on molecular properties that we predict using state-of-the-art \textit{ab initio} quantum-chemical methods.

\paragraph*{Ground-state interaction potential and electric properties} --- The ground electronic state of the YbCr molecule is described by the $^7\Sigma^+$ molecular term and can be well represented by a single Slater determinant. To calculate the corresponding potential energy curve within the Born-Oppenheimer approximation, we employ the coupled cluster methods with the Hartree--Fock reference. The interaction energy is obtained with the super-molecular approach, including the counterpoise correction,
$V_\textrm{int}(R)=E_\textrm{YbCr}(R)-E_\textrm{Yb}(R)-E_\textrm{Cr}(R)\,$,
where $E_\textrm{YbCr}(R)$ denotes the total energy of the YbCr dimer, and $E_\textrm{Yb}(R)$ and $E_\textrm{Cr}(R)$ are the energies of the atoms computed in the dimer basis~\cite{counterpoise}. To account for relativistic effects, inner-shell electrons of ytterbium and chromium are replaced by the effective scalar-relativistic pseudopotentials, ECP28MDF~\cite{newecpYb} and ECP10MDF~\cite{newecpCr}, respectively. The remaining 56 electrons (42 in Yb and 14 in Cr) are correlated and described with the recently developed augmented correlation consistent polarized weighted core-valence quintuple-$\zeta$ quality basis sets, aug-cc-pwCV5Z-PP~\cite{hill_new}. 

\begin{figure}[tb]
\centering
  \includegraphics[width=\columnwidth]{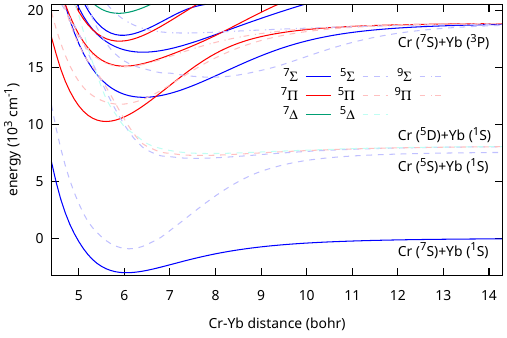}
  \caption{Potential energy curves of the ground and excited electronic states of the YbCr molecule.}
  \label{fgr:pecs}
\end{figure}

First, we use the spin-restricted open-shell coupled cluster method with single, double, and noniterative triple excitations, RCCSD(T)~\cite{KnowlesJCP93}, in the Molpro program~\cite{MOLPRO-WIREs,MOLPRO}. The depth of the potential well with this method is determined to be $ D_{e}=2775~\mathrm{cm}^{-1}$ with an equilibrium interatomic separation of $R_{e}=6.07$~bohr (in good agreement with the previous values of $2866~$cm$^{-1}$ and 6.05~bohr~\cite{PhysRevA.88.012519}). Next, in the composite approach~\cite{PhysRevA.109.052814}, we include the full triple correction within the CCSDT method calculated with the same pseudopotentials and aug-cc-pVTZ-PP basis sets, correlating $6s^2$ and $3d^54s^1$ electrons of ytterbium and chromium, respectively, in the MRCC program~\cite{MRCC}. The CCSDT correction makes the original potential well slightly deeper, resulting in the final parameters of $D_e = 3021~\mathrm{cm}^{-1}$, $R_e = 6.05~$bohr, $\omega_e=$ 90.2~cm$^{-1}$, and $B_e=$ 0.041$~$cm$^{-1}$ for $^{174}$Yb$^{52}$Cr. The obtained potential energy curve is presented in Fig.~\ref{fgr:pecs}. 

We also calculate the molecule-frame electric dipole moment using the finite-field method. At the CCSD(T) level of theory, the dipole moment at the equilibrium geometry is found to be 1.14~debye in agreement with the previous value of 1.19~debye~\cite{PhysRevA.88.012519}. Inclusion of a correction for full triple excitations increases this value to 1.24~debye.

\paragraph*{Excited electronic states} ---  Successful formation and further precision measurements with YbCr molecules require the efficient transfer of the weakly bound Feshbach dimers to the ground vibrational level via STIRAP. We calculate several electronically excited states of the YbCr molecule to guide this process. Due to the presence of an open $d$-shell in chromium and a $p$-shell in the excited state of ytterbium, one has to deal with a rich manifold of molecular electronic configurations. We consider three families of electronic states that arise from different asymptotic atomic configurations. The first family corresponds to Cr(${}^7S$) + Yb($^1S$), leading to the already-calculated ground $^7\Sigma^+$ molecular state. The second involves Cr($^5D$) + Yb($^1S$), resulting in the $^5\Sigma^+$, $^5\Pi$, and $^5\Delta$ states. The third family originates from Cr($^7S$) + Yb($^3P$) and gives rise to $^5\Sigma^+$, $^5\Pi$, $^7\Sigma^+$, $^7\Pi$, $^9\Sigma^+$, and $^9\Pi$ states.

We calculate the septet states using the equation-of-motion coupled cluster method with single and double excitations (EOM-CCSD)~\cite{10.1063/5.0004837}, the nonet states with the regular CCSD method, while the quintet states are obtained with the multireference configuration interaction method with single and double excitations (MRCISD)~\cite{WernerJCP88} due to numerical difficulties with EOM‑CCSD for low-spin states. The same effective core potentials as for the ground state and the aug-cc-pwCVTZ-PP basis sets \cite{hill_new} are used. The CFOUR program~\cite{MatthewsJCP20,cfour} is employed for the excited states. The final potential energy curves are shown in Fig.~\ref{fgr:pecs}.

The STIRAP scheme, using the excited $A^7\Pi$ or $B^7\Sigma^+$ state to support an intermediate level, appears to be the most promising. Following Ref.~\cite{PRXQuantum.5.020358}, we predict favorable Franck-Condon factors ($\gtrsim 10^{-4}$) for pump and Stokes transitions to vibrational levels of the $A^7\Pi_4$ state around 7000~cm$^{-1}$ below the intercombination line of Yb, but detailed studies of STIRAP are beyond the scope of this work. In the future, the spin-orbit coupling and other electronic states should be included to obtain a more complete prediction.
 
\paragraph*{eEDM sensitivity} --- Heavy, paramagnetic diatomic molecules composed of zero‑nuclear‑spin isotopes provide an ideal testbed for probing P,T-odd interactions in the (semi)leptonic sector. The dominant sources of CP‑violating energy shifts in molecular spectra are the electron electric dipole moment (eEDM) and the scalar-pseudoscalar electron-nucleon coupling (eN-SPS). Here, we focus on determining the molecular enhancement factors present in Eq.~\eqref{eq:HCPV}. Accurate theoretical modeling is indispensable because these factors cannot be determined experimentally. To this end, we first solve the many‑electron problem with the Dirac–Coulomb Hamiltonian with exact treatment of two-electron integrals over the small components of the wave function using the DIRAC24 package \cite{dirac24}. We generate molecular spinors using the Average of Configurations Dirac--Coulomb--Hartree--Fock (DCHF) method with six electrons distributed over six Kramers pairs (twelve spinors), yielding open‑shell spinors composed of ytterbium $6s$ and chromium $3d4s$ orbitals. Building on this DCHF reference, we then perform the multireference configuration interaction (MRCI) calculation using the Kramers‑Restricted Configuration Interaction (KRCI) module with the Generalized Active Space (GAS) formalism to capture the electron correlation \cite{knecht}.
MRCI with single and double excitations (MRCISD) and with single, double, and triple excitations (MRCISD) is used with different GASs. All computations employ the uncontracted Dyall basis sets \cite{dyall2012core}. Finally, we evaluate the expectation value of the eEDM Hamiltonian \cite{lindroth1989order} with respect to the electronic wave function $| \Psi_{^7\Sigma_{\Omega}} \rangle$, which is given by: 
\begin{equation}
    W_{\textrm{d}} = \frac{2ic}{\Omega}
    \left\langle \Psi_{^7\Sigma_{\Omega}} \left|
    \sum^{N_{\mathrm{e}}}_j  \gamma_j^0 \gamma_j^5 p_j^2
    \right| \Psi_{^7\Sigma_{\Omega}} \right\rangle,
\end{equation}
where $i$ is imaginary unit, $c$ is a speed of light, $\gamma_{j}^{k}$ are the Dirac matrices, $p_{j}$ is the
linear momentum operator of electron and $N_\text{e}$ is the number of electrons. Similarly, for the eN-SPS enhancement factor \cite{kozlov1985semiempirical}, we have the following expression:
\begin{equation}
    W_{\textrm{s}} = \frac{i G_\text{F} Z}{\sqrt{2} \Omega}
    \left\langle \Psi_{^7\Sigma_{\Omega}} \left|
    \sum_{j}^{N_{\mathrm{e}}} \gamma_j^0 \gamma_j^5 \rho(r_j)
    \right| \Psi_{^7\Sigma_{\Omega}} \right\rangle,
\end{equation}
where $G_\text{F}$ is the Fermi coupling constant, $Z$ is the atomic number, $\rho({r}_{j})$ is the nuclear charge distribution and $r_{j}$ is the position of $j$-th electron with respect to the nucleus. We compute both enhancement factors for the equilibrium geometry taken from our scalar-relativistic calculations ($R_e=6.05~$bohr). The final value of the eEDM factor is  $W_{\textrm{d}}=1.17\times 10^{24}~h\,\text{Hz}/({e\,\text{cm}})$ and the eN-SPS factor is $W_{s}=6.00~h$\,kHz. Detailed analysis of the convergence of those values concerning the number of correlated electrons, the size of active space, and the basis set cardinality is presented in Supplemental Material~\cite{SupplInfo}.

\paragraph*{Zero-field splitting} --- For molecules with at least two unpaired electrons ($S\geq 1$), even in a $\Sigma$ state with no first-order spin-orbit coupling, the degeneracy of magnetic sublevels is lifted by the spin-spin interaction, leading to the so-called zero-field splitting. This interaction can be described by the tensor that reduces to a single component $D$ for $\Sigma$-state diatomic molecules and is connected to the spin-spin coupling parameter $\lambda$ in Eq.~\eqref{eq:Heff0} via $D=2\lambda$. The corresponding effective molecule-frame Hamiltonian is  
\begin{equation}\label{eq:HSS}
        \mathcal{H}_{\textrm{SS}} =  D \left(\hat{\Sigma}^2 - {\mathbf{S}}^2/3\right).
\end{equation}
There are two contributions to this interaction: the direct spin-spin magnetic dipolar coupling between open-shell electrons and the spin-orbit coupling in the second order of perturbation theory.

We calculate the zero-field splitting parameter $D$ using wavefunction-based methods and density functional theory (DFT). We find that the magnetic dipolar spin-spin contribution within the DFT methodology is not significantly affected by the choice of the functional and is on the order of 0.009$~\mathrm{cm}^{-1}$. It is obtained as an expectation value with the uncontracted all-electron x2c-QZVPPall basis sets~\cite{franzke2020segmented} and scalar relativistic effects treated at the spin-free two-component formalism (X2C)~\cite{PengJCP2013} in the ORCA program~\cite{orca1,ORCA6}. In contrast, the second-order spin-orbit contribution, which dominates the total $D$ value (over 95\%), is highly sensitive to variations in the used exchange-correlation functional, basis set, and formalism applied in the calculations, thus, it should be treated with care. Because calculations as the sum over states do not provide sufficient accuracy, we extract the second-order spin-orbit contribution by fitting the splitting $D(\Omega^2-S(S+1)/3)$, resulting from Eq.~\eqref{eq:HSS}, to potential energies obtained from solving the many-electron Dirac-Coulomb Hamiltonian for different $^7\Sigma_\Omega$ states with the same methods and basis sets as for the eEDM sensitivity factors. Following extensive computational analyses based on the KRCI method, variations in basis set size, different reference spinors, and correlation treatment, we find the zero-field splitting parameter of $D=0.32(10)~\mathrm{cm}^{-1}$, in good agreement with a DFT value reported in Ref.~\cite{PhysRevResearch.6.023254}. A comprehensive numerical assessment and experimental implications are provided in the Supplemental Material~\cite{SupplInfo}.

%%%%%%%%%%%%%%%%%%%%%%%%%%%%%%%%%%%%%%%%%%%%%%%%%%%%%%%%%%%%%%
%%%%%%%%%%%%%%%%%%%%%%%BIBLIOGRAPHY%%%%%%%%%%%%%%%%%%%%%%%%%%%
%%%%%%%%%%%%%%%%%%%%%%%%%%%%%%%%%%%%%%%%%%%%%%%%%%%%%%%%%%%%%%
%\bibliography{references}

\begin{thebibliography}{113}%
\makeatletter
\providecommand \@ifxundefined [1]{%
 \@ifx{#1\undefined}
}%
\providecommand \@ifnum [1]{%
 \ifnum #1\expandafter \@firstoftwo
 \else \expandafter \@secondoftwo
 \fi
}%
\providecommand \@ifx [1]{%
 \ifx #1\expandafter \@firstoftwo
 \else \expandafter \@secondoftwo
 \fi
}%
\providecommand \natexlab [1]{#1}%
\providecommand \enquote  [1]{``#1''}%
\providecommand \bibnamefont  [1]{#1}%
\providecommand \bibfnamefont [1]{#1}%
\providecommand \citenamefont [1]{#1}%
\providecommand \href@noop [0]{\@secondoftwo}%
\providecommand \href [0]{\begingroup \@sanitize@url \@href}%
\providecommand \@href[1]{\@@startlink{#1}\@@href}%
\providecommand \@@href[1]{\endgroup#1\@@endlink}%
\providecommand \@sanitize@url [0]{\catcode `\\12\catcode `\$12\catcode `\&12\catcode `\#12\catcode `\^12\catcode `\_12\catcode `\%12\relax}%
\providecommand \@@startlink[1]{}%
\providecommand \@@endlink[0]{}%
\providecommand \url  [0]{\begingroup\@sanitize@url \@url }%
\providecommand \@url [1]{\endgroup\@href {#1}{\urlprefix }}%
\providecommand \urlprefix  [0]{URL }%
\providecommand \Eprint [0]{\href }%
\providecommand \doibase [0]{http://dx.doi.org/}%
\providecommand \selectlanguage [0]{\@gobble}%
\providecommand \bibinfo  [0]{\@secondoftwo}%
\providecommand \bibfield  [0]{\@secondoftwo}%
\providecommand \translation [1]{[#1]}%
\providecommand \BibitemOpen [0]{}%
\providecommand \bibitemStop [0]{}%
\providecommand \bibitemNoStop [0]{.\EOS\space}%
\providecommand \EOS [0]{\spacefactor3000\relax}%
\providecommand \BibitemShut  [1]{\csname bibitem#1\endcsname}%
\let\auto@bib@innerbib\@empty
%</preamble>
\bibitem [{\citenamefont {Safronova}\ \emph {et~al.}(2018)\citenamefont {Safronova}, \citenamefont {Budker}, \citenamefont {DeMille}, \citenamefont {Kimball}, \citenamefont {Derevianko},\ and\ \citenamefont {Clark}}]{RevModPhys.90.025008}%
  \BibitemOpen
  \bibfield  {author} {\bibinfo {author} {\bibfnamefont {M.~S.}\ \bibnamefont {Safronova}}, \bibinfo {author} {\bibfnamefont {D.}~\bibnamefont {Budker}}, \bibinfo {author} {\bibfnamefont {D.}~\bibnamefont {DeMille}}, \bibinfo {author} {\bibfnamefont {D.~F.~J.}\ \bibnamefont {Kimball}}, \bibinfo {author} {\bibfnamefont {A.}~\bibnamefont {Derevianko}}, \ and\ \bibinfo {author} {\bibfnamefont {C.~W.}\ \bibnamefont {Clark}},\ }\href {\doibase 10.1103/RevModPhys.90.025008} {\bibfield  {journal} {\bibinfo  {journal} {Rev. Mod. Phys.}\ }\textbf {\bibinfo {volume} {90}},\ \bibinfo {pages} {025008} (\bibinfo {year} {2018})}\BibitemShut {NoStop}%
\bibitem [{\citenamefont {Chupp}\ \emph {et~al.}(2019)\citenamefont {Chupp}, \citenamefont {Fierlinger}, \citenamefont {Ramsey-Musolf},\ and\ \citenamefont {Singh}}]{RevModPhys.91.015001}%
  \BibitemOpen
  \bibfield  {author} {\bibinfo {author} {\bibfnamefont {T.~E.}\ \bibnamefont {Chupp}}, \bibinfo {author} {\bibfnamefont {P.}~\bibnamefont {Fierlinger}}, \bibinfo {author} {\bibfnamefont {M.~J.}\ \bibnamefont {Ramsey-Musolf}}, \ and\ \bibinfo {author} {\bibfnamefont {J.~T.}\ \bibnamefont {Singh}},\ }\href {\doibase 10.1103/RevModPhys.91.015001} {\bibfield  {journal} {\bibinfo  {journal} {Rev. Mod. Phys.}\ }\textbf {\bibinfo {volume} {91}},\ \bibinfo {pages} {015001} (\bibinfo {year} {2019})}\BibitemShut {NoStop}%
\bibitem [{\citenamefont {Dine}\ and\ \citenamefont {Kusenko}(2003)}]{RevModPhys.76.1}%
  \BibitemOpen
  \bibfield  {author} {\bibinfo {author} {\bibfnamefont {M.}~\bibnamefont {Dine}}\ and\ \bibinfo {author} {\bibfnamefont {A.}~\bibnamefont {Kusenko}},\ }\href {\doibase 10.1103/RevModPhys.76.1} {\bibfield  {journal} {\bibinfo  {journal} {Rev. Mod. Phys.}\ }\textbf {\bibinfo {volume} {76}},\ \bibinfo {pages} {1} (\bibinfo {year} {2003})}\BibitemShut {NoStop}%
\bibitem [{\citenamefont {Canetti}\ \emph {et~al.}(2012)\citenamefont {Canetti}, \citenamefont {Drewes},\ and\ \citenamefont {Shaposhnikov}}]{Canetti_2012}%
  \BibitemOpen
  \bibfield  {author} {\bibinfo {author} {\bibfnamefont {L.}~\bibnamefont {Canetti}}, \bibinfo {author} {\bibfnamefont {M.}~\bibnamefont {Drewes}}, \ and\ \bibinfo {author} {\bibfnamefont {M.}~\bibnamefont {Shaposhnikov}},\ }\href {\doibase 10.1088/1367-2630/14/9/095012} {\bibfield  {journal} {\bibinfo  {journal} {New J. Phys.}\ }\textbf {\bibinfo {volume} {14}},\ \bibinfo {pages} {095012} (\bibinfo {year} {2012})}\BibitemShut {NoStop}%
\bibitem [{\citenamefont {Sakharov}(1967)}]{Sakharov1967}%
  \BibitemOpen
  \bibfield  {author} {\bibinfo {author} {\bibfnamefont {A.~D.}\ \bibnamefont {Sakharov}},\ }\href {\doibase 10.1070/PU1991v034n05ABEH002497} {\bibfield  {journal} {\bibinfo  {journal} {Pisma Zh.Eksp.Teor.Fiz.}\ }\textbf {\bibinfo {volume} {32}},\ \bibinfo {pages} {392} (\bibinfo {year} {1967})}\BibitemShut {NoStop}%
\bibitem [{\citenamefont {Patrignani}(2016)}]{Patrignani_2016}%
  \BibitemOpen
  \bibfield  {author} {\bibinfo {author} {\bibfnamefont {C.}~\bibnamefont {Patrignani}},\ }\href {\doibase 10.1088/1674-1137/40/10/100001} {\bibfield  {journal} {\bibinfo  {journal} {Chin. Phys. C.}\ }\textbf {\bibinfo {volume} {40}},\ \bibinfo {pages} {100001} (\bibinfo {year} {2016})}\BibitemShut {NoStop}%
\bibitem [{\citenamefont {Roussy}\ \emph {et~al.}(2023)\citenamefont {Roussy}, \citenamefont {Caldwell}, \citenamefont {Wright}, \citenamefont {Cairncross}, \citenamefont {Shagam}, \citenamefont {Ng}, \citenamefont {Schlossberger}, \citenamefont {Park}, \citenamefont {Wang}, \citenamefont {Ye},\ and\ \citenamefont {Cornell}}]{doi:10.1126/science.adg4084}%
  \BibitemOpen
  \bibfield  {author} {\bibinfo {author} {\bibfnamefont {T.~S.}\ \bibnamefont {Roussy}}, \bibinfo {author} {\bibfnamefont {L.}~\bibnamefont {Caldwell}}, \bibinfo {author} {\bibfnamefont {T.}~\bibnamefont {Wright}}, \bibinfo {author} {\bibfnamefont {W.~B.}\ \bibnamefont {Cairncross}}, \bibinfo {author} {\bibfnamefont {Y.}~\bibnamefont {Shagam}}, \bibinfo {author} {\bibfnamefont {K.~B.}\ \bibnamefont {Ng}}, \bibinfo {author} {\bibfnamefont {N.}~\bibnamefont {Schlossberger}}, \bibinfo {author} {\bibfnamefont {S.~Y.}\ \bibnamefont {Park}}, \bibinfo {author} {\bibfnamefont {A.}~\bibnamefont {Wang}}, \bibinfo {author} {\bibfnamefont {J.}~\bibnamefont {Ye}}, \ and\ \bibinfo {author} {\bibfnamefont {E.~A.}\ \bibnamefont {Cornell}},\ }\href {\doibase 10.1126/science.adg4084} {\bibfield  {journal} {\bibinfo  {journal} {Science}\ }\textbf {\bibinfo {volume} {381}},\ \bibinfo {pages} {46} (\bibinfo {year} {2023})}\BibitemShut {NoStop}%
\bibitem [{\citenamefont {Andreev}\ \emph {et~al.}(2018)\citenamefont {Andreev}, \citenamefont {Ang}, \citenamefont {DeMille}, \citenamefont {Doyle}, \citenamefont {Gabrielse}, \citenamefont {Haefner}, \citenamefont {Hutzler}, \citenamefont {Lasner}, \citenamefont {Meisenhelder}, \citenamefont {O'Leary}, \citenamefont {Panda}, \citenamefont {West}, \citenamefont {West}, \citenamefont {Wu},\ and\ \citenamefont {Collaboration}}]{Andreev2018}%
  \BibitemOpen
  \bibfield  {author} {\bibinfo {author} {\bibfnamefont {V.}~\bibnamefont {Andreev}}, \bibinfo {author} {\bibfnamefont {D.~G.}\ \bibnamefont {Ang}}, \bibinfo {author} {\bibfnamefont {D.}~\bibnamefont {DeMille}}, \bibinfo {author} {\bibfnamefont {J.~M.}\ \bibnamefont {Doyle}}, \bibinfo {author} {\bibfnamefont {G.}~\bibnamefont {Gabrielse}}, \bibinfo {author} {\bibfnamefont {J.}~\bibnamefont {Haefner}}, \bibinfo {author} {\bibfnamefont {N.~R.}\ \bibnamefont {Hutzler}}, \bibinfo {author} {\bibfnamefont {Z.}~\bibnamefont {Lasner}}, \bibinfo {author} {\bibfnamefont {C.}~\bibnamefont {Meisenhelder}}, \bibinfo {author} {\bibfnamefont {B.~R.}\ \bibnamefont {O'Leary}}, \bibinfo {author} {\bibfnamefont {C.~D.}\ \bibnamefont {Panda}}, \bibinfo {author} {\bibfnamefont {A.~D.}\ \bibnamefont {West}}, \bibinfo {author} {\bibfnamefont {E.~P.}\ \bibnamefont {West}}, \bibinfo {author} {\bibfnamefont {X.}~\bibnamefont {Wu}}, \ and\ \bibinfo {author} {\bibfnamefont {A.~C. M.~E.}\ \bibnamefont {Collaboration}},\ }\href {\doibase
  10.1038/s41586-018-0599-8} {\bibfield  {journal} {\bibinfo  {journal} {Nature}\ }\textbf {\bibinfo {volume} {562}},\ \bibinfo {pages} {355} (\bibinfo {year} {2018})}\BibitemShut {NoStop}%
\bibitem [{\citenamefont {Alarcon}\ \emph {et~al.}()\citenamefont {Alarcon}, \citenamefont {Alexander}, \citenamefont {Anastassopoulos},\ and\ \citenamefont {et~al.}}]{eEDMWhitePaper}%
  \BibitemOpen
  \bibfield  {author} {\bibinfo {author} {\bibfnamefont {R.}~\bibnamefont {Alarcon}}, \bibinfo {author} {\bibfnamefont {J.}~\bibnamefont {Alexander}}, \bibinfo {author} {\bibfnamefont {V.}~\bibnamefont {Anastassopoulos}}, \ and\ \bibinfo {author} {\bibnamefont {et~al.}},\ }\href {\doibase 10.48550/arXiv.2203.08103} {\ 10.48550/arXiv.2203.08103}\BibitemShut {NoStop}%
\bibitem [{\citenamefont {DeMille}\ \emph {et~al.}(2024)\citenamefont {DeMille}, \citenamefont {Hutzler}, \citenamefont {Rey},\ and\ \citenamefont {Zelevinsky}}]{DeMille2024}%
  \BibitemOpen
  \bibfield  {author} {\bibinfo {author} {\bibfnamefont {D.}~\bibnamefont {DeMille}}, \bibinfo {author} {\bibfnamefont {N.~R.}\ \bibnamefont {Hutzler}}, \bibinfo {author} {\bibfnamefont {A.~M.}\ \bibnamefont {Rey}}, \ and\ \bibinfo {author} {\bibfnamefont {T.}~\bibnamefont {Zelevinsky}},\ }\href {\doibase 10.1038/s41567-024-02499-9} {\bibfield  {journal} {\bibinfo  {journal} {Nat. Phys.}\ }\textbf {\bibinfo {volume} {20}},\ \bibinfo {pages} {741} (\bibinfo {year} {2024})}\BibitemShut {NoStop}%
\bibitem [{\citenamefont {Ye}\ and\ \citenamefont {Zoller}(2024)}]{PhysRevLett.132.190001}%
  \BibitemOpen
  \bibfield  {author} {\bibinfo {author} {\bibfnamefont {J.}~\bibnamefont {Ye}}\ and\ \bibinfo {author} {\bibfnamefont {P.}~\bibnamefont {Zoller}},\ }\href {\doibase 10.1103/PhysRevLett.132.190001} {\bibfield  {journal} {\bibinfo  {journal} {Phys. Rev. Lett.}\ }\textbf {\bibinfo {volume} {132}},\ \bibinfo {pages} {190001} (\bibinfo {year} {2024})}\BibitemShut {NoStop}%
\bibitem [{\citenamefont {Saffman}(2016)}]{Saffman_2016}%
  \BibitemOpen
  \bibfield  {author} {\bibinfo {author} {\bibfnamefont {M.}~\bibnamefont {Saffman}},\ }\href {\doibase 10.1088/0953-4075/49/20/202001} {\bibfield  {journal} {\bibinfo  {journal} {J. Phys. B:At., Mol. Opt.}\ }\textbf {\bibinfo {volume} {49}},\ \bibinfo {pages} {202001} (\bibinfo {year} {2016})}\BibitemShut {NoStop}%
\bibitem [{\citenamefont {Henriet}\ \emph {et~al.}(2020)\citenamefont {Henriet}, \citenamefont {Beguin}, \citenamefont {Signoles}, \citenamefont {Lahaye}, \citenamefont {Browaeys}, \citenamefont {Reymond},\ and\ \citenamefont {Jurczak}}]{Henriet2020quantumcomputing}%
  \BibitemOpen
  \bibfield  {author} {\bibinfo {author} {\bibfnamefont {L.}~\bibnamefont {Henriet}}, \bibinfo {author} {\bibfnamefont {L.}~\bibnamefont {Beguin}}, \bibinfo {author} {\bibfnamefont {A.}~\bibnamefont {Signoles}}, \bibinfo {author} {\bibfnamefont {T.}~\bibnamefont {Lahaye}}, \bibinfo {author} {\bibfnamefont {A.}~\bibnamefont {Browaeys}}, \bibinfo {author} {\bibfnamefont {G.-O.}\ \bibnamefont {Reymond}}, \ and\ \bibinfo {author} {\bibfnamefont {C.}~\bibnamefont {Jurczak}},\ }\href {\doibase 10.22331/q-2020-09-21-327} {\bibfield  {journal} {\bibinfo  {journal} {{Quantum}}\ }\textbf {\bibinfo {volume} {4}},\ \bibinfo {pages} {327} (\bibinfo {year} {2020})}\BibitemShut {NoStop}%
\bibitem [{\citenamefont {Browaeys}\ and\ \citenamefont {Lahaye}(2020)}]{Browaeys2020}%
  \BibitemOpen
  \bibfield  {author} {\bibinfo {author} {\bibfnamefont {A.}~\bibnamefont {Browaeys}}\ and\ \bibinfo {author} {\bibfnamefont {T.}~\bibnamefont {Lahaye}},\ }\href {\doibase 10.1038/s41567-019-0733-z} {\bibfield  {journal} {\bibinfo  {journal} {Nat. Phys.}\ }\textbf {\bibinfo {volume} {16}},\ \bibinfo {pages} {132} (\bibinfo {year} {2020})}\BibitemShut {NoStop}%
\bibitem [{\citenamefont {Kaufman}\ and\ \citenamefont {Ni}(2021)}]{Kaufman2021}%
  \BibitemOpen
  \bibfield  {author} {\bibinfo {author} {\bibfnamefont {A.~M.}\ \bibnamefont {Kaufman}}\ and\ \bibinfo {author} {\bibfnamefont {K.-K.}\ \bibnamefont {Ni}},\ }\href {\doibase 10.1038/s41567-021-01357-2} {\bibfield  {journal} {\bibinfo  {journal} {Nat. Phys.}\ }\textbf {\bibinfo {volume} {17}},\ \bibinfo {pages} {1324} (\bibinfo {year} {2021})}\BibitemShut {NoStop}%
\bibitem [{\citenamefont {Daley}\ \emph {et~al.}(2022)\citenamefont {Daley}, \citenamefont {Bloch}, \citenamefont {Kokail}, \citenamefont {Flannigan}, \citenamefont {Pearson}, \citenamefont {Troyer},\ and\ \citenamefont {Zoller}}]{Daley2022}%
  \BibitemOpen
  \bibfield  {author} {\bibinfo {author} {\bibfnamefont {A.~J.}\ \bibnamefont {Daley}}, \bibinfo {author} {\bibfnamefont {I.}~\bibnamefont {Bloch}}, \bibinfo {author} {\bibfnamefont {C.}~\bibnamefont {Kokail}}, \bibinfo {author} {\bibfnamefont {S.}~\bibnamefont {Flannigan}}, \bibinfo {author} {\bibfnamefont {N.}~\bibnamefont {Pearson}}, \bibinfo {author} {\bibfnamefont {M.}~\bibnamefont {Troyer}}, \ and\ \bibinfo {author} {\bibfnamefont {P.}~\bibnamefont {Zoller}},\ }\href {\doibase 10.1038/s41586-022-04940-6} {\bibfield  {journal} {\bibinfo  {journal} {Nature}\ }\textbf {\bibinfo {volume} {607}},\ \bibinfo {pages} {667} (\bibinfo {year} {2022})}\BibitemShut {NoStop}%
\bibitem [{\citenamefont {González-Cuadra}\ \emph {et~al.}(2023)\citenamefont {González-Cuadra}, \citenamefont {Bluvstein}, \citenamefont {Kalinowski}, \citenamefont {Kaubruegger}, \citenamefont {Maskara}, \citenamefont {Naldesi}, \citenamefont {Zache}, \citenamefont {Kaufman}, \citenamefont {Lukin}, \citenamefont {Pichler}, \citenamefont {Vermersch}, \citenamefont {Ye},\ and\ \citenamefont {Zoller}}]{doi:10.1073/pnas.2304294120}%
  \BibitemOpen
  \bibfield  {author} {\bibinfo {author} {\bibfnamefont {D.}~\bibnamefont {González-Cuadra}}, \bibinfo {author} {\bibfnamefont {D.}~\bibnamefont {Bluvstein}}, \bibinfo {author} {\bibfnamefont {M.}~\bibnamefont {Kalinowski}}, \bibinfo {author} {\bibfnamefont {R.}~\bibnamefont {Kaubruegger}}, \bibinfo {author} {\bibfnamefont {N.}~\bibnamefont {Maskara}}, \bibinfo {author} {\bibfnamefont {P.}~\bibnamefont {Naldesi}}, \bibinfo {author} {\bibfnamefont {T.~V.}\ \bibnamefont {Zache}}, \bibinfo {author} {\bibfnamefont {A.~M.}\ \bibnamefont {Kaufman}}, \bibinfo {author} {\bibfnamefont {M.~D.}\ \bibnamefont {Lukin}}, \bibinfo {author} {\bibfnamefont {H.}~\bibnamefont {Pichler}}, \bibinfo {author} {\bibfnamefont {B.}~\bibnamefont {Vermersch}}, \bibinfo {author} {\bibfnamefont {J.}~\bibnamefont {Ye}}, \ and\ \bibinfo {author} {\bibfnamefont {P.}~\bibnamefont {Zoller}},\ }\href {\doibase 10.1073/pnas.2304294120} {\bibfield  {journal} {\bibinfo  {journal} {Proc. Natl. Acad. Sci.}\ }\textbf {\bibinfo {volume} {120}},\
  \bibinfo {pages} {e2304294120} (\bibinfo {year} {2023})}\BibitemShut {NoStop}%
\bibitem [{\citenamefont {Kusano}\ \emph {et~al.}(2025)\citenamefont {Kusano}, \citenamefont {Nakamura}, \citenamefont {Yokoyama}, \citenamefont {Ozawa}, \citenamefont {Shibata}, \citenamefont {Takano}, \citenamefont {Takasu},\ and\ \citenamefont {Takahashi}}]{TakahashiArxiv}%
  \BibitemOpen
  \bibfield  {author} {\bibinfo {author} {\bibfnamefont {T.}~\bibnamefont {Kusano}}, \bibinfo {author} {\bibfnamefont {Y.}~\bibnamefont {Nakamura}}, \bibinfo {author} {\bibfnamefont {R.}~\bibnamefont {Yokoyama}}, \bibinfo {author} {\bibfnamefont {N.}~\bibnamefont {Ozawa}}, \bibinfo {author} {\bibfnamefont {K.}~\bibnamefont {Shibata}}, \bibinfo {author} {\bibfnamefont {T.}~\bibnamefont {Takano}}, \bibinfo {author} {\bibfnamefont {Y.}~\bibnamefont {Takasu}}, \ and\ \bibinfo {author} {\bibfnamefont {Y.}~\bibnamefont {Takahashi}},\ }\href {\doibase 10.1103/PhysRevResearch.7.L022045} {\bibfield  {journal} {\bibinfo  {journal} {Phys. Rev. Res.}\ }\textbf {\bibinfo {volume} {7}},\ \bibinfo {pages} {L022045} (\bibinfo {year} {2025})}\BibitemShut {NoStop}%
\bibitem [{\citenamefont {Ludlow}(2019)}]{Ludlow2019}%
  \BibitemOpen
  \bibfield  {author} {\bibinfo {author} {\bibfnamefont {A.}~\bibnamefont {Ludlow}},\ }\href {https://physics.aps.org/articles/v12/141} {\bibfield  {journal} {\bibinfo  {journal} {Physics}\ }\textbf {\bibinfo {volume} {12}} (\bibinfo {year} {2019})}\BibitemShut {NoStop}%
\bibitem [{\citenamefont {Pezz\`e}\ \emph {et~al.}(2018)\citenamefont {Pezz\`e}, \citenamefont {Smerzi}, \citenamefont {Oberthaler}, \citenamefont {Schmied},\ and\ \citenamefont {Treutlein}}]{RevModPhys.90.035005}%
  \BibitemOpen
  \bibfield  {author} {\bibinfo {author} {\bibfnamefont {L.}~\bibnamefont {Pezz\`e}}, \bibinfo {author} {\bibfnamefont {A.}~\bibnamefont {Smerzi}}, \bibinfo {author} {\bibfnamefont {M.~K.}\ \bibnamefont {Oberthaler}}, \bibinfo {author} {\bibfnamefont {R.}~\bibnamefont {Schmied}}, \ and\ \bibinfo {author} {\bibfnamefont {P.}~\bibnamefont {Treutlein}},\ }\href {\doibase 10.1103/RevModPhys.90.035005} {\bibfield  {journal} {\bibinfo  {journal} {Rev. Mod. Phys.}\ }\textbf {\bibinfo {volume} {90}},\ \bibinfo {pages} {035005} (\bibinfo {year} {2018})}\BibitemShut {NoStop}%
\bibitem [{\citenamefont {Malia}\ \emph {et~al.}(2022)\citenamefont {Malia}, \citenamefont {Wu}, \citenamefont {Mart{\'i}nez-Rinc{\'o}n},\ and\ \citenamefont {Kasevich}}]{Malia2022}%
  \BibitemOpen
  \bibfield  {author} {\bibinfo {author} {\bibfnamefont {B.~K.}\ \bibnamefont {Malia}}, \bibinfo {author} {\bibfnamefont {Y.}~\bibnamefont {Wu}}, \bibinfo {author} {\bibfnamefont {J.}~\bibnamefont {Mart{\'i}nez-Rinc{\'o}n}}, \ and\ \bibinfo {author} {\bibfnamefont {M.~A.}\ \bibnamefont {Kasevich}},\ }\href {\doibase 10.1038/s41586-022-05363-z} {\bibfield  {journal} {\bibinfo  {journal} {Nature}\ }\textbf {\bibinfo {volume} {612}},\ \bibinfo {pages} {661} (\bibinfo {year} {2022})}\BibitemShut {NoStop}%
\bibitem [{\citenamefont {Bornet}\ \emph {et~al.}(2023)\citenamefont {Bornet}, \citenamefont {Emperauger}, \citenamefont {Chen}, \citenamefont {Ye}, \citenamefont {Block}, \citenamefont {Bintz}, \citenamefont {Boyd}, \citenamefont {Barredo}, \citenamefont {Comparin}, \citenamefont {Mezzacapo}, \citenamefont {Roscilde}, \citenamefont {Lahaye}, \citenamefont {Yao},\ and\ \citenamefont {Browaeys}}]{Bornet2023}%
  \BibitemOpen
  \bibfield  {author} {\bibinfo {author} {\bibfnamefont {G.}~\bibnamefont {Bornet}}, \bibinfo {author} {\bibfnamefont {G.}~\bibnamefont {Emperauger}}, \bibinfo {author} {\bibfnamefont {C.}~\bibnamefont {Chen}}, \bibinfo {author} {\bibfnamefont {B.}~\bibnamefont {Ye}}, \bibinfo {author} {\bibfnamefont {M.}~\bibnamefont {Block}}, \bibinfo {author} {\bibfnamefont {M.}~\bibnamefont {Bintz}}, \bibinfo {author} {\bibfnamefont {J.~A.}\ \bibnamefont {Boyd}}, \bibinfo {author} {\bibfnamefont {D.}~\bibnamefont {Barredo}}, \bibinfo {author} {\bibfnamefont {T.}~\bibnamefont {Comparin}}, \bibinfo {author} {\bibfnamefont {F.}~\bibnamefont {Mezzacapo}}, \bibinfo {author} {\bibfnamefont {T.}~\bibnamefont {Roscilde}}, \bibinfo {author} {\bibfnamefont {T.}~\bibnamefont {Lahaye}}, \bibinfo {author} {\bibfnamefont {N.~Y.}\ \bibnamefont {Yao}}, \ and\ \bibinfo {author} {\bibfnamefont {A.}~\bibnamefont {Browaeys}},\ }\href {\doibase 10.1038/s41586-023-06414-9} {\bibfield  {journal} {\bibinfo  {journal} {Nature}\ }\textbf {\bibinfo
  {volume} {621}},\ \bibinfo {pages} {728} (\bibinfo {year} {2023})}\BibitemShut {NoStop}%
\bibitem [{\citenamefont {Eckner}\ \emph {et~al.}(2023)\citenamefont {Eckner}, \citenamefont {Darkwah~Oppong}, \citenamefont {Cao}, \citenamefont {Young}, \citenamefont {Milner}, \citenamefont {Robinson}, \citenamefont {Ye},\ and\ \citenamefont {Kaufman}}]{Eckner2023}%
  \BibitemOpen
  \bibfield  {author} {\bibinfo {author} {\bibfnamefont {W.~J.}\ \bibnamefont {Eckner}}, \bibinfo {author} {\bibfnamefont {N.}~\bibnamefont {Darkwah~Oppong}}, \bibinfo {author} {\bibfnamefont {A.}~\bibnamefont {Cao}}, \bibinfo {author} {\bibfnamefont {A.~W.}\ \bibnamefont {Young}}, \bibinfo {author} {\bibfnamefont {W.~R.}\ \bibnamefont {Milner}}, \bibinfo {author} {\bibfnamefont {J.~M.}\ \bibnamefont {Robinson}}, \bibinfo {author} {\bibfnamefont {J.}~\bibnamefont {Ye}}, \ and\ \bibinfo {author} {\bibfnamefont {A.~M.}\ \bibnamefont {Kaufman}},\ }\href {\doibase 10.1038/s41586-023-06360-6} {\bibfield  {journal} {\bibinfo  {journal} {Nature}\ }\textbf {\bibinfo {volume} {621}},\ \bibinfo {pages} {734} (\bibinfo {year} {2023})}\BibitemShut {NoStop}%
\bibitem [{\citenamefont {Robinson}\ \emph {et~al.}(2024)\citenamefont {Robinson}, \citenamefont {Miklos}, \citenamefont {Tso}, \citenamefont {Kennedy}, \citenamefont {Bothwell}, \citenamefont {Kedar}, \citenamefont {Thompson},\ and\ \citenamefont {Ye}}]{Robinson2024}%
  \BibitemOpen
  \bibfield  {author} {\bibinfo {author} {\bibfnamefont {J.~M.}\ \bibnamefont {Robinson}}, \bibinfo {author} {\bibfnamefont {M.}~\bibnamefont {Miklos}}, \bibinfo {author} {\bibfnamefont {Y.~M.}\ \bibnamefont {Tso}}, \bibinfo {author} {\bibfnamefont {C.~J.}\ \bibnamefont {Kennedy}}, \bibinfo {author} {\bibfnamefont {T.}~\bibnamefont {Bothwell}}, \bibinfo {author} {\bibfnamefont {D.}~\bibnamefont {Kedar}}, \bibinfo {author} {\bibfnamefont {J.~K.}\ \bibnamefont {Thompson}}, \ and\ \bibinfo {author} {\bibfnamefont {J.}~\bibnamefont {Ye}},\ }\href {\doibase 10.1038/s41567-023-02310-1} {\bibfield  {journal} {\bibinfo  {journal} {Nat. Phys.}\ }\textbf {\bibinfo {volume} {20}},\ \bibinfo {pages} {208} (\bibinfo {year} {2024})}\BibitemShut {NoStop}%
\bibitem [{\citenamefont {Zhang}\ \emph {et~al.}(2023)\citenamefont {Zhang}, \citenamefont {Yu}, \citenamefont {Jadbabaie},\ and\ \citenamefont {Hutzler}}]{PhysRevLett.131.193602}%
  \BibitemOpen
  \bibfield  {author} {\bibinfo {author} {\bibfnamefont {C.}~\bibnamefont {Zhang}}, \bibinfo {author} {\bibfnamefont {P.}~\bibnamefont {Yu}}, \bibinfo {author} {\bibfnamefont {A.}~\bibnamefont {Jadbabaie}}, \ and\ \bibinfo {author} {\bibfnamefont {N.~R.}\ \bibnamefont {Hutzler}},\ }\href {\doibase 10.1103/PhysRevLett.131.193602} {\bibfield  {journal} {\bibinfo  {journal} {Phys. Rev. Lett.}\ }\textbf {\bibinfo {volume} {131}},\ \bibinfo {pages} {193602} (\bibinfo {year} {2023})}\BibitemShut {NoStop}%
\bibitem [{\citenamefont {Lim}\ \emph {et~al.}(2018)\citenamefont {Lim}, \citenamefont {Almond}, \citenamefont {Trigatzis}, \citenamefont {Devlin}, \citenamefont {Fitch}, \citenamefont {Sauer}, \citenamefont {Tarbutt},\ and\ \citenamefont {Hinds}}]{PhysRevLett.120.123201}%
  \BibitemOpen
  \bibfield  {author} {\bibinfo {author} {\bibfnamefont {J.}~\bibnamefont {Lim}}, \bibinfo {author} {\bibfnamefont {J.~R.}\ \bibnamefont {Almond}}, \bibinfo {author} {\bibfnamefont {M.~A.}\ \bibnamefont {Trigatzis}}, \bibinfo {author} {\bibfnamefont {J.~A.}\ \bibnamefont {Devlin}}, \bibinfo {author} {\bibfnamefont {N.~J.}\ \bibnamefont {Fitch}}, \bibinfo {author} {\bibfnamefont {B.~E.}\ \bibnamefont {Sauer}}, \bibinfo {author} {\bibfnamefont {M.~R.}\ \bibnamefont {Tarbutt}}, \ and\ \bibinfo {author} {\bibfnamefont {E.~A.}\ \bibnamefont {Hinds}},\ }\href {\doibase 10.1103/PhysRevLett.120.123201} {\bibfield  {journal} {\bibinfo  {journal} {Phys. Rev. Lett.}\ }\textbf {\bibinfo {volume} {120}},\ \bibinfo {pages} {123201} (\bibinfo {year} {2018})}\BibitemShut {NoStop}%
\bibitem [{\citenamefont {Fitch}\ \emph {et~al.}(2020)\citenamefont {Fitch}, \citenamefont {Lim}, \citenamefont {Hinds}, \citenamefont {Sauer},\ and\ \citenamefont {Tarbutt}}]{Fitch_2021}%
  \BibitemOpen
  \bibfield  {author} {\bibinfo {author} {\bibfnamefont {N.~J.}\ \bibnamefont {Fitch}}, \bibinfo {author} {\bibfnamefont {J.}~\bibnamefont {Lim}}, \bibinfo {author} {\bibfnamefont {E.~A.}\ \bibnamefont {Hinds}}, \bibinfo {author} {\bibfnamefont {B.~E.}\ \bibnamefont {Sauer}}, \ and\ \bibinfo {author} {\bibfnamefont {M.~R.}\ \bibnamefont {Tarbutt}},\ }\href {\doibase 10.1088/2058-9565/abc931} {\bibfield  {journal} {\bibinfo  {journal} {Quantum Sci. Technol.}\ }\textbf {\bibinfo {volume} {6}},\ \bibinfo {pages} {014006} (\bibinfo {year} {2020})}\BibitemShut {NoStop}%
\bibitem [{\citenamefont {Kogel}\ \emph {et~al.}(2021)\citenamefont {Kogel}, \citenamefont {Rockenhäuser}, \citenamefont {Albrecht},\ and\ \citenamefont {Langen}}]{Kogel_2021}%
  \BibitemOpen
  \bibfield  {author} {\bibinfo {author} {\bibfnamefont {F.}~\bibnamefont {Kogel}}, \bibinfo {author} {\bibfnamefont {M.}~\bibnamefont {Rockenhäuser}}, \bibinfo {author} {\bibfnamefont {R.}~\bibnamefont {Albrecht}}, \ and\ \bibinfo {author} {\bibfnamefont {T.}~\bibnamefont {Langen}},\ }\href {\doibase 10.1088/1367-2630/ac1df2} {\bibfield  {journal} {\bibinfo  {journal} {New J. Phys.}\ }\textbf {\bibinfo {volume} {23}},\ \bibinfo {pages} {095003} (\bibinfo {year} {2021})}\BibitemShut {NoStop}%
\bibitem [{\citenamefont {Anderegg}\ \emph {et~al.}(2023)\citenamefont {Anderegg}, \citenamefont {Vilas}, \citenamefont {Hallas}, \citenamefont {Robichaud}, \citenamefont {Jadbabaie}, \citenamefont {Doyle},\ and\ \citenamefont {Hutzler}}]{doi:10.1126/science.adg8155}%
  \BibitemOpen
  \bibfield  {author} {\bibinfo {author} {\bibfnamefont {L.}~\bibnamefont {Anderegg}}, \bibinfo {author} {\bibfnamefont {N.~B.}\ \bibnamefont {Vilas}}, \bibinfo {author} {\bibfnamefont {C.}~\bibnamefont {Hallas}}, \bibinfo {author} {\bibfnamefont {P.}~\bibnamefont {Robichaud}}, \bibinfo {author} {\bibfnamefont {A.}~\bibnamefont {Jadbabaie}}, \bibinfo {author} {\bibfnamefont {J.~M.}\ \bibnamefont {Doyle}}, \ and\ \bibinfo {author} {\bibfnamefont {N.~R.}\ \bibnamefont {Hutzler}},\ }\href {\doibase 10.1126/science.adg8155} {\bibfield  {journal} {\bibinfo  {journal} {Science}\ }\textbf {\bibinfo {volume} {382}},\ \bibinfo {pages} {665} (\bibinfo {year} {2023})}\BibitemShut {NoStop}%
\bibitem [{\citenamefont {Chamorro}\ \emph {et~al.}(2022)\citenamefont {Chamorro}, \citenamefont {Borschevsky}, \citenamefont {Eliav}, \citenamefont {Hutzler}, \citenamefont {Hoekstra},\ and\ \citenamefont {Pa\ifmmode~\check{s}\else \v{s}\fi{}teka}}]{PhysRevA.106.052811}%
  \BibitemOpen
  \bibfield  {author} {\bibinfo {author} {\bibfnamefont {Y.}~\bibnamefont {Chamorro}}, \bibinfo {author} {\bibfnamefont {A.}~\bibnamefont {Borschevsky}}, \bibinfo {author} {\bibfnamefont {E.}~\bibnamefont {Eliav}}, \bibinfo {author} {\bibfnamefont {N.~R.}\ \bibnamefont {Hutzler}}, \bibinfo {author} {\bibfnamefont {S.}~\bibnamefont {Hoekstra}}, \ and\ \bibinfo {author} {\bibfnamefont {L.~c. v.~F.}\ \bibnamefont {Pa\ifmmode~\check{s}\else \v{s}\fi{}teka}},\ }\href {\doibase 10.1103/PhysRevA.106.052811} {\bibfield  {journal} {\bibinfo  {journal} {Phys. Rev. A}\ }\textbf {\bibinfo {volume} {106}},\ \bibinfo {pages} {052811} (\bibinfo {year} {2022})}\BibitemShut {NoStop}%
\bibitem [{\citenamefont {K\"ohler}\ \emph {et~al.}(2006)\citenamefont {K\"ohler}, \citenamefont {G\'oral},\ and\ \citenamefont {Julienne}}]{RevModPhys.78.1311}%
  \BibitemOpen
  \bibfield  {author} {\bibinfo {author} {\bibfnamefont {T.}~\bibnamefont {K\"ohler}}, \bibinfo {author} {\bibfnamefont {K.}~\bibnamefont {G\'oral}}, \ and\ \bibinfo {author} {\bibfnamefont {P.~S.}\ \bibnamefont {Julienne}},\ }\href {\doibase 10.1103/RevModPhys.78.1311} {\bibfield  {journal} {\bibinfo  {journal} {Rev. Mod. Phys.}\ }\textbf {\bibinfo {volume} {78}},\ \bibinfo {pages} {1311} (\bibinfo {year} {2006})}\BibitemShut {NoStop}%
\bibitem [{\citenamefont {Chin}\ \emph {et~al.}(2010)\citenamefont {Chin}, \citenamefont {Grimm}, \citenamefont {Julienne},\ and\ \citenamefont {Tiesinga}}]{RevModPhys.82.1225}%
  \BibitemOpen
  \bibfield  {author} {\bibinfo {author} {\bibfnamefont {C.}~\bibnamefont {Chin}}, \bibinfo {author} {\bibfnamefont {R.}~\bibnamefont {Grimm}}, \bibinfo {author} {\bibfnamefont {P.}~\bibnamefont {Julienne}}, \ and\ \bibinfo {author} {\bibfnamefont {E.}~\bibnamefont {Tiesinga}},\ }\href {\doibase 10.1103/RevModPhys.82.1225} {\bibfield  {journal} {\bibinfo  {journal} {Rev. Mod. Phys.}\ }\textbf {\bibinfo {volume} {82}},\ \bibinfo {pages} {1225} (\bibinfo {year} {2010})}\BibitemShut {NoStop}%
\bibitem [{\citenamefont {Vitanov}\ \emph {et~al.}(2017)\citenamefont {Vitanov}, \citenamefont {Rangelov}, \citenamefont {Shore},\ and\ \citenamefont {Bergmann}}]{RevModPhys.89.015006}%
  \BibitemOpen
  \bibfield  {author} {\bibinfo {author} {\bibfnamefont {N.~V.}\ \bibnamefont {Vitanov}}, \bibinfo {author} {\bibfnamefont {A.~A.}\ \bibnamefont {Rangelov}}, \bibinfo {author} {\bibfnamefont {B.~W.}\ \bibnamefont {Shore}}, \ and\ \bibinfo {author} {\bibfnamefont {K.}~\bibnamefont {Bergmann}},\ }\href {\doibase 10.1103/RevModPhys.89.015006} {\bibfield  {journal} {\bibinfo  {journal} {Rev. Mod. Phys.}\ }\textbf {\bibinfo {volume} {89}},\ \bibinfo {pages} {015006} (\bibinfo {year} {2017})}\BibitemShut {NoStop}%
\bibitem [{\citenamefont {Valtolina}\ \emph {et~al.}(2020)\citenamefont {Valtolina}, \citenamefont {Matsuda}, \citenamefont {Tobias}, \citenamefont {Li}, \citenamefont {De~Marco},\ and\ \citenamefont {Ye}}]{Valtolina2020}%
  \BibitemOpen
  \bibfield  {author} {\bibinfo {author} {\bibfnamefont {G.}~\bibnamefont {Valtolina}}, \bibinfo {author} {\bibfnamefont {K.}~\bibnamefont {Matsuda}}, \bibinfo {author} {\bibfnamefont {W.~G.}\ \bibnamefont {Tobias}}, \bibinfo {author} {\bibfnamefont {J.-R.}\ \bibnamefont {Li}}, \bibinfo {author} {\bibfnamefont {L.}~\bibnamefont {De~Marco}}, \ and\ \bibinfo {author} {\bibfnamefont {J.}~\bibnamefont {Ye}},\ }\href {\doibase 10.1038/s41586-020-2980-7} {\bibfield  {journal} {\bibinfo  {journal} {Nature}\ }\textbf {\bibinfo {volume} {588}},\ \bibinfo {pages} {239} (\bibinfo {year} {2020})}\BibitemShut {NoStop}%
\bibitem [{\citenamefont {Duda}\ \emph {et~al.}(2023)\citenamefont {Duda}, \citenamefont {Chen}, \citenamefont {Schindewolf}, \citenamefont {Bause}, \citenamefont {von Milczewski}, \citenamefont {Schmidt}, \citenamefont {Bloch},\ and\ \citenamefont {Luo}}]{Duda2023}%
  \BibitemOpen
  \bibfield  {author} {\bibinfo {author} {\bibfnamefont {M.}~\bibnamefont {Duda}}, \bibinfo {author} {\bibfnamefont {X.-Y.}\ \bibnamefont {Chen}}, \bibinfo {author} {\bibfnamefont {A.}~\bibnamefont {Schindewolf}}, \bibinfo {author} {\bibfnamefont {R.}~\bibnamefont {Bause}}, \bibinfo {author} {\bibfnamefont {J.}~\bibnamefont {von Milczewski}}, \bibinfo {author} {\bibfnamefont {R.}~\bibnamefont {Schmidt}}, \bibinfo {author} {\bibfnamefont {I.}~\bibnamefont {Bloch}}, \ and\ \bibinfo {author} {\bibfnamefont {X.-Y.}\ \bibnamefont {Luo}},\ }\href {\doibase 10.1038/s41567-023-01948-1} {\bibfield  {journal} {\bibinfo  {journal} {Nat. Phys.}\ }\textbf {\bibinfo {volume} {19}},\ \bibinfo {pages} {720} (\bibinfo {year} {2023})}\BibitemShut {NoStop}%
\bibitem [{\citenamefont {Bigagli}\ \emph {et~al.}(2024)\citenamefont {Bigagli}, \citenamefont {Yuan}, \citenamefont {Zhang}, \citenamefont {Bulatovic}, \citenamefont {Karman}, \citenamefont {Stevenson},\ and\ \citenamefont {Will}}]{Bigagli2024}%
  \BibitemOpen
  \bibfield  {author} {\bibinfo {author} {\bibfnamefont {N.}~\bibnamefont {Bigagli}}, \bibinfo {author} {\bibfnamefont {W.}~\bibnamefont {Yuan}}, \bibinfo {author} {\bibfnamefont {S.}~\bibnamefont {Zhang}}, \bibinfo {author} {\bibfnamefont {B.}~\bibnamefont {Bulatovic}}, \bibinfo {author} {\bibfnamefont {T.}~\bibnamefont {Karman}}, \bibinfo {author} {\bibfnamefont {I.}~\bibnamefont {Stevenson}}, \ and\ \bibinfo {author} {\bibfnamefont {S.}~\bibnamefont {Will}},\ }\href {\doibase 10.1038/s41586-024-07492-z} {\bibfield  {journal} {\bibinfo  {journal} {Nature}\ }\textbf {\bibinfo {volume} {631}},\ \bibinfo {pages} {289} (\bibinfo {year} {2024})}\BibitemShut {NoStop}%
\bibitem [{\citenamefont {Karman}\ \emph {et~al.}(2024)\citenamefont {Karman}, \citenamefont {Tomza},\ and\ \citenamefont {P{\'e}rez-R{\'i}os}}]{Karman2024}%
  \BibitemOpen
  \bibfield  {author} {\bibinfo {author} {\bibfnamefont {T.}~\bibnamefont {Karman}}, \bibinfo {author} {\bibfnamefont {M.}~\bibnamefont {Tomza}}, \ and\ \bibinfo {author} {\bibfnamefont {J.}~\bibnamefont {P{\'e}rez-R{\'i}os}},\ }\href {\doibase 10.1038/s41567-024-02467-3} {\bibfield  {journal} {\bibinfo  {journal} {Nat. Phys.}\ }\textbf {\bibinfo {volume} {20}},\ \bibinfo {pages} {722} (\bibinfo {year} {2024})}\BibitemShut {NoStop}%
\bibitem [{\citenamefont {Cornish}\ \emph {et~al.}(2024)\citenamefont {Cornish}, \citenamefont {Tarbutt},\ and\ \citenamefont {Hazzard}}]{Cornish2024}%
  \BibitemOpen
  \bibfield  {author} {\bibinfo {author} {\bibfnamefont {S.~L.}\ \bibnamefont {Cornish}}, \bibinfo {author} {\bibfnamefont {M.~R.}\ \bibnamefont {Tarbutt}}, \ and\ \bibinfo {author} {\bibfnamefont {K.~R.~A.}\ \bibnamefont {Hazzard}},\ }\href {\doibase 10.1038/s41567-024-02453-9} {\bibfield  {journal} {\bibinfo  {journal} {Nat. Phys.}\ }\textbf {\bibinfo {volume} {20}},\ \bibinfo {pages} {730} (\bibinfo {year} {2024})}\BibitemShut {NoStop}%
\bibitem [{\citenamefont {Holland}\ \emph {et~al.}(2023)\citenamefont {Holland}, \citenamefont {Yukai},\ and\ \citenamefont {Cheuk}}]{holland}%
  \BibitemOpen
  \bibfield  {author} {\bibinfo {author} {\bibfnamefont {C.~M.}\ \bibnamefont {Holland}}, \bibinfo {author} {\bibfnamefont {L.}~\bibnamefont {Yukai}}, \ and\ \bibinfo {author} {\bibfnamefont {L.~W.}\ \bibnamefont {Cheuk}},\ }\href {\doibase 10.1126/science.adf4272} {\bibfield  {journal} {\bibinfo  {journal} {Science}\ }\textbf {\bibinfo {volume} {382}},\ \bibinfo {pages} {1143} (\bibinfo {year} {2023})}\BibitemShut {NoStop}%
\bibitem [{\citenamefont {Meyer}\ and\ \citenamefont {Bohn}(2009)}]{PhysRevA.80.042508}%
  \BibitemOpen
  \bibfield  {author} {\bibinfo {author} {\bibfnamefont {E.~R.}\ \bibnamefont {Meyer}}\ and\ \bibinfo {author} {\bibfnamefont {J.~L.}\ \bibnamefont {Bohn}},\ }\href {\doibase 10.1103/PhysRevA.80.042508} {\bibfield  {journal} {\bibinfo  {journal} {Phys. Rev. A}\ }\textbf {\bibinfo {volume} {80}},\ \bibinfo {pages} {042508} (\bibinfo {year} {2009})}\BibitemShut {NoStop}%
\bibitem [{\citenamefont {Barb{\'e}}\ \emph {et~al.}(2018)\citenamefont {Barb{\'e}}, \citenamefont {Ciamei}, \citenamefont {Pasquiou}, \citenamefont {Reichs{\"o}llner}, \citenamefont {Schreck}, \citenamefont {{\.{Z}}uchowski},\ and\ \citenamefont {Hutson}}]{Barbé2018}%
  \BibitemOpen
  \bibfield  {author} {\bibinfo {author} {\bibfnamefont {V.}~\bibnamefont {Barb{\'e}}}, \bibinfo {author} {\bibfnamefont {A.}~\bibnamefont {Ciamei}}, \bibinfo {author} {\bibfnamefont {B.}~\bibnamefont {Pasquiou}}, \bibinfo {author} {\bibfnamefont {L.}~\bibnamefont {Reichs{\"o}llner}}, \bibinfo {author} {\bibfnamefont {F.}~\bibnamefont {Schreck}}, \bibinfo {author} {\bibfnamefont {P.~S.}\ \bibnamefont {{\.{Z}}uchowski}}, \ and\ \bibinfo {author} {\bibfnamefont {J.~M.}\ \bibnamefont {Hutson}},\ }\href {\doibase 10.1038/s41567-018-0169-x} {\bibfield  {journal} {\bibinfo  {journal} {Nat. Phys.}\ }\textbf {\bibinfo {volume} {14}},\ \bibinfo {pages} {881} (\bibinfo {year} {2018})}\BibitemShut {NoStop}%
\bibitem [{\citenamefont {Verma}\ \emph {et~al.}(2020)\citenamefont {Verma}, \citenamefont {Jayich},\ and\ \citenamefont {Vutha}}]{PhysRevLett.125.153201}%
  \BibitemOpen
  \bibfield  {author} {\bibinfo {author} {\bibfnamefont {M.}~\bibnamefont {Verma}}, \bibinfo {author} {\bibfnamefont {A.~M.}\ \bibnamefont {Jayich}}, \ and\ \bibinfo {author} {\bibfnamefont {A.~C.}\ \bibnamefont {Vutha}},\ }\href {\doibase 10.1103/PhysRevLett.125.153201} {\bibfield  {journal} {\bibinfo  {journal} {Phys. Rev. Lett.}\ }\textbf {\bibinfo {volume} {125}},\ \bibinfo {pages} {153201} (\bibinfo {year} {2020})}\BibitemShut {NoStop}%
\bibitem [{\citenamefont {Polet}\ \emph {et~al.}(2024)\citenamefont {Polet}, \citenamefont {Chamorro}, \citenamefont {Pašteka}, \citenamefont {Hoekstra}, \citenamefont {Tomza}, \citenamefont {Borschevsky},\ and\ \citenamefont {Aucar}}]{PoletJCP24}%
  \BibitemOpen
  \bibfield  {author} {\bibinfo {author} {\bibfnamefont {J.~D.}\ \bibnamefont {Polet}}, \bibinfo {author} {\bibfnamefont {Y.}~\bibnamefont {Chamorro}}, \bibinfo {author} {\bibfnamefont {L.~F.}\ \bibnamefont {Pašteka}}, \bibinfo {author} {\bibfnamefont {S.}~\bibnamefont {Hoekstra}}, \bibinfo {author} {\bibfnamefont {M.}~\bibnamefont {Tomza}}, \bibinfo {author} {\bibfnamefont {A.}~\bibnamefont {Borschevsky}}, \ and\ \bibinfo {author} {\bibfnamefont {I.~A.}\ \bibnamefont {Aucar}},\ }\href {\doibase 10.1063/5.0235522} {\bibfield  {journal} {\bibinfo  {journal} {J. Chem. Phys.}\ }\textbf {\bibinfo {volume} {161}},\ \bibinfo {pages} {234302} (\bibinfo {year} {2024})}\BibitemShut {NoStop}%
\bibitem [{\citenamefont {Fleig}\ and\ \citenamefont {DeMille}(2021)}]{Fleig_2021}%
  \BibitemOpen
  \bibfield  {author} {\bibinfo {author} {\bibfnamefont {T.}~\bibnamefont {Fleig}}\ and\ \bibinfo {author} {\bibfnamefont {D.}~\bibnamefont {DeMille}},\ }\href {\doibase 10.1088/1367-2630/ac3619} {\bibfield  {journal} {\bibinfo  {journal} {New J. Phys.}\ }\textbf {\bibinfo {volume} {23}},\ \bibinfo {pages} {113039} (\bibinfo {year} {2021})}\BibitemShut {NoStop}%
\bibitem [{\citenamefont {Khriplovich}\ and\ \citenamefont {Lamoreaux}(1997)}]{Khriplovich1997}%
  \BibitemOpen
  \bibfield  {author} {\bibinfo {author} {\bibfnamefont {I.~B.}\ \bibnamefont {Khriplovich}}\ and\ \bibinfo {author} {\bibfnamefont {S.~K.}\ \bibnamefont {Lamoreaux}},\ }\enquote {\bibinfo {title} {General features of edm experiments},}\ in\ \href {\doibase 10.1007/978-3-642-60838-4_3} {\emph {\bibinfo {booktitle} {CP Violation Without Strangeness: Electric Dipole Moments of Particles, Atoms, and Molecules}}}\ (\bibinfo  {publisher} {Springer Berlin Heidelberg},\ \bibinfo {address} {Berlin, Heidelberg},\ \bibinfo {year} {1997})\ pp.\ \bibinfo {pages} {19--51}\BibitemShut {NoStop}%
\bibitem [{\citenamefont {Vutha}(2011)}]{PhDVutha}%
  \BibitemOpen
  \bibfield  {author} {\bibinfo {author} {\bibfnamefont {A.}~\bibnamefont {Vutha}},\ }\emph {\bibinfo {title} {A search for the electric dipole moment of the electron using thorium monoxide}},\ \href@noop {} {Ph.D. thesis},\ \bibinfo  {school} {Yale University} (\bibinfo {year} {2011})\BibitemShut {NoStop}%
\bibitem [{\citenamefont {Baron}\ \emph {et~al.}(2017)\citenamefont {Baron}, \citenamefont {Campbell}, \citenamefont {DeMille}, \citenamefont {Doyle}, \citenamefont {Gabrielse}, \citenamefont {Gurevich}, \citenamefont {Hess}, \citenamefont {Hutzler}, \citenamefont {Kirilov}, \citenamefont {Kozyryev}, \citenamefont {O’Leary}, \citenamefont {Panda}, \citenamefont {Parsons}, \citenamefont {Spaun}, \citenamefont {Vutha}, \citenamefont {West}, \citenamefont {West},\ and\ \citenamefont {Collaboration}}]{Baron_2017}%
  \BibitemOpen
  \bibfield  {author} {\bibinfo {author} {\bibfnamefont {J.}~\bibnamefont {Baron}}, \bibinfo {author} {\bibfnamefont {W.~C.}\ \bibnamefont {Campbell}}, \bibinfo {author} {\bibfnamefont {D.}~\bibnamefont {DeMille}}, \bibinfo {author} {\bibfnamefont {J.~M.}\ \bibnamefont {Doyle}}, \bibinfo {author} {\bibfnamefont {G.}~\bibnamefont {Gabrielse}}, \bibinfo {author} {\bibfnamefont {Y.~V.}\ \bibnamefont {Gurevich}}, \bibinfo {author} {\bibfnamefont {P.~W.}\ \bibnamefont {Hess}}, \bibinfo {author} {\bibfnamefont {N.~R.}\ \bibnamefont {Hutzler}}, \bibinfo {author} {\bibfnamefont {E.}~\bibnamefont {Kirilov}}, \bibinfo {author} {\bibfnamefont {I.}~\bibnamefont {Kozyryev}}, \bibinfo {author} {\bibfnamefont {B.~R.}\ \bibnamefont {O’Leary}}, \bibinfo {author} {\bibfnamefont {C.~D.}\ \bibnamefont {Panda}}, \bibinfo {author} {\bibfnamefont {M.~F.}\ \bibnamefont {Parsons}}, \bibinfo {author} {\bibfnamefont {B.}~\bibnamefont {Spaun}}, \bibinfo {author} {\bibfnamefont {A.~C.}\ \bibnamefont {Vutha}}, \bibinfo {author}
  {\bibfnamefont {A.~D.}\ \bibnamefont {West}}, \bibinfo {author} {\bibfnamefont {E.~P.}\ \bibnamefont {West}}, \ and\ \bibinfo {author} {\bibfnamefont {A.}~\bibnamefont {Collaboration}},\ }\href {\doibase 10.1088/1367-2630/aa708e} {\bibfield  {journal} {\bibinfo  {journal} {New J. Phys.}\ }\textbf {\bibinfo {volume} {19}},\ \bibinfo {pages} {073029} (\bibinfo {year} {2017})}\BibitemShut {NoStop}%
\bibitem [{\citenamefont {Brown}\ and\ \citenamefont {Carrington}(2003)}]{Brown_Carrington_2003}%
  \BibitemOpen
  \bibfield  {author} {\bibinfo {author} {\bibfnamefont {J.~M.}\ \bibnamefont {Brown}}\ and\ \bibinfo {author} {\bibfnamefont {A.}~\bibnamefont {Carrington}},\ }\href@noop {} {\emph {\bibinfo {title} {Rotational Spectroscopy of Diatomic Molecules}}},\ Cambridge Molecular Science\ (\bibinfo  {publisher} {Cambridge University Press},\ \bibinfo {year} {2003})\BibitemShut {NoStop}%
\bibitem [{Sup()}]{SupplInfo}%
  \BibitemOpen
  \href@noop {} {}\bibinfo {note} {See Supplemental Material for detailed information and complementary results.}\BibitemShut {Stop}%
\bibitem [{\citenamefont {Kopp}\ and\ \citenamefont {Hougen}(1967)}]{doi:10.1139/p67-209}%
  \BibitemOpen
  \bibfield  {author} {\bibinfo {author} {\bibfnamefont {I.}~\bibnamefont {Kopp}}\ and\ \bibinfo {author} {\bibfnamefont {J.~T.}\ \bibnamefont {Hougen}},\ }\href {\doibase 10.1139/p67-209} {\bibfield  {journal} {\bibinfo  {journal} {Can. J. Phys.}\ }\textbf {\bibinfo {volume} {45}},\ \bibinfo {pages} {2581} (\bibinfo {year} {1967})}\BibitemShut {NoStop}%
\bibitem [{\citenamefont {Halfen}\ and\ \citenamefont {Ziurys}(2007)}]{Halfen_2008}%
  \BibitemOpen
  \bibfield  {author} {\bibinfo {author} {\bibfnamefont {D.~T.}\ \bibnamefont {Halfen}}\ and\ \bibinfo {author} {\bibfnamefont {L.~M.}\ \bibnamefont {Ziurys}},\ }\href {\doibase 10.1086/526399} {\bibfield  {journal} {\bibinfo  {journal} {Astrophys. J.}\ }\textbf {\bibinfo {volume} {672}},\ \bibinfo {pages} {L77} (\bibinfo {year} {2007})}\BibitemShut {NoStop}%
\bibitem [{\citenamefont {Halfen}\ and\ \citenamefont {Ziurys}(2005)}]{10.1063/1.1824036}%
  \BibitemOpen
  \bibfield  {author} {\bibinfo {author} {\bibfnamefont {D.~T.}\ \bibnamefont {Halfen}}\ and\ \bibinfo {author} {\bibfnamefont {L.~M.}\ \bibnamefont {Ziurys}},\ }\href {\doibase 10.1063/1.1824036} {\bibfield  {journal} {\bibinfo  {journal} {J. Chem. Phys.}\ }\textbf {\bibinfo {volume} {122}},\ \bibinfo {pages} {054309} (\bibinfo {year} {2005})}\BibitemShut {NoStop}%
\bibitem [{\citenamefont {Schreck}\ and\ \citenamefont {Druten}(2021)}]{Schreck2021}%
  \BibitemOpen
  \bibfield  {author} {\bibinfo {author} {\bibfnamefont {F.}~\bibnamefont {Schreck}}\ and\ \bibinfo {author} {\bibfnamefont {K.~v.}\ \bibnamefont {Druten}},\ }\href {\doibase 10.1038/s41567-021-01379-w} {\bibfield  {journal} {\bibinfo  {journal} {Nat. Phys.}\ }\textbf {\bibinfo {volume} {17}},\ \bibinfo {pages} {1296} (\bibinfo {year} {2021})}\BibitemShut {NoStop}%
\bibitem [{\citenamefont {He}\ \emph {et~al.}(2019)\citenamefont {He}, \citenamefont {Hajiyev}, \citenamefont {Ren}, \citenamefont {Song},\ and\ \citenamefont {Jo}}]{He_2019}%
  \BibitemOpen
  \bibfield  {author} {\bibinfo {author} {\bibfnamefont {C.}~\bibnamefont {He}}, \bibinfo {author} {\bibfnamefont {E.}~\bibnamefont {Hajiyev}}, \bibinfo {author} {\bibfnamefont {Z.}~\bibnamefont {Ren}}, \bibinfo {author} {\bibfnamefont {B.}~\bibnamefont {Song}}, \ and\ \bibinfo {author} {\bibfnamefont {G.-B.}\ \bibnamefont {Jo}},\ }\href {\doibase 10.1088/1361-6455/ab153e} {\bibfield  {journal} {\bibinfo  {journal} {J. Phys. B:At., Mol. Opt.}\ }\textbf {\bibinfo {volume} {52}},\ \bibinfo {pages} {102001} (\bibinfo {year} {2019})}\BibitemShut {NoStop}%
\bibitem [{\citenamefont {Chomaz}\ \emph {et~al.}(2022)\citenamefont {Chomaz}, \citenamefont {Ferrier-Barbut}, \citenamefont {Ferlaino}, \citenamefont {Laburthe-Tolra}, \citenamefont {Lev},\ and\ \citenamefont {Pfau}}]{Chomaz_2023}%
  \BibitemOpen
  \bibfield  {author} {\bibinfo {author} {\bibfnamefont {L.}~\bibnamefont {Chomaz}}, \bibinfo {author} {\bibfnamefont {I.}~\bibnamefont {Ferrier-Barbut}}, \bibinfo {author} {\bibfnamefont {F.}~\bibnamefont {Ferlaino}}, \bibinfo {author} {\bibfnamefont {B.}~\bibnamefont {Laburthe-Tolra}}, \bibinfo {author} {\bibfnamefont {B.~L.}\ \bibnamefont {Lev}}, \ and\ \bibinfo {author} {\bibfnamefont {T.}~\bibnamefont {Pfau}},\ }\href {\doibase 10.1088/1361-6633/aca814} {\bibfield  {journal} {\bibinfo  {journal} {Rep. Prog. Phys.}\ }\textbf {\bibinfo {volume} {86}},\ \bibinfo {pages} {026401} (\bibinfo {year} {2022})}\BibitemShut {NoStop}%
\bibitem [{\citenamefont {Lee}\ \emph {et~al.}(2015)\citenamefont {Lee}, \citenamefont {Lee}, \citenamefont {Noh},\ and\ \citenamefont {Mun}}]{PhysRevA.91.053405}%
  \BibitemOpen
  \bibfield  {author} {\bibinfo {author} {\bibfnamefont {J.}~\bibnamefont {Lee}}, \bibinfo {author} {\bibfnamefont {J.~H.}\ \bibnamefont {Lee}}, \bibinfo {author} {\bibfnamefont {J.}~\bibnamefont {Noh}}, \ and\ \bibinfo {author} {\bibfnamefont {J.}~\bibnamefont {Mun}},\ }\href {\doibase 10.1103/PhysRevA.91.053405} {\bibfield  {journal} {\bibinfo  {journal} {Phys. Rev. A}\ }\textbf {\bibinfo {volume} {91}},\ \bibinfo {pages} {053405} (\bibinfo {year} {2015})}\BibitemShut {NoStop}%
\bibitem [{\citenamefont {Gabardos}\ \emph {et~al.}(2019)\citenamefont {Gabardos}, \citenamefont {Lepoutre}, \citenamefont {Gorceix}, \citenamefont {Vernac},\ and\ \citenamefont {Laburthe-Tolra}}]{PhysRevA.99.023607}%
  \BibitemOpen
  \bibfield  {author} {\bibinfo {author} {\bibfnamefont {L.}~\bibnamefont {Gabardos}}, \bibinfo {author} {\bibfnamefont {S.}~\bibnamefont {Lepoutre}}, \bibinfo {author} {\bibfnamefont {O.}~\bibnamefont {Gorceix}}, \bibinfo {author} {\bibfnamefont {L.}~\bibnamefont {Vernac}}, \ and\ \bibinfo {author} {\bibfnamefont {B.}~\bibnamefont {Laburthe-Tolra}},\ }\href {\doibase 10.1103/PhysRevA.99.023607} {\bibfield  {journal} {\bibinfo  {journal} {Phys. Rev. A}\ }\textbf {\bibinfo {volume} {99}},\ \bibinfo {pages} {023607} (\bibinfo {year} {2019})}\BibitemShut {NoStop}%
\bibitem [{\citenamefont {Ciamei}\ \emph {et~al.}(2022)\citenamefont {Ciamei}, \citenamefont {Finelli}, \citenamefont {Cosco}, \citenamefont {Inguscio}, \citenamefont {Trenkwalder},\ and\ \citenamefont {Zaccanti}}]{PhysRevA.106.053318}%
  \BibitemOpen
  \bibfield  {author} {\bibinfo {author} {\bibfnamefont {A.}~\bibnamefont {Ciamei}}, \bibinfo {author} {\bibfnamefont {S.}~\bibnamefont {Finelli}}, \bibinfo {author} {\bibfnamefont {A.}~\bibnamefont {Cosco}}, \bibinfo {author} {\bibfnamefont {M.}~\bibnamefont {Inguscio}}, \bibinfo {author} {\bibfnamefont {A.}~\bibnamefont {Trenkwalder}}, \ and\ \bibinfo {author} {\bibfnamefont {M.}~\bibnamefont {Zaccanti}},\ }\href {\doibase 10.1103/PhysRevA.106.053318} {\bibfield  {journal} {\bibinfo  {journal} {Phys. Rev. A}\ }\textbf {\bibinfo {volume} {106}},\ \bibinfo {pages} {053318} (\bibinfo {year} {2022})}\BibitemShut {NoStop}%
\bibitem [{\citenamefont {Kitagawa}\ \emph {et~al.}(2008)\citenamefont {Kitagawa}, \citenamefont {Enomoto}, \citenamefont {Kasa}, \citenamefont {Takahashi}, \citenamefont {Ciury\l{}o}, \citenamefont {Naidon},\ and\ \citenamefont {Julienne}}]{PhysRevA.77.012719}%
  \BibitemOpen
  \bibfield  {author} {\bibinfo {author} {\bibfnamefont {M.}~\bibnamefont {Kitagawa}}, \bibinfo {author} {\bibfnamefont {K.}~\bibnamefont {Enomoto}}, \bibinfo {author} {\bibfnamefont {K.}~\bibnamefont {Kasa}}, \bibinfo {author} {\bibfnamefont {Y.}~\bibnamefont {Takahashi}}, \bibinfo {author} {\bibfnamefont {R.}~\bibnamefont {Ciury\l{}o}}, \bibinfo {author} {\bibfnamefont {P.}~\bibnamefont {Naidon}}, \ and\ \bibinfo {author} {\bibfnamefont {P.~S.}\ \bibnamefont {Julienne}},\ }\href {\doibase 10.1103/PhysRevA.77.012719} {\bibfield  {journal} {\bibinfo  {journal} {Phys. Rev. A}\ }\textbf {\bibinfo {volume} {77}},\ \bibinfo {pages} {012719} (\bibinfo {year} {2008})}\BibitemShut {NoStop}%
\bibitem [{\citenamefont {Werner}\ \emph {et~al.}(2005)\citenamefont {Werner}, \citenamefont {Griesmaier}, \citenamefont {Hensler}, \citenamefont {Stuhler}, \citenamefont {Pfau}, \citenamefont {Simoni},\ and\ \citenamefont {Tiesinga}}]{PhysRevLett.94.183201}%
  \BibitemOpen
  \bibfield  {author} {\bibinfo {author} {\bibfnamefont {J.}~\bibnamefont {Werner}}, \bibinfo {author} {\bibfnamefont {A.}~\bibnamefont {Griesmaier}}, \bibinfo {author} {\bibfnamefont {S.}~\bibnamefont {Hensler}}, \bibinfo {author} {\bibfnamefont {J.}~\bibnamefont {Stuhler}}, \bibinfo {author} {\bibfnamefont {T.}~\bibnamefont {Pfau}}, \bibinfo {author} {\bibfnamefont {A.}~\bibnamefont {Simoni}}, \ and\ \bibinfo {author} {\bibfnamefont {E.}~\bibnamefont {Tiesinga}},\ }\href {\doibase 10.1103/PhysRevLett.94.183201} {\bibfield  {journal} {\bibinfo  {journal} {Phys. Rev. Lett.}\ }\textbf {\bibinfo {volume} {94}},\ \bibinfo {pages} {183201} (\bibinfo {year} {2005})}\BibitemShut {NoStop}%
\bibitem [{\citenamefont {Pasquiou}\ \emph {et~al.}(2010)\citenamefont {Pasquiou}, \citenamefont {Bismut}, \citenamefont {Beaufils}, \citenamefont {Crubellier}, \citenamefont {Mar\'echal}, \citenamefont {Pedri}, \citenamefont {Vernac}, \citenamefont {Gorceix},\ and\ \citenamefont {Laburthe-Tolra}}]{PhysRevA.81.042716}%
  \BibitemOpen
  \bibfield  {author} {\bibinfo {author} {\bibfnamefont {B.}~\bibnamefont {Pasquiou}}, \bibinfo {author} {\bibfnamefont {G.}~\bibnamefont {Bismut}}, \bibinfo {author} {\bibfnamefont {Q.}~\bibnamefont {Beaufils}}, \bibinfo {author} {\bibfnamefont {A.}~\bibnamefont {Crubellier}}, \bibinfo {author} {\bibfnamefont {E.}~\bibnamefont {Mar\'echal}}, \bibinfo {author} {\bibfnamefont {P.}~\bibnamefont {Pedri}}, \bibinfo {author} {\bibfnamefont {L.}~\bibnamefont {Vernac}}, \bibinfo {author} {\bibfnamefont {O.}~\bibnamefont {Gorceix}}, \ and\ \bibinfo {author} {\bibfnamefont {B.}~\bibnamefont {Laburthe-Tolra}},\ }\href {\doibase 10.1103/PhysRevA.81.042716} {\bibfield  {journal} {\bibinfo  {journal} {Phys. Rev. A}\ }\textbf {\bibinfo {volume} {81}},\ \bibinfo {pages} {042716} (\bibinfo {year} {2010})}\BibitemShut {NoStop}%
\bibitem [{\citenamefont {Roy}\ \emph {et~al.}(2016)\citenamefont {Roy}, \citenamefont {Green}, \citenamefont {Bowler},\ and\ \citenamefont {Gupta}}]{PhysRevA.93.043403}%
  \BibitemOpen
  \bibfield  {author} {\bibinfo {author} {\bibfnamefont {R.}~\bibnamefont {Roy}}, \bibinfo {author} {\bibfnamefont {A.}~\bibnamefont {Green}}, \bibinfo {author} {\bibfnamefont {R.}~\bibnamefont {Bowler}}, \ and\ \bibinfo {author} {\bibfnamefont {S.}~\bibnamefont {Gupta}},\ }\href {\doibase 10.1103/PhysRevA.93.043403} {\bibfield  {journal} {\bibinfo  {journal} {Phys. Rev. A}\ }\textbf {\bibinfo {volume} {93}},\ \bibinfo {pages} {043403} (\bibinfo {year} {2016})}\BibitemShut {NoStop}%
\bibitem [{\citenamefont {Frye}\ \emph {et~al.}(2024)\citenamefont {Frye}, \citenamefont {\ifmmode~\dot{Z}\else \.{Z}\fi{}uchowski},\ and\ \citenamefont {Tomza}}]{PhysRevResearch.6.023254}%
  \BibitemOpen
  \bibfield  {author} {\bibinfo {author} {\bibfnamefont {M.~D.}\ \bibnamefont {Frye}}, \bibinfo {author} {\bibfnamefont {P.~S.}\ \bibnamefont {\ifmmode~\dot{Z}\else \.{Z}\fi{}uchowski}}, \ and\ \bibinfo {author} {\bibfnamefont {M.}~\bibnamefont {Tomza}},\ }\href {\doibase 10.1103/PhysRevResearch.6.023254} {\bibfield  {journal} {\bibinfo  {journal} {Phys. Rev. Res.}\ }\textbf {\bibinfo {volume} {6}},\ \bibinfo {pages} {023254} (\bibinfo {year} {2024})}\BibitemShut {NoStop}%
\bibitem [{\citenamefont {Farolfi}\ \emph {et~al.}(2019)\citenamefont {Farolfi}, \citenamefont {Trypogeorgos}, \citenamefont {Colzi}, \citenamefont {Fava}, \citenamefont {Lamporesi},\ and\ \citenamefont {Ferrari}}]{10.1063/1.5119915}%
  \BibitemOpen
  \bibfield  {author} {\bibinfo {author} {\bibfnamefont {A.}~\bibnamefont {Farolfi}}, \bibinfo {author} {\bibfnamefont {D.}~\bibnamefont {Trypogeorgos}}, \bibinfo {author} {\bibfnamefont {G.}~\bibnamefont {Colzi}}, \bibinfo {author} {\bibfnamefont {E.}~\bibnamefont {Fava}}, \bibinfo {author} {\bibfnamefont {G.}~\bibnamefont {Lamporesi}}, \ and\ \bibinfo {author} {\bibfnamefont {G.}~\bibnamefont {Ferrari}},\ }\href {\doibase 10.1063/1.5119915} {\bibfield  {journal} {\bibinfo  {journal} {Rev. Sci. Instrum.}\ }\textbf {\bibinfo {volume} {90}},\ \bibinfo {pages} {115114} (\bibinfo {year} {2019})}\BibitemShut {NoStop}%
\bibitem [{\citenamefont {Colzi}\ \emph {et~al.}(2018)\citenamefont {Colzi}, \citenamefont {Fava}, \citenamefont {Barbiero}, \citenamefont {Mordini}, \citenamefont {Lamporesi},\ and\ \citenamefont {Ferrari}}]{PhysRevA.97.053625}%
  \BibitemOpen
  \bibfield  {author} {\bibinfo {author} {\bibfnamefont {G.}~\bibnamefont {Colzi}}, \bibinfo {author} {\bibfnamefont {E.}~\bibnamefont {Fava}}, \bibinfo {author} {\bibfnamefont {M.}~\bibnamefont {Barbiero}}, \bibinfo {author} {\bibfnamefont {C.}~\bibnamefont {Mordini}}, \bibinfo {author} {\bibfnamefont {G.}~\bibnamefont {Lamporesi}}, \ and\ \bibinfo {author} {\bibfnamefont {G.}~\bibnamefont {Ferrari}},\ }\href {\doibase 10.1103/PhysRevA.97.053625} {\bibfield  {journal} {\bibinfo  {journal} {Phys. Rev. A}\ }\textbf {\bibinfo {volume} {97}},\ \bibinfo {pages} {053625} (\bibinfo {year} {2018})}\BibitemShut {NoStop}%
\bibitem [{\citenamefont {Farolfi}(2021)}]{PhDFarolfi}%
  \BibitemOpen
  \bibfield  {author} {\bibinfo {author} {\bibfnamefont {A.}~\bibnamefont {Farolfi}},\ }\emph {\bibinfo {title} {Spin dynamics in two-component Bose-Einstein condensates}},\ \href@noop {} {Ph.D. thesis},\ \bibinfo  {school} {University of Trento} (\bibinfo {year} {2021})\BibitemShut {NoStop}%
\bibitem [{\citenamefont {Fava}(2018)}]{PhDFava}%
  \BibitemOpen
  \bibfield  {author} {\bibinfo {author} {\bibfnamefont {E.}~\bibnamefont {Fava}},\ }\emph {\bibinfo {title} {Static and dynamics properties of a miscible two-component Bose–Einstein Condensate}},\ \href@noop {} {Ph.D. thesis},\ \bibinfo  {school} {University of Trento} (\bibinfo {year} {2018})\BibitemShut {NoStop}%
\bibitem [{\citenamefont {Eckel}\ \emph {et~al.}(2013)\citenamefont {Eckel}, \citenamefont {Hamilton}, \citenamefont {Kirilov}, \citenamefont {Smith},\ and\ \citenamefont {DeMille}}]{PhysRevA.87.052130}%
  \BibitemOpen
  \bibfield  {author} {\bibinfo {author} {\bibfnamefont {S.}~\bibnamefont {Eckel}}, \bibinfo {author} {\bibfnamefont {P.}~\bibnamefont {Hamilton}}, \bibinfo {author} {\bibfnamefont {E.}~\bibnamefont {Kirilov}}, \bibinfo {author} {\bibfnamefont {H.~W.}\ \bibnamefont {Smith}}, \ and\ \bibinfo {author} {\bibfnamefont {D.}~\bibnamefont {DeMille}},\ }\href {\doibase 10.1103/PhysRevA.87.052130} {\bibfield  {journal} {\bibinfo  {journal} {Phys. Rev. A}\ }\textbf {\bibinfo {volume} {87}},\ \bibinfo {pages} {052130} (\bibinfo {year} {2013})}\BibitemShut {NoStop}%
\bibitem [{\citenamefont {Bause}\ \emph {et~al.}(2025)\citenamefont {Bause}, \citenamefont {Balasubramanian}, \citenamefont {Fikkers}, \citenamefont {Prinsen}, \citenamefont {Steinebach}, \citenamefont {Jadbabaie}, \citenamefont {Hutzler}, \citenamefont {Aucar}, \citenamefont {Pa\ifmmode~\check{s}\else \v{s}\fi{}teka}, \citenamefont {Borschevsky},\ and\ \citenamefont {Hoekstra}}]{HoekstraOL}%
  \BibitemOpen
  \bibfield  {author} {\bibinfo {author} {\bibfnamefont {R.}~\bibnamefont {Bause}}, \bibinfo {author} {\bibfnamefont {N.}~\bibnamefont {Balasubramanian}}, \bibinfo {author} {\bibfnamefont {T.}~\bibnamefont {Fikkers}}, \bibinfo {author} {\bibfnamefont {E.~H.}\ \bibnamefont {Prinsen}}, \bibinfo {author} {\bibfnamefont {K.}~\bibnamefont {Steinebach}}, \bibinfo {author} {\bibfnamefont {A.}~\bibnamefont {Jadbabaie}}, \bibinfo {author} {\bibfnamefont {N.~R.}\ \bibnamefont {Hutzler}}, \bibinfo {author} {\bibfnamefont {I.~A.}\ \bibnamefont {Aucar}}, \bibinfo {author} {\bibfnamefont {L.~c. v.~F.}\ \bibnamefont {Pa\ifmmode~\check{s}\else \v{s}\fi{}teka}}, \bibinfo {author} {\bibfnamefont {A.}~\bibnamefont {Borschevsky}}, \ and\ \bibinfo {author} {\bibfnamefont {S.}~\bibnamefont {Hoekstra}},\ }\href {\doibase 10.1103/8ltl-7wsb} {\bibfield  {journal} {\bibinfo  {journal} {Phys. Rev. A}\ }\textbf {\bibinfo {volume} {111}},\ \bibinfo {pages} {062815} (\bibinfo {year} {2025})}\BibitemShut {NoStop}%
\bibitem [{\citenamefont {Tomza}(2013)}]{PhysRevA.88.012519}%
  \BibitemOpen
  \bibfield  {author} {\bibinfo {author} {\bibfnamefont {M.}~\bibnamefont {Tomza}},\ }\href {\doibase 10.1103/PhysRevA.88.012519} {\bibfield  {journal} {\bibinfo  {journal} {Phys. Rev. A}\ }\textbf {\bibinfo {volume} {88}},\ \bibinfo {pages} {012519} (\bibinfo {year} {2013})}\BibitemShut {NoStop}%
\bibitem [{\citenamefont {Zhu}\ \emph {et~al.}(2013)\citenamefont {Zhu}, \citenamefont {Solmeyer}, \citenamefont {Tang},\ and\ \citenamefont {Weiss}}]{PhysRevLett.111.243006}%
  \BibitemOpen
  \bibfield  {author} {\bibinfo {author} {\bibfnamefont {K.}~\bibnamefont {Zhu}}, \bibinfo {author} {\bibfnamefont {N.}~\bibnamefont {Solmeyer}}, \bibinfo {author} {\bibfnamefont {C.}~\bibnamefont {Tang}}, \ and\ \bibinfo {author} {\bibfnamefont {D.~S.}\ \bibnamefont {Weiss}},\ }\href {\doibase 10.1103/PhysRevLett.111.243006} {\bibfield  {journal} {\bibinfo  {journal} {Phys. Rev. Lett.}\ }\textbf {\bibinfo {volume} {111}},\ \bibinfo {pages} {243006} (\bibinfo {year} {2013})}\BibitemShut {NoStop}%
\bibitem [{\citenamefont {Vanhaecke}\ and\ \citenamefont {Dulieu}(2007)}]{doi:10.1080/00268970701466261}%
  \BibitemOpen
  \bibfield  {author} {\bibinfo {author} {\bibfnamefont {N.}~\bibnamefont {Vanhaecke}}\ and\ \bibinfo {author} {\bibfnamefont {O.}~\bibnamefont {Dulieu}},\ }\href {\doibase 10.1080/00268970701466261} {\bibfield  {journal} {\bibinfo  {journal} {Mol. Phys.}\ }\textbf {\bibinfo {volume} {105}},\ \bibinfo {pages} {1723} (\bibinfo {year} {2007})}\BibitemShut {NoStop}%
\bibitem [{\citenamefont {Takahashi}\ \emph {et~al.}(2023)\citenamefont {Takahashi}, \citenamefont {Zhang}, \citenamefont {Jadbabaie},\ and\ \citenamefont {Hutzler}}]{PhysRevLett.131.183003}%
  \BibitemOpen
  \bibfield  {author} {\bibinfo {author} {\bibfnamefont {Y.}~\bibnamefont {Takahashi}}, \bibinfo {author} {\bibfnamefont {C.}~\bibnamefont {Zhang}}, \bibinfo {author} {\bibfnamefont {A.}~\bibnamefont {Jadbabaie}}, \ and\ \bibinfo {author} {\bibfnamefont {N.~R.}\ \bibnamefont {Hutzler}},\ }\href {\doibase 10.1103/PhysRevLett.131.183003} {\bibfield  {journal} {\bibinfo  {journal} {Phys. Rev. Lett.}\ }\textbf {\bibinfo {volume} {131}},\ \bibinfo {pages} {183003} (\bibinfo {year} {2023})}\BibitemShut {NoStop}%
\bibitem [{\citenamefont {{N. Hutzler}}(2024)}]{NickHutzlerComm}%
  \BibitemOpen
  \bibfield  {author} {\bibinfo {author} {\bibnamefont {{N. Hutzler}}},\ }\href@noop {} {\enquote {\bibinfo {title} {private communication},}\ } (\bibinfo {year} {2024})\BibitemShut {NoStop}%
\bibitem [{\citenamefont {Tomza}(2014)}]{PhysRevA.90.022514}%
  \BibitemOpen
  \bibfield  {author} {\bibinfo {author} {\bibfnamefont {M.}~\bibnamefont {Tomza}},\ }\href {\doibase 10.1103/PhysRevA.90.022514} {\bibfield  {journal} {\bibinfo  {journal} {Phys. Rev. A}\ }\textbf {\bibinfo {volume} {90}},\ \bibinfo {pages} {022514} (\bibinfo {year} {2014})}\BibitemShut {NoStop}%
\bibitem [{\citenamefont {Boys}\ and\ \citenamefont {Bernardi}(1970)}]{counterpoise}%
  \BibitemOpen
  \bibfield  {author} {\bibinfo {author} {\bibfnamefont {S.}~\bibnamefont {Boys}}\ and\ \bibinfo {author} {\bibfnamefont {F.}~\bibnamefont {Bernardi}},\ }\href {\doibase 10.1080/00268977000101561} {\bibfield  {journal} {\bibinfo  {journal} {Mol. Phys.}\ }\textbf {\bibinfo {volume} {19}},\ \bibinfo {pages} {553} (\bibinfo {year} {1970})}\BibitemShut {NoStop}%
\bibitem [{\citenamefont {Wang}\ and\ \citenamefont {Dolg}(1998)}]{newecpYb}%
  \BibitemOpen
  \bibfield  {author} {\bibinfo {author} {\bibfnamefont {Y.}~\bibnamefont {Wang}}\ and\ \bibinfo {author} {\bibfnamefont {M.}~\bibnamefont {Dolg}},\ }\href {https://link.springer.com/article/10.1007/s002140050373#citeas} {\bibfield  {journal} {\bibinfo  {journal} {Theor. Chem. Acc.}\ }\textbf {\bibinfo {volume} {100}},\ \bibinfo {pages} {124} (\bibinfo {year} {1998})}\BibitemShut {NoStop}%
\bibitem [{\citenamefont {Dolg}\ \emph {et~al.}(1987)\citenamefont {Dolg}, \citenamefont {Wedig}, \citenamefont {Stoll},\ and\ \citenamefont {Preuss}}]{newecpCr}%
  \BibitemOpen
  \bibfield  {author} {\bibinfo {author} {\bibfnamefont {M.}~\bibnamefont {Dolg}}, \bibinfo {author} {\bibfnamefont {U.}~\bibnamefont {Wedig}}, \bibinfo {author} {\bibfnamefont {H.}~\bibnamefont {Stoll}}, \ and\ \bibinfo {author} {\bibfnamefont {H.}~\bibnamefont {Preuss}},\ }\href {\doibase https://doi.org/10.1063/1.452288} {\bibfield  {journal} {\bibinfo  {journal} {J. Chem. Phys.}\ }\textbf {\bibinfo {volume} {86}},\ \bibinfo {pages} {866} (\bibinfo {year} {1987})}\BibitemShut {NoStop}%
\bibitem [{\citenamefont {Hill}\ \emph {et~al.}(2025)\citenamefont {Hill}, \citenamefont {Gronowski},\ and\ \citenamefont {Tomza}}]{hill_new}%
  \BibitemOpen
  \bibfield  {author} {\bibinfo {author} {\bibfnamefont {J.~G.}\ \bibnamefont {Hill}}, \bibinfo {author} {\bibfnamefont {M.}~\bibnamefont {Gronowski}}, \ and\ \bibinfo {author} {\bibfnamefont {M.}~\bibnamefont {Tomza}},\ }\href@noop {} {\enquote {\bibinfo {title} {{to be published}},}\ } (\bibinfo {year} {2025})\BibitemShut {NoStop}%
\bibitem [{\citenamefont {Knowles}\ \emph {et~al.}(1993)\citenamefont {Knowles}, \citenamefont {Hampel},\ and\ \citenamefont {Werner}}]{KnowlesJCP93}%
  \BibitemOpen
  \bibfield  {author} {\bibinfo {author} {\bibfnamefont {P.~J.}\ \bibnamefont {Knowles}}, \bibinfo {author} {\bibfnamefont {C.}~\bibnamefont {Hampel}}, \ and\ \bibinfo {author} {\bibfnamefont {H.-J.}\ \bibnamefont {Werner}},\ }\href {\doibase 10.1063/1.465990} {\bibfield  {journal} {\bibinfo  {journal} {J. Chem. Phys.}\ }\textbf {\bibinfo {volume} {99}},\ \bibinfo {pages} {5219} (\bibinfo {year} {1993})}\BibitemShut {NoStop}%
\bibitem [{\citenamefont {Werner}\ \emph {et~al.}(2012)\citenamefont {Werner}, \citenamefont {Knowles}, \citenamefont {Knizia}, \citenamefont {Manby},\ and\ \citenamefont {Sch{\"u}tz}}]{MOLPRO-WIREs}%
  \BibitemOpen
  \bibfield  {author} {\bibinfo {author} {\bibfnamefont {H.-J.}\ \bibnamefont {Werner}}, \bibinfo {author} {\bibfnamefont {P.~J.}\ \bibnamefont {Knowles}}, \bibinfo {author} {\bibfnamefont {G.}~\bibnamefont {Knizia}}, \bibinfo {author} {\bibfnamefont {F.~R.}\ \bibnamefont {Manby}}, \ and\ \bibinfo {author} {\bibfnamefont {M.}~\bibnamefont {Sch{\"u}tz}},\ }\href {\doibase 10.1002/wcms.82} {\bibfield  {journal} {\bibinfo  {journal} {WIREs Comput Mol Sci}\ }\textbf {\bibinfo {volume} {2}},\ \bibinfo {pages} {242} (\bibinfo {year} {2012})}\BibitemShut {NoStop}%
\bibitem [{\citenamefont {Werner}\ \emph {et~al.}()\citenamefont {Werner}, \citenamefont {Knowles}, \citenamefont {Knizia}, \citenamefont {Manby}, \citenamefont {{Sch\"{u}tz}} \emph {et~al.}}]{MOLPRO}%
  \BibitemOpen
  \bibfield  {author} {\bibinfo {author} {\bibfnamefont {H.-J.}\ \bibnamefont {Werner}}, \bibinfo {author} {\bibfnamefont {P.~J.}\ \bibnamefont {Knowles}}, \bibinfo {author} {\bibfnamefont {G.}~\bibnamefont {Knizia}}, \bibinfo {author} {\bibfnamefont {F.~R.}\ \bibnamefont {Manby}}, \bibinfo {author} {\bibfnamefont {M.}~\bibnamefont {{Sch\"{u}tz}}},  \emph {et~al.},\ }\href@noop {} {\enquote {\bibinfo {title} {Molpro, version 2022, a package of ab initio programs},}\ }\bibinfo {note} {See https://www.molpro.net.}\BibitemShut {Stop}%
\bibitem [{\citenamefont {Ladjimi}\ and\ \citenamefont {Tomza}(2024)}]{PhysRevA.109.052814}%
  \BibitemOpen
  \bibfield  {author} {\bibinfo {author} {\bibfnamefont {H.}~\bibnamefont {Ladjimi}}\ and\ \bibinfo {author} {\bibfnamefont {M.}~\bibnamefont {Tomza}},\ }\href {\doibase 10.1103/PhysRevA.109.052814} {\bibfield  {journal} {\bibinfo  {journal} {Phys. Rev. A}\ }\textbf {\bibinfo {volume} {109}},\ \bibinfo {pages} {052814} (\bibinfo {year} {2024})}\BibitemShut {NoStop}%
\bibitem [{\citenamefont {Kállay}\ \emph {et~al.}()\citenamefont {Kállay}, \citenamefont {Nagy}, \citenamefont {Mester}, \citenamefont {Rolik}, \citenamefont {Samu}, \citenamefont {Csontos}, \citenamefont {Csóka}, \citenamefont {Szabó}, \citenamefont {Gyevi-Nagy}, \citenamefont {Hégely}, \citenamefont {Ladjánszki}, \citenamefont {Szegedy}, \citenamefont {Ladóczki}, \citenamefont {Petrov}, \citenamefont {Farkas}, \citenamefont {Mezei},\ and\ \citenamefont {Ganyecz}}]{MRCC}%
  \BibitemOpen
  \bibfield  {author} {\bibinfo {author} {\bibfnamefont {M.}~\bibnamefont {Kállay}}, \bibinfo {author} {\bibfnamefont {P.~R.}\ \bibnamefont {Nagy}}, \bibinfo {author} {\bibfnamefont {D.}~\bibnamefont {Mester}}, \bibinfo {author} {\bibfnamefont {Z.}~\bibnamefont {Rolik}}, \bibinfo {author} {\bibfnamefont {G.}~\bibnamefont {Samu}}, \bibinfo {author} {\bibfnamefont {J.}~\bibnamefont {Csontos}}, \bibinfo {author} {\bibfnamefont {J.}~\bibnamefont {Csóka}}, \bibinfo {author} {\bibfnamefont {P.~B.}\ \bibnamefont {Szabó}}, \bibinfo {author} {\bibfnamefont {L.}~\bibnamefont {Gyevi-Nagy}}, \bibinfo {author} {\bibfnamefont {B.}~\bibnamefont {Hégely}}, \bibinfo {author} {\bibfnamefont {I.}~\bibnamefont {Ladjánszki}}, \bibinfo {author} {\bibfnamefont {L.}~\bibnamefont {Szegedy}}, \bibinfo {author} {\bibfnamefont {B.}~\bibnamefont {Ladóczki}}, \bibinfo {author} {\bibfnamefont {K.}~\bibnamefont {Petrov}}, \bibinfo {author} {\bibfnamefont {M.}~\bibnamefont {Farkas}}, \bibinfo {author} {\bibfnamefont {P.~D.}\
  \bibnamefont {Mezei}}, \ and\ \bibinfo {author} {\bibfnamefont {A.}~\bibnamefont {Ganyecz}},\ }\href@noop {} {\enquote {\bibinfo {title} {Mrcc, a quantum chemical program},}\ }\bibinfo {note} {See https://www.mrcc.hu}\BibitemShut {NoStop}%
\bibitem [{\citenamefont {Matthews}\ \emph {et~al.}(2020{\natexlab{a}})\citenamefont {Matthews}, \citenamefont {Cheng}, \citenamefont {Harding}, \citenamefont {Lipparini}, \citenamefont {Stopkowicz}, \citenamefont {Jagau}, \citenamefont {Szalay}, \citenamefont {Gauss},\ and\ \citenamefont {Stanton}}]{10.1063/5.0004837}%
  \BibitemOpen
  \bibfield  {author} {\bibinfo {author} {\bibfnamefont {D.~A.}\ \bibnamefont {Matthews}}, \bibinfo {author} {\bibfnamefont {L.}~\bibnamefont {Cheng}}, \bibinfo {author} {\bibfnamefont {M.~E.}\ \bibnamefont {Harding}}, \bibinfo {author} {\bibfnamefont {F.}~\bibnamefont {Lipparini}}, \bibinfo {author} {\bibfnamefont {S.}~\bibnamefont {Stopkowicz}}, \bibinfo {author} {\bibfnamefont {T.-C.}\ \bibnamefont {Jagau}}, \bibinfo {author} {\bibfnamefont {P.~G.}\ \bibnamefont {Szalay}}, \bibinfo {author} {\bibfnamefont {J.}~\bibnamefont {Gauss}}, \ and\ \bibinfo {author} {\bibfnamefont {J.~F.}\ \bibnamefont {Stanton}},\ }\href {\doibase 10.1063/5.0004837} {\bibfield  {journal} {\bibinfo  {journal} {J. Chem. Phys.}\ }\textbf {\bibinfo {volume} {152}},\ \bibinfo {pages} {214108} (\bibinfo {year} {2020}{\natexlab{a}})}\BibitemShut {NoStop}%
\bibitem [{\citenamefont {Werner}\ and\ \citenamefont {Knowles}(1988)}]{WernerJCP88}%
  \BibitemOpen
  \bibfield  {author} {\bibinfo {author} {\bibfnamefont {H.}~\bibnamefont {Werner}}\ and\ \bibinfo {author} {\bibfnamefont {P.~J.}\ \bibnamefont {Knowles}},\ }\href {\doibase 10.1063/1.455556} {\bibfield  {journal} {\bibinfo  {journal} {J. Chem. Phys.}\ }\textbf {\bibinfo {volume} {89}},\ \bibinfo {pages} {5803} (\bibinfo {year} {1988})}\BibitemShut {NoStop}%
\bibitem [{\citenamefont {Matthews}\ \emph {et~al.}(2020{\natexlab{b}})\citenamefont {Matthews}, \citenamefont {Cheng}, \citenamefont {Harding}, \citenamefont {Lipparini}, \citenamefont {Stopkowicz}, \citenamefont {Jagau}, \citenamefont {Szalay}, \citenamefont {Gauss},\ and\ \citenamefont {Stanton}}]{MatthewsJCP20}%
  \BibitemOpen
  \bibfield  {author} {\bibinfo {author} {\bibfnamefont {D.~A.}\ \bibnamefont {Matthews}}, \bibinfo {author} {\bibfnamefont {L.}~\bibnamefont {Cheng}}, \bibinfo {author} {\bibfnamefont {M.~E.}\ \bibnamefont {Harding}}, \bibinfo {author} {\bibfnamefont {F.}~\bibnamefont {Lipparini}}, \bibinfo {author} {\bibfnamefont {S.}~\bibnamefont {Stopkowicz}}, \bibinfo {author} {\bibfnamefont {T.-C.}\ \bibnamefont {Jagau}}, \bibinfo {author} {\bibfnamefont {P.~G.}\ \bibnamefont {Szalay}}, \bibinfo {author} {\bibfnamefont {J.}~\bibnamefont {Gauss}}, \ and\ \bibinfo {author} {\bibfnamefont {J.~F.}\ \bibnamefont {Stanton}},\ }\href {\doibase 10.1063/5.0004837} {\bibfield  {journal} {\bibinfo  {journal} {J. Chem. Phys.}\ }\textbf {\bibinfo {volume} {152}},\ \bibinfo {pages} {214108} (\bibinfo {year} {2020}{\natexlab{b}})}\BibitemShut {NoStop}%
\bibitem [{\citenamefont {Stanton}\ \emph {et~al.}()\citenamefont {Stanton}, \citenamefont {Gauss}, \citenamefont {Cheng}, \citenamefont {Harding}, \citenamefont {Matthews},\ and\ \citenamefont {Szalay}}]{cfour}%
  \BibitemOpen
  \bibfield  {author} {\bibinfo {author} {\bibfnamefont {J.~F.}\ \bibnamefont {Stanton}}, \bibinfo {author} {\bibfnamefont {J.}~\bibnamefont {Gauss}}, \bibinfo {author} {\bibfnamefont {L.}~\bibnamefont {Cheng}}, \bibinfo {author} {\bibfnamefont {M.~E.}\ \bibnamefont {Harding}}, \bibinfo {author} {\bibfnamefont {D.~A.}\ \bibnamefont {Matthews}}, \ and\ \bibinfo {author} {\bibfnamefont {P.~G.}\ \bibnamefont {Szalay}},\ }\href@noop {} {\enquote {\bibinfo {title} {{CFOUR, Coupled-Cluster techniques for Computational Chemistry, a quantum-chemical program package}},}\ }\bibinfo {note} {{W}ith contributions from {A}. {A}sthana, {A}.{A}. {A}uer, {R}.{J}. {B}artlett, {U}. {B}enedikt, {C}. {B}erger, {D}.{E}. {B}ernholdt, {S}. {B}laschke, {Y}. {J}. {B}omble, {S}. {B}urger, {O}. {C}hristiansen, et al., and the integral packages {MOLECULE} ({J}. {A}lml{\"o}f and {P}.{R}. {T}aylor), {PROPS} ({P}.{R}. {T}aylor), {ABACUS} ({T}. {H}elgaker, {H}.{J}. {A}a. {J}ensen, {P}. {J}{\o}rgensen, and {J}. {O}lsen), and {ECP} routines by
  {A}. {V}. {M}itin and {C}. van {W}{\"u}llen. {F}or the current version, see http://www.cfour.de.}\BibitemShut {Stop}%
\bibitem [{\citenamefont {Finelli}\ \emph {et~al.}(2024)\citenamefont {Finelli}, \citenamefont {Ciamei}, \citenamefont {Restivo}, \citenamefont {Schemmer}, \citenamefont {Cosco}, \citenamefont {Inguscio}, \citenamefont {Trenkwalder}, \citenamefont {Zaremba-Kopczyk}, \citenamefont {Gronowski}, \citenamefont {Tomza},\ and\ \citenamefont {Zaccanti}}]{PRXQuantum.5.020358}%
  \BibitemOpen
  \bibfield  {author} {\bibinfo {author} {\bibfnamefont {S.}~\bibnamefont {Finelli}}, \bibinfo {author} {\bibfnamefont {A.}~\bibnamefont {Ciamei}}, \bibinfo {author} {\bibfnamefont {B.}~\bibnamefont {Restivo}}, \bibinfo {author} {\bibfnamefont {M.}~\bibnamefont {Schemmer}}, \bibinfo {author} {\bibfnamefont {A.}~\bibnamefont {Cosco}}, \bibinfo {author} {\bibfnamefont {M.}~\bibnamefont {Inguscio}}, \bibinfo {author} {\bibfnamefont {A.}~\bibnamefont {Trenkwalder}}, \bibinfo {author} {\bibfnamefont {K.}~\bibnamefont {Zaremba-Kopczyk}}, \bibinfo {author} {\bibfnamefont {M.}~\bibnamefont {Gronowski}}, \bibinfo {author} {\bibfnamefont {M.}~\bibnamefont {Tomza}}, \ and\ \bibinfo {author} {\bibfnamefont {M.}~\bibnamefont {Zaccanti}},\ }\href {\doibase 10.1103/PRXQuantum.5.020358} {\bibfield  {journal} {\bibinfo  {journal} {PRX Quantum}\ }\textbf {\bibinfo {volume} {5}},\ \bibinfo {pages} {020358} (\bibinfo {year} {2024})}\BibitemShut {NoStop}%
\bibitem [{\citenamefont {Saue}\ \emph {et~al.}(2020)\citenamefont {Saue}, \citenamefont {Bast}, \citenamefont {Gomes}, \citenamefont {Jensen}, \citenamefont {Visscher}, \citenamefont {Aucar}, \citenamefont {Di~Remigio}, \citenamefont {Dyall}, \citenamefont {Eliav}, \citenamefont {Fasshauer}, \citenamefont {Fleig}, \citenamefont {Halbert}, \citenamefont {Hedegård}, \citenamefont {Helmich-Paris}, \citenamefont {Iliaš}, \citenamefont {Jacob}, \citenamefont {Knecht}, \citenamefont {Laerdahl}, \citenamefont {Vidal}, \citenamefont {Nayak}, \citenamefont {Olejniczak}, \citenamefont {Olsen}, \citenamefont {Pernpointner}, \citenamefont {Senjean}, \citenamefont {Shee}, \citenamefont {Sunaga},\ and\ \citenamefont {van Stralen}}]{dirac24}%
  \BibitemOpen
  \bibfield  {author} {\bibinfo {author} {\bibfnamefont {T.}~\bibnamefont {Saue}}, \bibinfo {author} {\bibfnamefont {R.}~\bibnamefont {Bast}}, \bibinfo {author} {\bibfnamefont {A.~S.~P.}\ \bibnamefont {Gomes}}, \bibinfo {author} {\bibfnamefont {H.~J.~A.}\ \bibnamefont {Jensen}}, \bibinfo {author} {\bibfnamefont {L.}~\bibnamefont {Visscher}}, \bibinfo {author} {\bibfnamefont {I.~A.}\ \bibnamefont {Aucar}}, \bibinfo {author} {\bibfnamefont {R.}~\bibnamefont {Di~Remigio}}, \bibinfo {author} {\bibfnamefont {K.~G.}\ \bibnamefont {Dyall}}, \bibinfo {author} {\bibfnamefont {E.}~\bibnamefont {Eliav}}, \bibinfo {author} {\bibfnamefont {E.}~\bibnamefont {Fasshauer}}, \bibinfo {author} {\bibfnamefont {T.}~\bibnamefont {Fleig}}, \bibinfo {author} {\bibfnamefont {L.}~\bibnamefont {Halbert}}, \bibinfo {author} {\bibfnamefont {E.~D.}\ \bibnamefont {Hedegård}}, \bibinfo {author} {\bibfnamefont {B.}~\bibnamefont {Helmich-Paris}}, \bibinfo {author} {\bibfnamefont {M.}~\bibnamefont {Iliaš}}, \bibinfo {author} {\bibfnamefont
  {C.~R.}\ \bibnamefont {Jacob}}, \bibinfo {author} {\bibfnamefont {S.}~\bibnamefont {Knecht}}, \bibinfo {author} {\bibfnamefont {J.~K.}\ \bibnamefont {Laerdahl}}, \bibinfo {author} {\bibfnamefont {M.~L.}\ \bibnamefont {Vidal}}, \bibinfo {author} {\bibfnamefont {M.~K.}\ \bibnamefont {Nayak}}, \bibinfo {author} {\bibfnamefont {M.}~\bibnamefont {Olejniczak}}, \bibinfo {author} {\bibfnamefont {J.~M.~H.}\ \bibnamefont {Olsen}}, \bibinfo {author} {\bibfnamefont {M.}~\bibnamefont {Pernpointner}}, \bibinfo {author} {\bibfnamefont {B.}~\bibnamefont {Senjean}}, \bibinfo {author} {\bibfnamefont {A.}~\bibnamefont {Shee}}, \bibinfo {author} {\bibfnamefont {A.}~\bibnamefont {Sunaga}}, \ and\ \bibinfo {author} {\bibfnamefont {J.~N.~P.}\ \bibnamefont {van Stralen}},\ }\href {\doibase 10.1063/5.0004844} {\bibfield  {journal} {\bibinfo  {journal} {J. Chem. Phys.}\ }\textbf {\bibinfo {volume} {152}},\ \bibinfo {pages} {204104} (\bibinfo {year} {2020})}\BibitemShut {NoStop}%
\bibitem [{\citenamefont {Knecht}\ \emph {et~al.}(2010)\citenamefont {Knecht}, \citenamefont {Jensen},\ and\ \citenamefont {Fleig}}]{knecht}%
  \BibitemOpen
  \bibfield  {author} {\bibinfo {author} {\bibfnamefont {S.}~\bibnamefont {Knecht}}, \bibinfo {author} {\bibfnamefont {H.~J.~A.}\ \bibnamefont {Jensen}}, \ and\ \bibinfo {author} {\bibfnamefont {T.}~\bibnamefont {Fleig}},\ }\href {\doibase 10.1063/1.3276157} {\bibfield  {journal} {\bibinfo  {journal} {J. Chem. Phys.}\ }\textbf {\bibinfo {volume} {132}},\ \bibinfo {pages} {014108} (\bibinfo {year} {2010})}\BibitemShut {NoStop}%
\bibitem [{\citenamefont {Dyall}(2012)}]{dyall2012core}%
  \BibitemOpen
  \bibfield  {author} {\bibinfo {author} {\bibfnamefont {K.~G.}\ \bibnamefont {Dyall}},\ }\href {https://link.springer.com/article/10.1007/s00214-012-1217-8} {\bibfield  {journal} {\bibinfo  {journal} {Theor. Chem. Acc.}\ }\textbf {\bibinfo {volume} {131}},\ \bibinfo {pages} {1} (\bibinfo {year} {2012})}\BibitemShut {NoStop}%
\bibitem [{\citenamefont {Lindroth}\ \emph {et~al.}(1989)\citenamefont {Lindroth}, \citenamefont {Lynn},\ and\ \citenamefont {Sandars}}]{lindroth1989order}%
  \BibitemOpen
  \bibfield  {author} {\bibinfo {author} {\bibfnamefont {E.}~\bibnamefont {Lindroth}}, \bibinfo {author} {\bibfnamefont {B.}~\bibnamefont {Lynn}}, \ and\ \bibinfo {author} {\bibfnamefont {P.}~\bibnamefont {Sandars}},\ }\href {\doibase 10.1088/0953-4075/22/4/004} {\bibfield  {journal} {\bibinfo  {journal} {J. Phys. B:At. Mol. Opt. Phys.}\ }\textbf {\bibinfo {volume} {22}},\ \bibinfo {pages} {559} (\bibinfo {year} {1989})}\BibitemShut {NoStop}%
\bibitem [{\citenamefont {Kozlov}(1985)}]{kozlov1985semiempirical}%
  \BibitemOpen
  \bibfield  {author} {\bibinfo {author} {\bibfnamefont {M.}~\bibnamefont {Kozlov}},\ }\href {http://www.jetp.ras.ru/cgi-bin/dn/e_062_06_1114.pdf} {\bibfield  {journal} {\bibinfo  {journal} {Zh. Eksp. Teor. Fiz.}\ }\textbf {\bibinfo {volume} {89}},\ \bibinfo {pages} {1933} (\bibinfo {year} {1985})}\BibitemShut {NoStop}%
\bibitem [{\citenamefont {Franzke}\ \emph {et~al.}(2020)\citenamefont {Franzke}, \citenamefont {Spiske}, \citenamefont {Pollak},\ and\ \citenamefont {Weigend}}]{franzke2020segmented}%
  \BibitemOpen
  \bibfield  {author} {\bibinfo {author} {\bibfnamefont {Y.~J.}\ \bibnamefont {Franzke}}, \bibinfo {author} {\bibfnamefont {L.}~\bibnamefont {Spiske}}, \bibinfo {author} {\bibfnamefont {P.}~\bibnamefont {Pollak}}, \ and\ \bibinfo {author} {\bibfnamefont {F.}~\bibnamefont {Weigend}},\ }\href {https://pubs.acs.org/doi/10.1021/acs.jctc.0c00546} {\bibfield  {journal} {\bibinfo  {journal} {J. Chem. Theory Comput.}\ }\textbf {\bibinfo {volume} {16}},\ \bibinfo {pages} {5658} (\bibinfo {year} {2020})}\BibitemShut {NoStop}%
\bibitem [{\citenamefont {Peng}\ \emph {et~al.}(2013)\citenamefont {Peng}, \citenamefont {Middendorf}, \citenamefont {Weigend},\ and\ \citenamefont {Reiher}}]{PengJCP2013}%
  \BibitemOpen
  \bibfield  {author} {\bibinfo {author} {\bibfnamefont {D.}~\bibnamefont {Peng}}, \bibinfo {author} {\bibfnamefont {N.}~\bibnamefont {Middendorf}}, \bibinfo {author} {\bibfnamefont {F.}~\bibnamefont {Weigend}}, \ and\ \bibinfo {author} {\bibfnamefont {M.}~\bibnamefont {Reiher}},\ }\href {\doibase 10.1063/1.4803693} {\bibfield  {journal} {\bibinfo  {journal} {J. Chem. Phys.}\ }\textbf {\bibinfo {volume} {138}},\ \bibinfo {pages} {184105} (\bibinfo {year} {2013})}\BibitemShut {NoStop}%
\bibitem [{\citenamefont {Neese}(2012)}]{orca1}%
  \BibitemOpen
  \bibfield  {author} {\bibinfo {author} {\bibfnamefont {F.}~\bibnamefont {Neese}},\ }\href {\doibase https://doi.org/10.1002/wcms.81} {\bibfield  {journal} {\bibinfo  {journal} {WIREs Comput. Mol. Sci.}\ }\textbf {\bibinfo {volume} {2}},\ \bibinfo {pages} {73} (\bibinfo {year} {2012})}\BibitemShut {NoStop}%
\bibitem [{\citenamefont {Neese}(2025)}]{ORCA6}%
  \BibitemOpen
  \bibfield  {author} {\bibinfo {author} {\bibfnamefont {F.}~\bibnamefont {Neese}},\ }\href {\doibase 10.1002/wcms.70019} {\bibfield  {journal} {\bibinfo  {journal} {Wiley Interdiscip. Rev. Comput. Mol. Sci.}\ }\textbf {\bibinfo {volume} {15}},\ \bibinfo {pages} {e70019} (\bibinfo {year} {2025})}\BibitemShut {NoStop}%
\bibitem [{\citenamefont {Sunaga}\ \emph {et~al.}(2019)\citenamefont {Sunaga}, \citenamefont {Prasannaa}, \citenamefont {Abe}, \citenamefont {Hada},\ and\ \citenamefont {Das}}]{PhysRevA.99.040501}%
  \BibitemOpen
  \bibfield  {author} {\bibinfo {author} {\bibfnamefont {A.}~\bibnamefont {Sunaga}}, \bibinfo {author} {\bibfnamefont {V.~S.}\ \bibnamefont {Prasannaa}}, \bibinfo {author} {\bibfnamefont {M.}~\bibnamefont {Abe}}, \bibinfo {author} {\bibfnamefont {M.}~\bibnamefont {Hada}}, \ and\ \bibinfo {author} {\bibfnamefont {B.~P.}\ \bibnamefont {Das}},\ }\href {\doibase 10.1103/PhysRevA.99.040501} {\bibfield  {journal} {\bibinfo  {journal} {Phys. Rev. A}\ }\textbf {\bibinfo {volume} {99}},\ \bibinfo {pages} {040501} (\bibinfo {year} {2019})}\BibitemShut {NoStop}%
\bibitem [{\citenamefont {Baroni}\ \emph {et~al.}(2024)\citenamefont {Baroni}, \citenamefont {Lamporesi},\ and\ \citenamefont {Zaccanti}}]{Baroni2024}%
  \BibitemOpen
  \bibfield  {author} {\bibinfo {author} {\bibfnamefont {C.}~\bibnamefont {Baroni}}, \bibinfo {author} {\bibfnamefont {G.}~\bibnamefont {Lamporesi}}, \ and\ \bibinfo {author} {\bibfnamefont {M.}~\bibnamefont {Zaccanti}},\ }\href {\doibase 10.1038/s42254-024-00773-6} {\bibfield  {journal} {\bibinfo  {journal} {Nat. Rev. Phys.}\ }\textbf {\bibinfo {volume} {6}},\ \bibinfo {pages} {736} (\bibinfo {year} {2024})}\BibitemShut {NoStop}%
\bibitem [{\citenamefont {Larsson}\ \emph {et~al.}(2022)\citenamefont {Larsson}, \citenamefont {Zhai}, \citenamefont {Umrigar},\ and\ \citenamefont {Chan}}]{Larsson2022}%
  \BibitemOpen
  \bibfield  {author} {\bibinfo {author} {\bibfnamefont {H.~R.}\ \bibnamefont {Larsson}}, \bibinfo {author} {\bibfnamefont {H.}~\bibnamefont {Zhai}}, \bibinfo {author} {\bibfnamefont {C.~J.}\ \bibnamefont {Umrigar}}, \ and\ \bibinfo {author} {\bibfnamefont {G.~K.-L.}\ \bibnamefont {Chan}},\ }\href {\doibase 10.1021/jacs.2c06357} {\bibfield  {journal} {\bibinfo  {journal} {J. Am. Chem. Soc.}\ }\textbf {\bibinfo {volume} {144}},\ \bibinfo {pages} {15932} (\bibinfo {year} {2022})}\BibitemShut {NoStop}%
\bibitem [{\citenamefont {Tiesinga}\ \emph {et~al.}(2021)\citenamefont {Tiesinga}, \citenamefont {Kłos}, \citenamefont {Li}, \citenamefont {Petrov},\ and\ \citenamefont {Kotochigova}}]{TiesingaNJP21}%
  \BibitemOpen
  \bibfield  {author} {\bibinfo {author} {\bibfnamefont {E.}~\bibnamefont {Tiesinga}}, \bibinfo {author} {\bibfnamefont {J.}~\bibnamefont {Kłos}}, \bibinfo {author} {\bibfnamefont {M.}~\bibnamefont {Li}}, \bibinfo {author} {\bibfnamefont {A.}~\bibnamefont {Petrov}}, \ and\ \bibinfo {author} {\bibfnamefont {S.}~\bibnamefont {Kotochigova}},\ }\href {\doibase 10.1088/1367-2630/ac1a9a} {\bibfield  {journal} {\bibinfo  {journal} {New J. Phys.}\ }\textbf {\bibinfo {volume} {23}},\ \bibinfo {pages} {085007} (\bibinfo {year} {2021})}\BibitemShut {NoStop}%
\bibitem [{\citenamefont {Gonz\'alez-Mart\'{\i}nez}\ and\ \citenamefont {\ifmmode~\dot{Z}\else \.{Z}\fi{}uchowski}(2015)}]{GonzalezPRA15}%
  \BibitemOpen
  \bibfield  {author} {\bibinfo {author} {\bibfnamefont {M.~L.}\ \bibnamefont {Gonz\'alez-Mart\'{\i}nez}}\ and\ \bibinfo {author} {\bibfnamefont {P.~S.}\ \bibnamefont {\ifmmode~\dot{Z}\else \.{Z}\fi{}uchowski}},\ }\href {\doibase 10.1103/PhysRevA.92.022708} {\bibfield  {journal} {\bibinfo  {journal} {Phys. Rev. A}\ }\textbf {\bibinfo {volume} {92}},\ \bibinfo {pages} {022708} (\bibinfo {year} {2015})}\BibitemShut {NoStop}%
\bibitem [{\citenamefont {Kosicki}\ \emph {et~al.}(2020)\citenamefont {Kosicki}, \citenamefont {Borkowski},\ and\ \citenamefont {Żuchowski}}]{KosickiNJP20}%
  \BibitemOpen
  \bibfield  {author} {\bibinfo {author} {\bibfnamefont {M.~B.}\ \bibnamefont {Kosicki}}, \bibinfo {author} {\bibfnamefont {M.}~\bibnamefont {Borkowski}}, \ and\ \bibinfo {author} {\bibfnamefont {P.~S.}\ \bibnamefont {Żuchowski}},\ }\href {\doibase 10.1088/1367-2630/ab6c36} {\bibfield  {journal} {\bibinfo  {journal} {New J. Phys.}\ }\textbf {\bibinfo {volume} {22}},\ \bibinfo {pages} {023024} (\bibinfo {year} {2020})}\BibitemShut {NoStop}%
\bibitem [{\citenamefont {Hamilton}(2010)}]{HamiltonPhD}%
  \BibitemOpen
  \bibfield  {author} {\bibinfo {author} {\bibfnamefont {P.}~\bibnamefont {Hamilton}},\ }\emph {\bibinfo {title} {Preliminary results in the search for the electron electric dipole moment in PbO*}},\ \href@noop {} {Ph.D. thesis},\ \bibinfo  {school} {Yale University} (\bibinfo {year} {2010})\BibitemShut {NoStop}%
\bibitem [{\citenamefont {Bickman}\ \emph {et~al.}(2009)\citenamefont {Bickman}, \citenamefont {Hamilton}, \citenamefont {Jiang},\ and\ \citenamefont {DeMille}}]{PhysRevA.80.023418}%
  \BibitemOpen
  \bibfield  {author} {\bibinfo {author} {\bibfnamefont {S.}~\bibnamefont {Bickman}}, \bibinfo {author} {\bibfnamefont {P.}~\bibnamefont {Hamilton}}, \bibinfo {author} {\bibfnamefont {Y.}~\bibnamefont {Jiang}}, \ and\ \bibinfo {author} {\bibfnamefont {D.}~\bibnamefont {DeMille}},\ }\href {\doibase 10.1103/PhysRevA.80.023418} {\bibfield  {journal} {\bibinfo  {journal} {Phys. Rev. A}\ }\textbf {\bibinfo {volume} {80}},\ \bibinfo {pages} {023418} (\bibinfo {year} {2009})}\BibitemShut {NoStop}%
\bibitem [{\citenamefont {Curl}(1965)}]{Curl01011965}%
  \BibitemOpen
  \bibfield  {author} {\bibinfo {author} {\bibfnamefont {R.~F.}\ \bibnamefont {Curl}},\ }\href {\doibase 10.1080/00268976500100761} {\bibfield  {journal} {\bibinfo  {journal} {Mol. Phys.}\ }\textbf {\bibinfo {volume} {9}},\ \bibinfo {pages} {585} (\bibinfo {year} {1965})}\BibitemShut {NoStop}%
\bibitem [{\citenamefont {Neese}(2001)}]{neese_g10.1063/1.1419058}%
  \BibitemOpen
  \bibfield  {author} {\bibinfo {author} {\bibfnamefont {F.}~\bibnamefont {Neese}},\ }\href {\doibase 10.1063/1.1419058} {\bibfield  {journal} {\bibinfo  {journal} {J. Chem. Phys.}\ }\textbf {\bibinfo {volume} {115}},\ \bibinfo {pages} {11080} (\bibinfo {year} {2001})}\BibitemShut {NoStop}%
\bibitem [{\citenamefont {Flambaum}(1994)}]{FLAMBAUM1994211}%
  \BibitemOpen
  \bibfield  {author} {\bibinfo {author} {\bibfnamefont {V.}~\bibnamefont {Flambaum}},\ }\href {\doibase https://doi.org/10.1016/0370-2693(94)90646-7} {\bibfield  {journal} {\bibinfo  {journal} {Phys. Lett. B}\ }\textbf {\bibinfo {volume} {320}},\ \bibinfo {pages} {211} (\bibinfo {year} {1994})}\BibitemShut {NoStop}%
\bibitem [{\citenamefont {Skripnikov}\ \emph {et~al.}(2017)\citenamefont {Skripnikov}, \citenamefont {Titov},\ and\ \citenamefont {Flambaum}}]{skripnikovNQMM}%
  \BibitemOpen
  \bibfield  {author} {\bibinfo {author} {\bibfnamefont {L.~V.}\ \bibnamefont {Skripnikov}}, \bibinfo {author} {\bibfnamefont {A.~V.}\ \bibnamefont {Titov}}, \ and\ \bibinfo {author} {\bibfnamefont {V.~V.}\ \bibnamefont {Flambaum}},\ }\href {\doibase 10.1103/PhysRevA.95.022512} {\bibfield  {journal} {\bibinfo  {journal} {Phys. Rev. A}\ }\textbf {\bibinfo {volume} {95}},\ \bibinfo {pages} {022512} (\bibinfo {year} {2017})}\BibitemShut {NoStop}%
\bibitem [{\citenamefont {Denis}\ \emph {et~al.}(2020)\citenamefont {Denis}, \citenamefont {Hao}, \citenamefont {Eliav}, \citenamefont {Hutzler}, \citenamefont {Nayak}, \citenamefont {Timmermans},\ and\ \citenamefont {Borschesvky}}]{denis_nqmm}%
  \BibitemOpen
  \bibfield  {author} {\bibinfo {author} {\bibfnamefont {M.}~\bibnamefont {Denis}}, \bibinfo {author} {\bibfnamefont {Y.}~\bibnamefont {Hao}}, \bibinfo {author} {\bibfnamefont {E.}~\bibnamefont {Eliav}}, \bibinfo {author} {\bibfnamefont {N.~R.}\ \bibnamefont {Hutzler}}, \bibinfo {author} {\bibfnamefont {M.~K.}\ \bibnamefont {Nayak}}, \bibinfo {author} {\bibfnamefont {R.~G.~E.}\ \bibnamefont {Timmermans}}, \ and\ \bibinfo {author} {\bibfnamefont {A.}~\bibnamefont {Borschesvky}},\ }\href {\doibase 10.1063/1.5141065} {\bibfield  {journal} {\bibinfo  {journal} {J. Chem. Phys.}\ }\textbf {\bibinfo {volume} {152}},\ \bibinfo {pages} {084303} (\bibinfo {year} {2020})}\BibitemShut {NoStop}%
\bibitem [{\citenamefont {Fleig}\ and\ \citenamefont {Nayak}(2014)}]{FLEIG201416}%
  \BibitemOpen
  \bibfield  {author} {\bibinfo {author} {\bibfnamefont {T.}~\bibnamefont {Fleig}}\ and\ \bibinfo {author} {\bibfnamefont {M.~K.}\ \bibnamefont {Nayak}},\ }\href {\doibase https://doi.org/10.1016/j.jms.2014.03.017} {\bibfield  {journal} {\bibinfo  {journal} {J. Mol. Spectrosc.}\ }\textbf {\bibinfo {volume} {300}},\ \bibinfo {pages} {16} (\bibinfo {year} {2014})}\BibitemShut {NoStop}%
\bibitem [{\citenamefont {Wu}\ \emph {et~al.}(2022)\citenamefont {Wu}, \citenamefont {Zhou}, \citenamefont {Bao}, \citenamefont {Gagliardi},\ and\ \citenamefont {Truhlar}}]{wu2022zero}%
  \BibitemOpen
  \bibfield  {author} {\bibinfo {author} {\bibfnamefont {D.}~\bibnamefont {Wu}}, \bibinfo {author} {\bibfnamefont {C.}~\bibnamefont {Zhou}}, \bibinfo {author} {\bibfnamefont {J.~J.}\ \bibnamefont {Bao}}, \bibinfo {author} {\bibfnamefont {L.}~\bibnamefont {Gagliardi}}, \ and\ \bibinfo {author} {\bibfnamefont {D.~G.}\ \bibnamefont {Truhlar}},\ }\href {https://pubs.acs.org/doi/abs/10.1021/acs.jctc.1c01115} {\bibfield  {journal} {\bibinfo  {journal} {J. Chem. Theory Comput.}\ }\textbf {\bibinfo {volume} {18}},\ \bibinfo {pages} {2199} (\bibinfo {year} {2022})}\BibitemShut {NoStop}%
\end{thebibliography}


%%%%%%%%%%%%%%%%%%%%%%%%%%%%%%%%%%%%%%%%%%%%%%%%%%%%%%%%%%%%%%
%%%%%%%%%%%%%START OF SUPPLEMENTAL MATERIAL%%%%%%%%%%%%%%%%%%%
%%%%%%%%%%%%%%%%%%%%%%%%%%%%%%%%%%%%%%%%%%%%%%%%%%%%%%%%%%%%%%

\onecolumngrid

%%%%%%%%%%%%%%%%%%%%%%%%%%%%%%%%%%%%%%%%%%%%%%%%%%%%%%%%%%%%%%
%%%%%%%%%%%%%%%%%%%%%%%authors & affiliation%%%%%%%%%%%%%%%%%%
%%%%%%%%%%%%%%%%%%%%%%%%%%%%%%%%%%%%%%%%%%%%%%%%%%%%%%%%%%%%%%
\clearpage

\begin{center}
\textbf{Supplemental Material\\[4mm]
\large Ultracold high-spin $\Sigma$-state polar molecules for new physics searches}\\

\vspace{4mm}
A.~Ciamei,$^{\ddagger,1,2,*}$ 
A.~Koza,$^{\ddagger,3}$
M.~Gronowski,$^{3}$
M.~Tomza,$^{3,\dag}$\\
\vspace{2mm}
{\em \small
$^1$Istituto Nazionale di Ottica del Consiglio Nazionale delle Ricerche (INO-CNR), 50019 Sesto Fiorentino, Italy\\
$^2$\mbox{LENS and Dipartimento di Fisica e Astronomia, Universit\`{a} di Firenze, 50019 Sesto Fiorentino, Italy}\\
$^3$Faculty of Physics, University of Warsaw, Pasteura 5, 02-093 Warsaw, Poland\\}
{\small$^*$ E-mail: ciamei@lens.unifi.it}\\
{\small$^{\dag}$ E-mail: michal.tomza@fuw.edu.pl}

\def\thefootnote{$^\ddagger$}\footnotetext{These authors contributed equally to this work}\def\thefootnote{\arabic{footnote}}

\end{center}

%%%%%%%%%% Prefix a "S" to all equations, figures, tables and reset the counter %%%%%%%%%%
\setcounter{equation}{0}
\setcounter{figure}{0}
\setcounter{table}{0}
\setcounter{section}{0}
\setcounter{page}{1}
\makeatletter
\renewcommand{\theequation}{S.\arabic{equation}}
\renewcommand{\thefigure}{S\arabic{figure}}
\renewcommand{\thetable}{S\arabic{table}}
\renewcommand{\thesection}{S.\arabic{section}}
\newcommand{\vv}[1]{\boldsymbol{\mathbf{#1}}}

\begin{figure*}[t]
\centering
  \includegraphics[width=\textwidth]
  {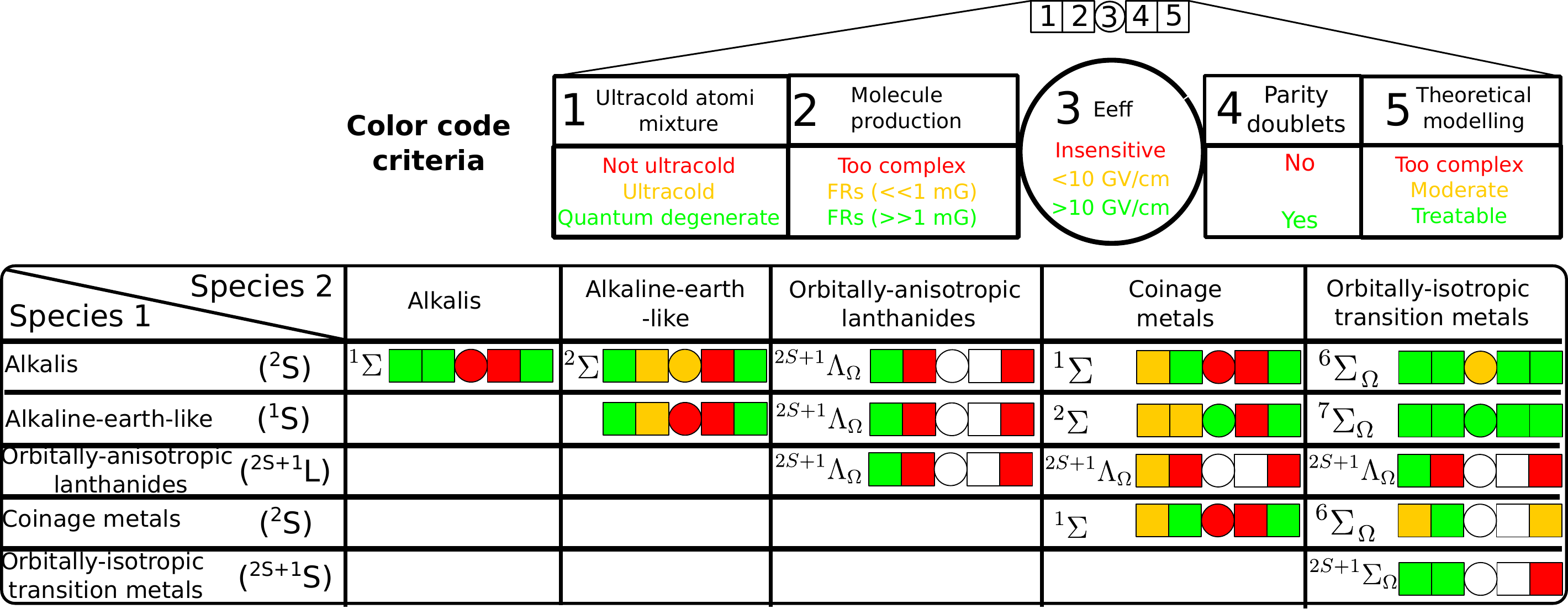}
  \caption{Overview of available combinations of laser-cooled atomic species, within which the best candidates are assessed according to requirements 1-5. The color code explained in the legend is used, and white if unknown. Mark 1 is based on Ref. \cite{Schreck2021}, mark 2 on Refs. \cite{RevModPhys.82.1225, Baroni2024}, mark 3 on Refs. \cite{PhysRevA.80.042508,Fleig_2021,PoletJCP24}, mark 4 follows from electronic angular momentum coupling and mark 5 on available literature \cite{PhysRevA.109.052814,Larsson2022,PRXQuantum.5.020358,TiesingaNJP21,GonzalezPRA15,KosickiNJP20,PhysRevA.88.012519}.} 
  \label{fgr:MolSpecies}
\end{figure*}

\twocolumngrid
%==============================================================================
\paragraph{Useful molecular species via ultracold atom assembly} --- It is instructive to classify all molecular species obtainable, in principle, combining all laser cooled atoms divided into isovalent classes as presented in Fig.~\ref{fgr:MolSpecies}. This table should be read as a feasibility check for eEDM searches via ultracold diatomics in their electronic ground state. We assign marks to all pairwise combinations on 5 different requirements:
\begin{enumerate}
    \item Currently achievable ultralow temperature/degeneracy of the parent atomic mixture. 
    \item Applicability of magneto-association to a weakly-bound molecular state amenable to quantum number assignment. 
    \item Strong internal electric field $E_{\mathrm{int}}$, which requires at least one heavy nucleus and one unpaired electron in an $s-p$ hybridized orbital. 
    
    \item High $|\langle\hat{\Sigma} \rangle|$ and rejection of systematic effects, offered by highly-polarizable, parity-doubled molecules. 
    
    \item  Accuracy of theoretical methods, which is crucial both for calculating $E_{\mathrm{int}}$ and for investigating specialized molecules, for which spectroscopy data is not available.
    
\end{enumerate}

We go over a few notable classes, while referring the interested reader to the table. Starting from the simplest choice, we can consider bi-alkali dimers, which count more than a dozen experiments worldwide: the ultracold atomic mixtures are easily realized, the molecule production process is highly efficient, and theory methods are available to guide the experiments. However, their ground-state electronic structure, corresponding to $^1\Sigma^+$ symmetry, results in no sensitivity to the eEDM. A potential improvement beyond standard bi-alkali species, albeit with relatively poor sensitivities, was identified in alkali--alkaline-earth compounds \cite{PhysRevA.80.042508}. Following theoretical predictions, magnetic Feshbach resonances (FRs) induced by novel coupling mechanisms were recently discovered in this class \cite{Barbé2018}, however their ultra-narrow character has so far hindered magneto-association. Diatomics isovalent to alkali--alkaline-earth compounds have been proposed for eEDM searches, notably YbAg and RaAg, which thanks to the high electron affinity of Ag feature a large $E_{\mathrm{eff}}$, as shown by the green circle for this combination \cite{Fleig_2021,PhysRevLett.125.153201,PhysRevA.99.040501}. However, this molecular class presents several challenges at the atomic cooling stage and will feature FRs of similar ultra-narrow character. Very recently, the experimental demonstration of efficient magneto-association of LiCr molecules has paved the way to the creation of ultracold polar paramagnetic molecules~\cite{PRXQuantum.5.020358}. Despite LiCr not being well-suited for eEDM searches because of low scores on criteria (3-4), this result shows the potential of transition metal elements in this context. Finally, the combination of orbitally-isotropic transition metal Cr with alkaline-earth-like, heavy Yb stands out as an ideal candidate for eEDM searches as it compares favorably against all criteria:
 \begin{enumerate}
    \item Both  Yb and Cr are well known, and large quantum gases are routinely produced \cite{PhysRevA.91.053405,PhysRevA.106.053318}.
    \item Efficient magneto-association is enabled by novel FRs with predicted magnetic-field widths similar to those of standard bi-alkali systems at convenient magnetic fields~\cite{PhysRevResearch.6.023254}. The implementation of STIRAP transfer to the rovibrational ground state appears feasible and comparable to previous results on bi-alkali systems.
    \item Heavy Yb leads to strong relativistic effects, and occupation of the YbCr bonding orbital by one of the $6 s^2$ valence electrons leaves one unpaired electron in an \emph{s-p} hybridized orbital centered on the Yb$^+$ core.
    \item YbCr features an extremely high electric polarizability and the possibility of spectroscopic inversion of eEDM shifts, thanks to the $\Omega$-like doubling in its electronic ground state.
    \item  Theory support is available and has already been exploited in the present proposal.
\end{enumerate}

\paragraph{Useful matrix elements} --- We report here as reference the relevant matrix elements employed in this work to perform calculations on the effective Hamiltonian in Hund's case (a) basis $|\Lambda=0,S=3,\Omega=\Sigma,J,M_J\rangle$, where $\Lambda$ and $S$ are dropped in the following.

The zero-field Hamiltonian $\mathcal{H}_0$, obtained by summing the rotational energy $B_0 (\mathbf{J}-\mathbf{S})^2$ to spin-rotation and spin-spin interactions, involves the following terms~\cite{Brown_Carrington_2003}:
\begin{itemize}
\item The rotational term:
\begin{multline}
\label{eq:RotTerm}
\langle \Sigma,J,M_J |B_0 T^1(\mathbf{J})\cdot T^1(\mathbf{J})| \Sigma',J',M_J' \rangle=\\B_0 J (J+1) \delta_{\Sigma,\Sigma'} \delta_{M_J,M_J'}.
\end{multline}

\item The Coriolis mixing and spin-rotation term:
\begin{multline}
\label{eq:SpinRot}
\langle \Sigma,J,M_J |(\gamma-2B_0)T^1(\mathbf{J})\cdot T^1(\mathbf{S})| \Sigma',J',M_J' \rangle=\\(\gamma-2B_0)\delta_{J,J'} \delta_{M_J,M_J'} \\ \sum_q (-1)^{J+S-2 \Sigma} (J(J+1)(2J+1)S(S+1)(2S+1))^{1/2} \times \\ 
\begin{pmatrix}
J & 1 & J\\
-\Sigma & q & \Sigma'
\end{pmatrix}\times
\begin{pmatrix}
S & 1 & S\\
-\Sigma & q & \Sigma'
\end{pmatrix}.
\end{multline}

\item The spin-spin term:
\begin{multline}
\label{eq:SpinSpin}
\langle \Sigma,J,M_J |(2\sqrt{6}/3)\lambda T^2_0(\mathbf{S})| \Sigma',J',M_J' \rangle=\\ \delta_{\Sigma,\Sigma'} \delta_{J,J'} \delta_{M_J,M_J'} \frac{2}{3}\lambda (3 \Sigma^2 - S(S+1)).
\end{multline}

\item The bias spin-multiplicity term:
\begin{multline}
\label{eq:SpinSpin}
\langle \Sigma,J,M_J |(B_0-\gamma_0)T^1(\mathbf{S})\cdot T^1(\mathbf{S})| \Sigma',J',M_J' \rangle=\\ (B_0-\gamma_0) \delta_{\Sigma,\Sigma'} \delta_{J,J'} \delta_{M_J,M_J'} S(S+1).
\end{multline}

\end{itemize}

The Stark Hamiltonian $\mathcal{H}_{\mathrm{St}}$ for vertically aligned electric field $T^1_0(\mathbf{E_{\mathrm{lab}}})=E_z$ and $T^1_{\pm 1}(\mathbf{E_{\mathrm{lab}}})=0$ contains:
\begin{multline}
\label{eq:Zeeman}
\langle \Sigma,J,M_J |-T^1(\mathbf{D})\cdot T^1(\mathbf{E_{\mathrm{lab}}})| \Sigma',J',M_J' \rangle=\\ -\mu_E E_Z (-1)^{M_J-\Sigma} ((2J+1)(2J'+1))^{1/2}  \\ \delta_{\Sigma,\Sigma'} \delta_{M_J,M_J'} \begin{pmatrix}
J & 1 & J'\\
-\Sigma & 0 & \Sigma'
\end{pmatrix} \begin{pmatrix}
J & 1 & J'\\
-M_J & 0 & M_J'
\end{pmatrix}.
\end{multline}

The Zeeman Hamiltonian $\mathcal{H}_{\mathrm{Z}}$ for vertically aligned magnetic field $T^1_0(\mathbf{B})=B_z$ and $T^1_{\pm 1}(\mathbf{B})=0$ contains:
\begin{multline}
\label{eq:Stark}
\langle \Sigma,J,M_J |-g_s \mu_{\mathrm{B}} T^1(\mathbf{S})\cdot T^1(\mathbf{B})| \Sigma',J',M_J' \rangle= -g_s \mu_{\mathrm{B}} B_z \\\sum_q (-1)^{S+M_J-2 \Sigma} (S(S+1)(2S+1))^{1/2} ((2J+1)(2J'+1))^{1/2}\\\delta_{M_J,M_J'} \begin{pmatrix}
S & 1 & S'\\
-\Sigma & q & \Sigma'
\end{pmatrix}
\begin{pmatrix}
J & 1 & J'\\
-\Sigma & q & \Sigma'
\end{pmatrix} \begin{pmatrix}
J & 1 & J'\\
-M_J & 0 & M_J'
\end{pmatrix}.
\end{multline}

\paragraph{Hund's case transition in YbCr} --- The ratio $\lambda / B$ between the spin-spin coupling parameter $\lambda$ and the rotational constant $B$, appearing in the effective Hamiltonian of Eq.\,(2) in the main text, is crucial as it dictates the angular momentum coupling scheme for diatomics in $^{2S+1}\Sigma$ electronic states with $S>1/2$. The transition from Hund's case (a) to case (b) in terms of $\lambda / B$ is well known, and it is an instructive exercise to look at Cr-bearing high-spin diatomics. For $\lambda/B \gg 1$, e.g., ground-state $X^4\Sigma^-$ CrN, the electron spin locks to the internuclear axis with projection $\Omega$, and leads to Hund's case (a) molecule for high $\Omega>1$ and low total angular momentum $J$ (we follow the literature in using the quantum number $\Omega=\Sigma$ with $\Lambda=0$). Hence, the zero-field eigenstates are labelled in terms of the quantum numbers $|\Omega,J,M_J,e(f)\rangle$, where $e(f)$ denotes $\pm 1$ parity according to spectroscopic notation. For $\lambda/B < 1$ instead, e.g., ground-state $X^6\Sigma^+$ CrH, the electronic spin is mostly decoupled from the internuclear axis, leading to a perturbative splitting of the underlying rotational series, with good basis $|J, N, M_J\rangle$. Generally speaking, heavy constituents tend to lead to lower $B$ and, concurrently, to larger $\lambda$, dominated by second-order spin-orbit coupling rather than direct spin-spin interaction for elements beyond first row of the periodic table. Hence, heavier species more often approximate Hund's case (a), compared to lighter species, which usually fall into case (b).

\begin{figure}[t]
\centering
\includegraphics[width=\columnwidth]{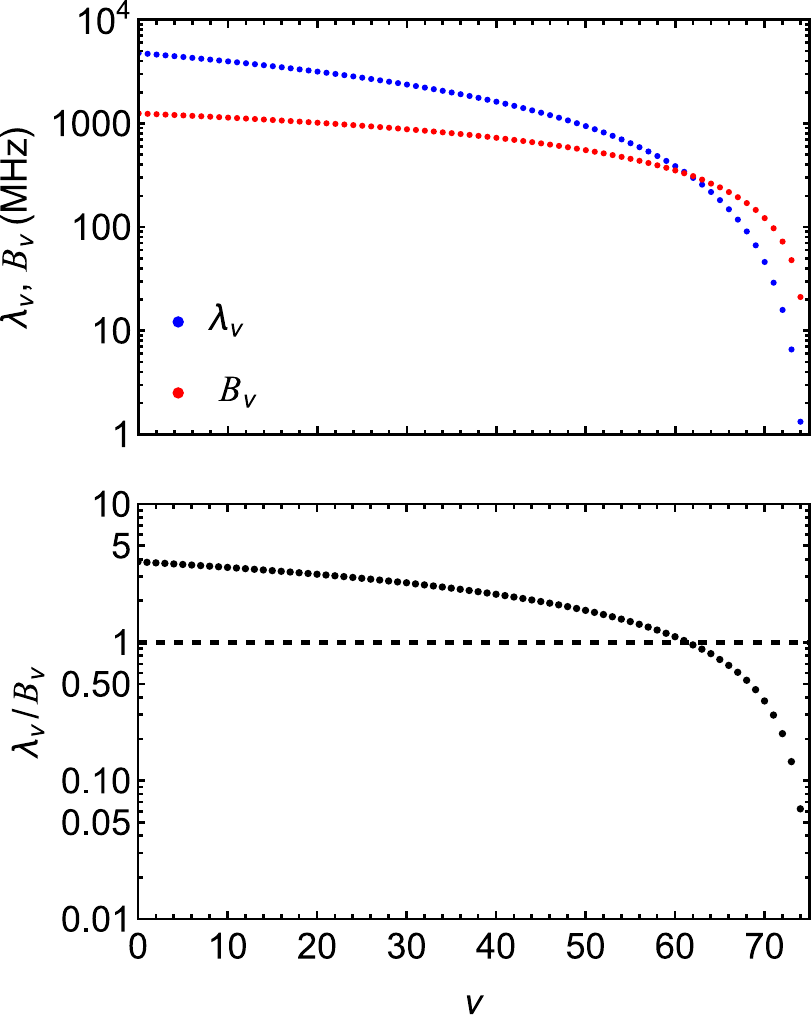}
%   \vskip -5 pt
  \caption{Hund's case transition in YbCr. Top panel: spin-spin coupling parameter (blue) and rotational constant (red) as a function of the vibrational quantum number $v$. Bottom panel: ratio between them as function of the vibrational quantum number $v$. Values much higher than unity (dashed line) tend to case (a) while points much lower than that tend to case (b).}
% \vskip -5 pt
  \label{fgr:SpinLocking}
\end{figure}

However, in a diatomic molecule, $\lambda$ and $B$ also depend on the vibrational level of interest for which the effective Hamiltonian is employed. Thus, in principle, a transition between cases (a) and (b) is possible, and indeed occurs for YbCr. This is mostly due to the different scaling of these two quantities, with $\lambda$ typically featuring a double exponential decay with the internuclear distance $r$ and $B$ a simple power law $\propto r^{-2}$, which suggests that Hund's case (a) is privileged in the vibrational ground state while case (b) in excited levels. In order to show this point we exploit the \emph{ab initio} $\lambda$ \cite{PhysRevResearch.6.023254} and the potential energy curve \cite{PhysRevA.88.012519} to extract $B$, see also additional, consistent \emph{ab initio} results below. The irreducible tensor representation $2 \lambda (r) T^2(\mathbf{S},\mathbf{S}) \cdot T^2(\mathbf{n},\mathbf{n})$ used in Ref.~\cite{PhysRevResearch.6.023254}, of which we consider the diagonal $q=0$ component, is identical to our effective Hamiltonian, with the same numerical value of $\lambda_0$, knowing that the average $\langle T^2_0 (\mathbf{n},\mathbf{n})\rangle=2/\sqrt{6}$ \cite{Brown_Carrington_2003}. We introduce the notation $\lambda_{v}=\langle v |\lambda (r)| v \rangle$ and $B_{v}=\langle v |1/(2\mu r^2)| v \rangle$ to make the dependence on the vibrational degree of freedom explicit. Fig.~\ref{fgr:SpinLocking} shows $\lambda_{v}$ and $B_{v}$ (top panel), as well as their ratio (bottom panel) for the ground-state YbCr as a function of the vibrational quantum number. As explained in the main text, Hund's case (a) becomes a good approximation towards the vibrational ground state, especially for high $\Omega$ and low $J$.

\begin{figure}[t]
\centering
  \includegraphics[width=\columnwidth]{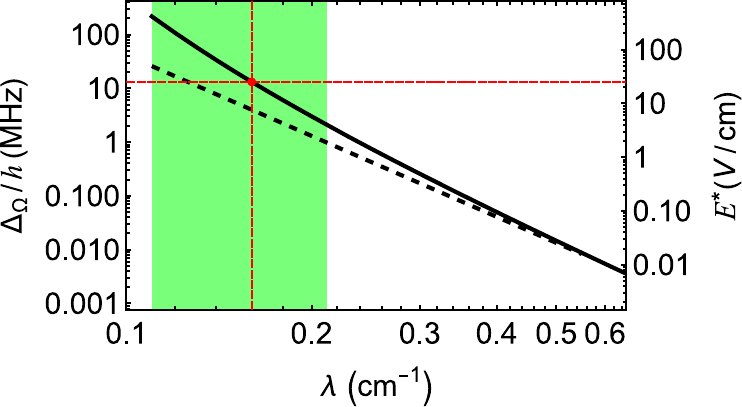}
%   \vskip -5 pt
  \caption{Effect of the variation of the spin-spin parameter $\lambda_0$ on the $\Omega$-doublet splitting $\Delta_{\Omega}$ and polarizing field $E^*$ for $J=|\Omega|=3$. The green shaded area marks the the confidence region predicted by our \emph{ab initio} methods, and the red circle is the mean expected value $\lambda_0=0.16\,\textrm{cm}^{-1}$. The solid black curves is the numerical simulation, while the black dashed line shows the asymptotic trend, well captured by perturbation theory fitted to large-$\lambda$ numerics.}
% \vskip -5 pt
  \label{fgr:PolarizabilityVsLambda}
\end{figure}

To which extent YbCr is a pure Hund's case (a) molecule in its vibrational ground state is determined by $\lambda_0 / B_0$. In particular, the splitting between parity doublets $\Delta_{\Omega}$ originates from Coriolis-mixing as a $2 |\Omega|$th-order effect in perturbation theory leading to $\Delta_{\Omega} / \lambda_0 \propto (B_0/\lambda_0)^{2 |\Omega|}$.  While $B_0$ is predicted with a \% -level relative uncertainty by \emph{ab initio} calculations, the calculation of $\lambda_0$ is challenging since it involves contributions from electronically excited states, see below. That is why the main uncertainty on the rotational spectrum of YbCr stems from $\lambda_0=0.16(5)\,\textrm{cm}^{-1}$. Within this range, and focusing on the relevant for $J=|\Omega|=3$ states, the proposed experiment, protocols and projected sensitivity are valid. In Fig.~\ref{fgr:PolarizabilityVsLambda}, we show the effect of the variation of $\lambda_0$ within our confidence region on the $\Delta_{\Omega}$ ($\Omega=3$) and on the required electric field $E^*$ for 50-\% polarization, i.e., $P=|\langle \hat{\Omega} \rangle|/3=0.5$.

\begin{figure}[t]
\centering
  \includegraphics[width=\columnwidth]{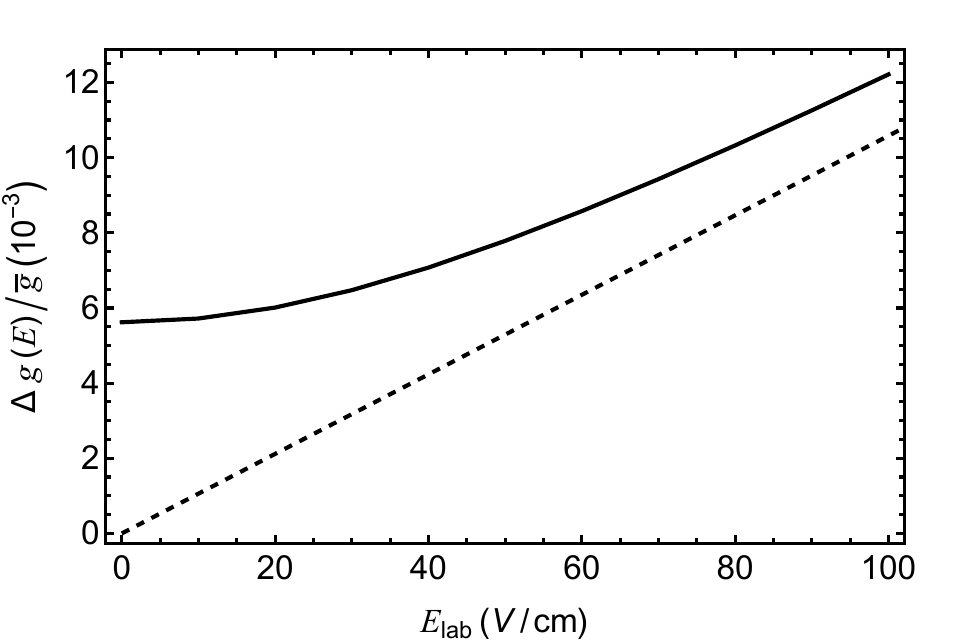}
%   \vskip -5 pt
  \caption{Relative difference between $g$-factors of top and bottom doublets $\Delta g/\bar{g}$ as a function of the applied electric field $E_{\mathrm{lab}}$: the solid line shows the fully numerical results ($\bar{g}=1.24$), the dashed line shows the effect of rotational mixing alone to first order from Eq. (\ref{eq:Deltag}) ($\bar{g}=1.5$).}
% \vskip -5 pt
  \label{fgr:Comagnetometer}
\end{figure}

Our model also allows to quantitatively assess the goodness of the internal co-magnetometer in YbCr (co-magnetometry on atomic Cr is instead straightforward). While ideally the top ($\tilde{e}$) and bottom ($\tilde{f}$) doublets feature identical $g$-factors, even in the case of negligible magnetic anisotropy, two effects will make the molecule slightly deviate from this condition. To quantify these effects we define $\bar{g}=(g_e + g_f)/2$ and $\Delta g=(g_e - g_f)$, where $g_{e}$ ($g_{f}$) is the g-factor of the $\tilde{e}$ ($\tilde{f}$) doublet. Firstly, even at zero electric field, Coriolis mixing between different $\Omega$ at fixed $J$, responsible for parity-doubling, leads to $\Delta g\neq0$ (on top of a correction on $\bar{g}\simeq1.24$). Moreover, application of the external electric field, induces a small but non-negligible rotational mixing between $J \leftrightarrow J+1$ levels at fixed $\Omega$. It can be shown that rotational mixing alone yields the following first-order correction for high-spin molecules: 
\begin{equation}
\label{eq:Deltag}
%\frac{\Delta g(E)}{2 \bar{g}}=\frac{D E_{\mathrm{lab}}}{B_0}\times\frac{J}{|M \Omega|}\times\frac{((J+1)^2-M^2)((J+1)^2-\Omega^2)}{(2J+1)(2J+3)(J+1)^2},
\frac{\Delta g(E)}{2 \bar{g}}=\frac{D E_{\mathrm{lab}}}{B_0}\times\frac{J ((J+1)^2-M^2)((J+1)^2-\Omega^2)}{|M| \Omega (2J+1)(2J+3)(J+1)^2},
\end{equation}
with $\bar{g}=g_{e,f}(0)=g_s \Omega^2/(J(J+1))$, which generalizes the result from Refs.~\cite{HamiltonPhD,PhysRevA.80.023418}. In Fig.~\ref{fgr:Comagnetometer}, we show $\Delta g(E)/\bar{g}$ from numerical calculations together with the analytic first-order correction due only to rotational mixing of Eq.\,(\ref{eq:Deltag}).

\paragraph{Dipolar interactions} --- Effects of intermolecular interactions should be addressed since, unlike for quantum metrology protocols, they should be avoided in standard spin-precession experiments \cite{Fitch_2021}. 
In this case we are sensitive to differential shifts within top ($|3,3,\pm1,\tilde{e}\rangle$) and bottom ($|3,3,\pm1,\tilde{f}\rangle$) doublets, induced by either electric or magnetic interactions. The  interaction energy between electric (E) and magnetic (M) dipoles reads:
\begin{equation}
\label{eq:DipolarInteraction}
V^\text{dd}_{\mathrm{E(M)}}=\alpha_{\mathrm{E(M)}}\frac{\boldsymbol{\mu}^1\cdot\boldsymbol{\mu}^2-3(\boldsymbol{\mu}^1\cdot\hat{R})(\boldsymbol{\mu}^2\cdot\hat{R})}{R^3},
\end{equation}
where the electric or magnetic dipoles $\boldsymbol{\mu}^{1(2)}=\boldsymbol{\mu}^{1(2)}_{\mathrm{E(M)}}$ are separated by the cartesian vector $\mathbf{R}=R \hat{R}$, and the constants are either $\alpha_E=(4 \pi \epsilon_0)^{-1}$ or $\alpha_M=(4 \pi)^{-1} \mu_0$, where $\epsilon_0$ is the vacuum permittivity and $\mu_0$ is the vacuum magnetic permeability. Reverting to the irreducible spherical tensor representation in the laboratory frame, we obtain the following expressions for the dipolar Hamiltonian \cite{Brown_Carrington_2003}:
\begin{equation}
\label{eq:DipoleInteraction}
\mathcal{H}^\text{dd}_{\mathrm{E(M)}} = - \alpha_{\mathrm{E(M)}} \sqrt{6} T^2(\boldsymbol{\mu}_1,\boldsymbol{\mu}_2)\cdot T^2(C),
\end{equation}
where $T^2(\boldsymbol{\mu}_1,\boldsymbol{\mu}_2)$ is the second-rank tensor composed of first-rank tensors $T^1_p(\boldsymbol{\mu}_{1(2)})$ and $T^2(C)$ denotes second-rank reduced spherical harmonics. The laboratory-frame $T^1_p(\boldsymbol{\mu}_{1(2)})$ are conveniently expressed in terms of molecule-frame matrix elements: 
\begin{equation}
\label{eq:LabFrameDipoles}
T^1_p(\boldsymbol{\mu}) = \sum_q\mathcal{D}^1_{p,q}(w)^* T^1_q (\boldsymbol{\mu}),
\end{equation}
with $\mathcal{D}^1_{p,q}(w)^*$ being the complex-conjugate of the $p,q$ element of the first-rank Wigner $\mathcal{D}$ rotation matrix.

We first consider the electric dipolar interaction between a pair of YbCr molecules:
\begin{equation}
\label{eq:ElectricDipoleInteraction}
\mathcal{H}^\text{dd}_{\mathrm{E}} = - \alpha_{\mathrm{E}} \sqrt{6} T^2(\mathbf{D_1},\mathbf{D_2})\cdot T^2(C),
\end{equation}
with molecule-frame matrix elements $\langle \Sigma' |T^1_0(\mathbf{D})|\Sigma\rangle = \delta_{\Sigma,\Sigma'} D$ and $T^1_{\pm1}(\mathbf{D})=0$. To first order, the differential shift within a doublet $\tilde{e}$ or $\tilde{f}$, can be evaluated considering the molecule-pair states $\{|M_1,M_2\rangle\}$, where $M_i=\pm 1$ refer to single molecule states. Diagonal matrix elements of $\mathcal{H}^\text{dd}_{\mathrm{E}}$ are identical, hence the differential shift within a doublet is zero to lowest order. This result is expected, since the laboratory-frame orientation of the molecules within a doublet is identical and the expectation value of $\mathcal{H}^\text{dd}_{\mathrm{E}}$ is simply the dipole-dipole interaction of two vertically-aligned dipoles $\boldsymbol{\mu}^{1(2)}_{\mathrm{E}}=P D |\Omega|(J(J+1))^{-1} \hat{z}$, where $P$ is the degree of polarization. To second order, we need instead to consider coupling to virtual states outside the basis spanned by the spin-precession doublets. We find that the dominant contributions come from within the subspace of the same "parity" $\tilde{e}$ (or $\tilde{f}$), e.g., $|+1,-1\rangle \leftrightarrow |0,0\rangle$ and $|+1,-1\rangle \leftrightarrow |\pm 2,0\rangle$, which are smaller than or comparable to single-molecule shot noise ($\lesssim 100~h\,$mHz) even for nearest-neighbor molecules in a 3D optical lattice with $\lambda=1064\,$nm.

We now move to the magnetic dipolar interaction dominated by the electron spins:
\begin{equation}
\label{eq:MagneticDipoleInteraction}
\mathcal{H}^\text{dd}_{\mathrm{M}}=- \alpha_{\mathrm{M}} g_s^2 \mu_{\mathrm{B}}^2 \sqrt{6}  T^2(\mathbf{S_1},\mathbf{S_2})\cdot T^2(C),
\end{equation}
with the relevant molecule-frame matrix elements being:
\begin{equation}
\label{eq:MagDIpoleMAtrixElements}
\langle \Sigma|T^1_q(\mathbf{S})|\Sigma' \rangle= (-1)^{S-\Sigma} \begin{pmatrix}
S & 1 & S\\
-\Sigma & q & \Sigma'
\end{pmatrix}
\sqrt{S(S+1)(2S+1)}.
\end{equation}
The differential shift originating from magnetic dipole-dipole is already well below the single-molecule shot noise to first order even for nearest-neighbor molecules in the same optical lattice.

\paragraph{Electron spin-rotation interaction} --- To model the electron spin-rotation interaction, we calculate the coupling constant $\varepsilon$, connected by the so-called Curl's relation with the electronic $g$-tensor \cite{Curl01011965}. For diatomic molecules, it takes a simple form:
\begin{equation}
    \varepsilon=2B_{0}\Delta g_{\perp},
\end{equation}
where $\Delta g_{\perp}$ is a difference between perpendicular component of the electronic $g$-tensor and the free electron $g$-factor. We first calculate the $g$-tensor at the equilibrium interatomic separation of the YbCr molecule at the DFT level with quasirestricted orbitals using the ORCA software \cite{neese_g10.1063/1.1419058}. We use the scalar x2c formalism to include relativistic effects and the all-electron x2c-QZVPPall basis set \cite{franzke2020segmented}. We obtain $\varepsilon=30.3\,$MHz and $\varepsilon=29.2\,$MHz, using the B3LYP and PBE0 functionals, respectively. The B3LYP value is used in our effective Hamiltonian.

\paragraph{Details on correlated relativistic four-component wave-function method} --- As stated in the Appendix, molecular spinors are generated with the DCHF method by averaging six valence electrons over twelve spinors. Subsequently, we split spinors into four generalized active spaces (GASs):
\begin{itemize}
    \item GAS1 -- $5p^{6}$ ytterbium orbitals,
    \item GAS2 -- $4f^{14}$ ytterbium orbitals,
    \item GAS3 -- $6s^2$ ytterbium and $3d^54s^1$ chromium orbitals,
    \item GAS4 -- virtual orbitals of ytterbium and chromium.
\end{itemize}
Next, we perform relativistic MRCI calculations correlating twenty-two electrons MRCISD(22), where the number in the bracket gives the number of correlated electrons. This model is obtained by keeping GAS1 closed (frozen) and allowing for single excitation from GAS2 and single and double excitations from GAS3. We also perform a test with the MRCISD(28) model where we additionally allow for single excitations from GAS1. For both computational models, we test three values of virtual orbitals cutoff in GAS4: 2.5 hartree, 5 hartree and 8 hartree. For each calculation, we also test the basis set dependence. Additionally, we analyze the contribution from triple excitations to the molecular properties within the MRCISDT method by correlating electrons occupying GAS3 only. All calculations accounting for triple excitations are performed using a Dyall-v3z basis set \cite{{dyall2012core}} and a virtual cutoff of 5 hartree. In this way, we evaluate the expectation values for the eEDM, eN-SPS, and nuclear magnetic quadrupole moment (NMQM) enhancement factors, and hyperfine constants (HFS) using these computational models. Numerical results are presented in the respective paragraphs below.

\paragraph{eEDM senstivity} --- To analyze the sensitivity of the enhancement factors to the electron correlation treatment and basis set cardinality, we carry out additional calculations in which both the correlation model and the basis set cardinality are systematically varied. In Table \ref{tab:eedm_virtuals_table}, we examine a potential uncertainty in $W_{\mathrm{d}}$ and $W_{\mathrm{s}}$ coming from the incompleteness of virtual active space. One can notice a relatively negligible dependence of the final results on the number of virtual orbitals included in the correlation treatment. Due to the substantial computational cost, we are unable to perform calculations with a larger virtual space. However, the final values of both factors exhibit good convergence, and they should not change much with a larger number of correlated virtual spinors. In Table \ref{tab:eEDM_electrons_table}, we check the impact of the number of explicitly correlated electrons, basis set size and excitation level in MRCI on $W_{\mathrm{d}}$ and $W_{\mathrm{s}}$. Including $5p$ orbitals of ytterbium within MRCISD(28) causes a slight increase in the value. Moreover, applying a larger basis set also enhances the sensitivity of the YbCr molecule to eEDM effects. Finally, the inclusion of triple excitations significantly increases its sensitivity.

The final recommended values of $W_{\textrm{d}}=1.17\times 10^{24}\, h\text{ Hz}/({e\text{ cm}})$  and $W_{s}=6.00$ {}$h$ kHz{} are obtained as a sum of the value calculated with the MRCISD(28) model with the Dyall-cv3z basis set and the correction for triple excitations estimated as the difference between the values from the MRCISDT(8) and MRCISD(8) approaches with the Dyall-v3z basis set. 

\begin{table}[tb]
  \centering
  \caption{Influence of a virtual active space size on $W_{\mathrm{d}}$ and $W_{\mathrm{s}}$ for the YbCr molecule at the equilibrium geometry. The MRCISD(28) model with the Dyall-cv3z basis set is used.}
  \label{tab:eedm_virtuals_table}
    \begin{ruledtabular}
      \begin{tabular}{ccc}
        Virtual cutoff  & $W_{\mathrm{d}}\,$($10^{24} \frac{h\text{ Hz}}{e\text{ cm}}$)  & $W_{\mathrm{s}}\,$($h$ kHz)  \\ 
        \hline
         8.0 $E_\text{h}$   & 0.82 &  4.18  \\ 
         5.0 $E_\text{h}$   & 0.82 &  4.20  \\ 
         2.5 $E_\text{h}$   & 0.83 &  4.25  \\ 
       \end{tabular}
    \end{ruledtabular}%
 % }
\end{table}

\begin{table}[tb!]
  \centering
  \caption{Influence of a number of correlated electrons, basis set size and excitation level in MRCI on $W_{\mathrm{d}}$ and $W_{\mathrm{s}}$ for the YbCr molecule at the equilibrium geometry. The virtual cutoff is set to 8~$E_\text{h}$.
   }
  \label{tab:eEDM_electrons_table}
    \begin{ruledtabular}
      \begin{tabular}{ccc}
        Method, basis  & $W_{\mathrm{d}}\,$($10^{24} \frac{h\text{ Hz}}{e\text{ cm}}$)  & $W_{\mathrm{s}}\,$  ($h$ kHz)   \\ 
        \hline
          MRCISD(28), cv3z   &  0.82   &  4.18  \\
          MRCISD(28), cv2z   &  0.78  &   3.79  \\
          MRCISD(22), cv3z   &  0.79 &  4.06  \\
          MRCISD(22), cv2z   &  0.75   &  3.69  \\
          MRCISDT(8), v3z   &  1.17   &  6.00  \\
          MRCISD(8), v3z   &  0.82   &  4.18  \\
      \end{tabular}
    \end{ruledtabular}%
 % }
\end{table}

\paragraph{Hadronic CP-violation} --- The YbCr molecule can also be an interesting candidate for studying CP violation in the hadronic sector. Here, symmetry-violating effects can arise from  P,T-odd electromagnetic moments in nuclei. One is the nuclear magnetic quadrupole moment, which has a non-zero value for systems with a nuclear spin $I>1/2$. The NMQM couples to the magnetic field gradient produced by unpaired electrons, leading to the experimentally observable shifts in spectra. The sensitivity of potential experiments can be significantly enhanced by using molecules containing deformed nuclei due to the collective effects \cite{FLAMBAUM1994211}. Notably, among stable ytterbium isotopes, $\,^{173}\mathrm{Yb}$ (with $I=5/2$) exhibits significant deformation, making $\,^{173}\mathrm{YbCr}$ a potentially attractive candidate for molecular NMQM searches. Theoretically, the NMQM can be modelled by the following Hamiltonian \cite{skripnikovNQMM}:
 \begin{equation}
    \mathcal{H}_{\mathrm{NQMM}}=-\frac{M}{2I(2I-1)}T_{jk} \sum_i^{N_{\mathrm{e}}}\frac{3}{2}\frac{[{\vv\alpha}_i \times \vv{r}_i]_j}{r_i^5}[r_i]_k,
    \label{eq:H_M}
\end{equation}
where $M$ is the NMQM, $T_{jk}$ are the components of the second-rank tensor $T_{jk} = I_jI_k + I_kI_j - \frac{2}{3}I(I+1)\delta_{jk}$ with $I$ the nuclear spin, ${\vv\alpha_i}$ are the Dirac matrices and $\vv r$ is the position of an electron with respect to the nucleus. For diatomic molecules, this Hamiltonian can be simplified and written in the following way:
\begin{equation}\label{eq:H_diatom}
        \mathcal{H}_{\text{eff-NQMM}}=-\frac{M W_{\mathrm{M}}}{2 I(2 I-1)} \vv{J} \cdot{T} \cdot{\hat{\vv{n}}} 
\end{equation}
where $\vv{J}$, $I$, and $\vv{\hat{n}}$ are the total electronic angular momentum, the nuclear spin and the unit vector along the internuclear axis, respectively. $W_{\mathrm{M}}$ is the molecular enhancement factor that needs to be calculated theoretically. In the present work, we evaluate the following matrix element to obtain a numerical value of $W_{\mathrm{M}}$ for Yb in the YbCr molecule:   
\begin{equation}
        W_{\mathrm{M}}=\frac{3}{2\Omega}\left\langle \Psi_{^7\Sigma_{\Omega}} \left|
\sum_i^{N_{\mathrm{e}}}\left(\frac{\vv{\alpha}_i \times \vv{r}_i}{r_i^5}\right)_z r_z
      \right|  \Psi_{^7\Sigma_{\Omega}} \right\rangle  ,
\end{equation}
where $\Omega$ is the projection of the total electronic angular momentum $\vv{J}$ on the molecular axis, and $z$ means projection on the molecular axis. 

In Table \ref{tab:NQMM_virtuals_table}, we report calculated enhancement factors with the KRCI method for the Yb atom in the YbCr molecule. The $W_{\mathrm{M}}$ parameter reveals a weak sensitivity to the size of virtual active space. 
As with the eEDM factors, the value of molecular enhancement for NMQM slightly increases with the number of correlated electrons and basis-set cardinality, as detailed in Table \ref{tab:NQMM_electrons_table}.

\begin{table}[tb]
  \centering
  \caption{Influence of a virtual active space size on $W_{\mathrm{M}}$ for the YbCr molecule at the equilibrium geometry. The MRCISD(28) model with the Dyall-cv3z basis set is used.
   }
  \label{tab:NQMM_virtuals_table}
    \begin{ruledtabular}
      \begin{tabular}{cc}
        Virtual cutoff  & $W_{\mathrm{M}}\,(10^{31} \frac{h\text{ Hz}}{e \text{ cm}^2})$ \\ 
        \hline
         8.0 $E_\text{h}$  & 7.23  \\
         5.0 $E_\text{h}$   & 7.16  \\
         2.5 $E_\text{h}$   & 7.24  \\
   
      \end{tabular}
    \end{ruledtabular}%
 % }
\end{table}

\begin{table}[tb!]
  \centering
  \caption{Influence of a number of correlated electrons, basis set size and excitation level in MRCI on $W_{\mathrm{M}}$ for the YbCr molecule at the equilibrium geometry.  The virtual cutoff is set to 8~$E_\text{h}$. %The MR-CISD(28) model with Dyall-cv3z was used.
   }
  \label{tab:NQMM_electrons_table}
    \begin{ruledtabular}
      \begin{tabular}{cc}
        Method, basis  & $W_{\mathrm{M}\,}(10^{31} \frac{h\text{ Hz}}{e \text{ cm}^2})$ \\ 
        \hline
          MRCISD(28), cv3z   & 7.23  \\
          MRCISD(28), cv2z   & 6.85  \\
          MRCISD(22), cv3z   & 6.88  \\
          MRCISD(22), cv2z   & 6.54  \\
          MRCISDT(8), v3z    & 9.29  \\
          MRCISD(8),  v3z    & 7.04  \\
      \end{tabular}
    \end{ruledtabular}%
 % }
\end{table}

The final value of  $W_\mathrm{M}=9.48 \times 10^{31}\,{h\text{ Hz}}/(e \text{ cm}^2)$ is obtained in the same way as the eEDM factors by summing different contributions. While $W_{\mathrm{M}}$ for YbCr is a bit smaller than that for other molecules containing Yb, such as YbF or YbOH~\cite{denis_nqmm}, this limitation can be potentially compensated by greater quantum control over ultracold YbCr molecules trapped in optical lattices. A more detailed analysis of high-spin $\Sigma$-state diatomics for CP violation in the hadronic sector will be the subject of our future research.

\paragraph{Hyperfine structure constants} --- We also calculate hyperfine structure (HFS) constants  for $^{173}$Yb and $^{53}$Cr nuclei, with spin $5/2$ and $3/2$, respectively, within a ground-state YbCr molecule. To accomplish this, we evaluate the expectation value of the operator given by:
\begin{equation}
        A_{\parallel}= \frac{\mu_{K}}{I \Omega 2c m_{p}} \left\langle\Psi_{^7\Sigma_{\Omega}} \left|
\sum_i^{N_{\mathrm{e}}}\left(\frac{\vv{\alpha}_i \times \vv{r}_i}{r_i^3}\right)_z \right|
        \Psi_{^7\Sigma_{\Omega}} \right\rangle ,
\label{EQ:hyperPARA}
\end{equation}
where  $\mu_{K}$ is the magnetic moment of nucleus $K$, $c$ is the speed of light and $m_{p}$ is the proton mass~\cite{FLEIG201416}. Table \ref{tab:hsc_virtuals_table} shows that an increasing number of correlated virtual spinors results in opposite trends for both atomic centers. The HFS constant decreases for ytterbium, whereas it increases for chromium. In both cases, however, the change remains below 5\%. 
Table~\ref{tab:hsc_electrons_table} shows that the effect of the number of correlated electrons parallels those seen for virtual‐space size: increasing the number of correlated electrons decreases the HFS constant for Yb while increasing it for Cr. Additionally, one can see a minor effect of the basis set cardinality in the case of chromium atoms.
Notably, the inclusion of triple excitations within MRCISDT(8) significantly impacts the HFS constant for Yb, whereas the effect for Cr is minor. 

The final recommended values obtained in the same way as for P,T-odd enhancement factors are $A_{\parallel}=69.5$~MHz for $^{173}$Yb and $A_{\parallel}=104.8$~MHz for $^{53}$Cr.

\begin{table}[tb]
  \centering
  \caption{Influence of a virtual active space size on the hyperfine structure constants $A_{\parallel}\,$(in MHz) for $^{173}\text{Yb}^{53}\text{Cr}$ at the equilibrium geometry. 
  %We used $-0.6483\mu_{\mathrm{N}}$ for Yb and $-0.47454\mu_{\mathrm{N}}$ for Cr. 
  The MRCISD(28) model with the Dyall-cv3z basis set is used.
   }
  \label{tab:hsc_virtuals_table}
    \begin{ruledtabular}
      \begin{tabular}{ccc}
        Virtual cutoff  &  $^{173}\text{Yb}$ & $^{53}\text{Cr}$ \\ 
        \hline
         8.0 $E_\text{h}$   & 49.32  & 108.07 \\
         5.0 $E_\text{h}$   & 49.58  & 107.41 \\
         2.5 $E_\text{h}$   & 52.27  & 104.63 \\
   
      \end{tabular}
    \end{ruledtabular}%
 % }
\end{table}

\begin{table}[tb]
  \centering
  \caption{Influence of a number of correlated electrons, basis set size and  excitation level in MRCI on the hyperfine structure constants $A_{\parallel}\,$(in MHz) for $^{173}\text{Yb}^{53}\text{Cr}$ at equilibrium geometry. The virtual cutoff is set to 8 $E_\text{h}$.}
  \label{tab:hsc_electrons_table}    \begin{ruledtabular}
      \begin{tabular}{ccc}
        Method, basis  &  $^{173}\text{Yb}$ & $^{53}\text{Cr}$ \\ 
        \hline
         MRCISD(28), cv3z   & 49.32   & 108.07 \\
         MRCISD(28), cv2z   & 44.94   & 107.28 \\
         MRCISD(22), cv3z   & 53.34   & 104.79 \\
         MRCISD(22), cv2z   & 48.48   & 104.39 \\
         MRCISDT(8), v3z   & 78.53   & 97.08 \\
         MRCISD(8), v3z   & 58.33   & 101.07 \\
      \end{tabular}
    \end{ruledtabular}%
 % }
\end{table}

\paragraph{Zero-field splitting} --- We find determining the zero-field splitting (ZFS) parameter for ground-state YbCr to be very challenging due to its high sensitivity to the details of the calculations. To address this as precisely as possible, we employ several methodologies with results summarized in Table \ref{tab:zfs_table}. First, we can notice a good agreement between the four-component relativistic MRCISD computational scheme (based on the splitting between $\Omega$ states) and the Schrödinger equation-based weighted-state-averaged complete active space self-consistent field approach (WSA-CASSCF)~\cite{wu2022zero} (based on the sum over states), resulting in a $D$ value of 0.22 $\textrm{cm}^{-1}$. However, this value is around 30\% smaller than the previously reported DFT-based $D=0.32$~$\textrm{cm}^{-1}$~\cite{PhysRevResearch.6.023254}. In our weighted-state averaged calculations, the $^7\Pi$ excited state dominates the second-order spin-orbit-coupling contribution to the ZFS. We also verify that the value of $D$ is relatively insensitive to the weight of the ground $^7\Sigma$ state in the final CASSCF wavefunction. Additionally, we found that the triple excitations (MRCISDT) lead to a further decrease of $D$. 

The importance of the contribution of the excited $^7\Pi$ state to the sum-over-states WSA-CASSCF value motivated us to extend the reference spinors in the MRCISD calculations. So far, our molecular wavefunction has been optimized to describe the ground electronic state of the YbCr molecule adequately (6 electrons distributed over 12 spinors (6in12) in DCHF). To account for the excited states, we perform MRCISD calculations based on the DCHF reference with 8 electrons distributed over 20 spinors (8in20) composed of $6s6p$ orbitals of ytterbium and the $3d4s$ orbitals of chromium. In this way, we obtain $D=0.38\,\textrm{cm}^{-1}$. 

The final recommended values of $D=0.32\,\textrm{cm}^{-1}$ is obtained as a sum of the leading MRCISD(28) contribution, corrected for triple excitations from the difference between the MRCISDT(8) and MRCISD(8) values and corrected for the reference spinor quality from the difference between the MRCISD(8) values with the 8in20 and 6in12 DCHF references.

\begin{table}[tb]
  \centering
  \caption{Zero-field splitting parameter calculated using different computational approaches, with the spin-spin contribution neglected.}
  \label{tab:zfs_table}
    \begin{ruledtabular}
      \begin{tabular}{cc}
        Method  &  $D\,$(cm$^{-1}$)  \\ 
        \hline
        MRCISD(28), 8 $E_\text{h}$   & 0.22 \\
        MRCISD(28), 5 $E_\text{h}$   & 0.23 \\
       MRCISD(28), 2.5 $E_\text{h}$  & 0.23 \\
        MRCISDT(8), 5 $E_\text{h}$  & 0.17 \\
        MRCISD(8), 5 $E_\text{h}$  & 0.22 \\
         MRCISD(8), 8in20, 5 $E_\text{h}$  & 0.38 \\
   % WSA-CASSCF, 100\% & Wartość 8 \\
        WSA-CASSCF, 90\% &  0.23 \\
        WSA-CASSCF, 50\% &  0.21 \\
        WSA-CASSCF, 25\% &  0.21  \\
        
      %  Wartość 15 & Wartość 16 \\
      \end{tabular}
    \end{ruledtabular}%
 % }
\end{table}

\end{document}